\documentclass[pdftex,twocolumn,epjc3]{svjour3}          

\RequirePackage[T1]{fontenc}

\smartqed

\RequirePackage{graphicx}
\RequirePackage{flushend}
\RequirePackage{amsfonts}
\RequirePackage{amssymb}
\RequirePackage[numbers,sort&compress]{natbib}

\usepackage{graphics}
\usepackage{subfig}
\usepackage{epstopdf}
\usepackage{pstricks}
\usepackage[T1]{fontenc}
\usepackage{rotating}
\usepackage{color}
\usepackage{amsmath}
\usepackage{amstext}
\usepackage{amsfonts}
\usepackage{amssymb}
\usepackage{mathtools}
\usepackage{latexsym}
\usepackage{epsfig,graphics}
\usepackage{bm}
\usepackage{graphicx}
\usepackage{epsf}
\usepackage{multirow}
\usepackage{dcolumn}     
\usepackage{slashed}
\usepackage{float}
\usepackage{appendix}

\newcommand{\MSb}{\overline{\textrm{MS}}}

\def\PM{\mathbb{P}_M}
\def\dirac#1{\gamma_{#1}}
\def\Tr{{\rm Tr}}
\newcommand{\be}{\begin{equation}}
\newcommand{\ee}{\end{equation}}

\newcommand{\deter}{\rm det}
\newcommand{\idnty}{\hbox{1$\!\!$1}}

\def\eq#1{Eq.~(\ref{#1})}
\def\fig#1{Fig. \ref{#1}}
\def\tbl#1{Table \ref{#1}}
\def\sec#1{Section \ref{#1}}

\newcommand{\bea}{\begin{eqnarray}}
\newcommand{\eea}{\end{eqnarray}}

\journalname{Eur. Phys. J. C}

\begin{document}

\title{Comparison of topological charge definitions in Lattice QCD}

\author{Constantia Alexandrou\thanksref{addr1,addr2} \and Andreas Athenodorou \thanksref{e1,addr2,addr1}
 \and \mbox{Krzysztof Cichy}\thanksref{e2,addr4} \and Arthur Dromard\thanksref{addr5} \and Elena Garcia-Ramos\thanksref{addr6,addr7} \and Karl Jansen\thanksref{addr6} \and Urs Wenger\thanksref{addr8} \and Falk Zimmermann\thanksref{addr9}
}                     
\institute{Department of Physics, University of Cyprus, P.O. Box 20537, 1678 Nicosia, Cyprus. \label{addr1}
  \and
  Computation-based Science and Technology Research Research, The Cyprus Institute, 20 Kavafi Str., Nicosia 2121, Cyprus. \label{addr2}
  \and
  Adam Mickiewicz University, Faculty of Physics, Umultowska 85, 61-614 Pozna\'n, Poland. \label{addr4}
  \and
  Institute for Theoretical Physics, Regensburg University, D-93053 Regensburg, Germany. \label{addr5}
  \and
  NIC, DESY, Platanenallee 6, 15738 Zeuthen, Germany. \label{addr6}
  \and
  Humboldt Universit\"at zu Berlin, Newtonstr. 15, 12489 Berlin, Germany. \label{addr7}
  \and
  Universit\"at Bern, Albert Einstein Center for Fundamental Physics, Institute for
  Theoretical Physics, Sidlerstrasse 5, 3012 Bern, Switzerland. \label{addr8}
  \and
  Helmholtz-Institut f\"ur Strahlen- und Kernphysik (Theorie) and Bethe Center for
  Theoretical Physics, Universit\"at Bonn, 53115 Bonn, Germany. \label{addr9}
}
\thankstext{e1}{e-mail:  andreas.athinodorou@pi.infn.it}
\thankstext{e2}{e-mail:  kcichy@amu.edu.pl}

\date{Received: date / Accepted: date}

\maketitle
\begin{abstract}
In this paper, we show a comparison of different definitions of the topological charge on the lattice. 
We concentrate on one small-volume ensemble with 2 flavours of dynamical, maximally twisted mass fermions and use three more ensembles to analyze the approach to the continuum limit. 
We investigate several fermionic and gluonic definitions.
The former include the index of the overlap Dirac operator, the spectral flow of the Wilson--Dirac operator and the spectral projectors.
For the latter, we take into account different discretizations of the topological charge operator and various smoothing schemes to filter out ultraviolet fluctuations: the gradient flow, stout smearing, APE smearing, HYP smearing and cooling.
We show that it is possible to perturbatively match different smoothing schemes and provide a well-defined smoothing scale. 
We relate the smoothing parameters for cooling, stout and APE smearing  to the gradient flow time $\tau$.  In the case of hypercubic smearing the matching is performed numerically. 
We investigate which conditions have to be met to obtain a valid definition of the topological charge and susceptibility and we argue that all valid definitions are highly correlated and allow good control over topology on the lattice. 
\end{abstract}

\section{Introduction}
\label{intro}
QCD gauge fields can have non-trivial topological properties, manifested in non-zero and integer values
of the so-called topological charge. Such topological properties are believed to have important
phenomenological implications, e.g. the fluctuations of the topological charge are related to the mass
of the flavour-singlet pseudoscalar $\eta'$ meson \cite{Witten:1979vv,Veneziano:1979ec}.
The topology of QCD gauge fields is a fully non-perturbative issue, hence lattice methods are
well-suited to investigate it. Naively, lattice gauge fields are topologically
trivial, since they can always be continuously deformed to the unit gauge field.
However, it can be shown that for smooth enough gauge fields (sufficiently close to the continuum
limit), the notion of topology of lattice gauge fields is still meaningful \cite{Luscher:1981zq}.
Historically, the first investigation aimed at studying the topological properties of a non-Abelian gauge theory was reported in Refs.~\cite{DiVecchia:1981aev,DiVecchia:1981hh} for the SU(2) gauge group case and then extended to SU(3) in Ref.~\cite{Fabricius:1983nj}.

Over the years, many definitions of the topological charge of a lattice gauge field were proposed \cite{Cichy:2014qta,Muller-Preussker:2015daa}. It is
clear that the definitions differ in terms of their computational cost and convenience, but also theoretical appeal.
These definitions can be characterized either as fermionic or gluonic. 
During the last decade, a number of efforts 
have revealed important aspects of the topological susceptibility, which reflects the fluctuations of the topological charge. Universality of the topological susceptibility in fermionic
definitions has been demonstrated \cite{Luscher:2010ik}, giving an insight in the basic properties a definition has to obey 
so that the topological susceptibility is free of short-distance singularities, which have plagued some of the earlier attempts at a proper and computationally affordable fermionic definition \cite{Giusti:2004qd,Luscher:2004fu}. On the other hand, considering 
the gluonic definition, a theoretically clean understanding on how the topological sectors emerge in the 
continuum limit has been attained through the gradient flow \cite{Narayanan:2006rf,Luscher:2010iy,Luscher:2011bx,Lohmayer:2011si,Luscher:2013vga}. Further projects have also divulged the numerical equivalence
of the gradient flow with the smoothing technique of cooling at finite lattice spacing. Previous investigations, on the other hand, have shown numerically
that the field theoretic topological susceptibility extracted with several smoothing techniques such as cooling, APE and HYP smearing
give the same continuum limit. Although no solid theoretical argument suggests so, it is believed that all different definitions of
the topological charge  agree in the continuum limit.
Good agreement has also been found between the gluonic definiton and a fermionic one in finite-temperature studies \cite{Taniguchi:2016tjc}.
For an overview of topology-related issues on the lattice, we refer to the review paper by M\"uller-Preussker \cite{Muller-Preussker:2015daa}.

This aim of the paper is two-fold. The first purpose is to investigate the perturbative equivalence between the smoothing schemes of the gradient flow, cooling as well as APE, stout and HYP smearing. In Refs.~\cite{Bonati:2014tqa,Alexandrou:2015yba} it was demonstrated that gradient flow and cooling are equivalent if the gradient flow time $\tau$ and the number of cooling steps $n_c$ are appropriately matched. By expanding the gauge links perturbatively in the lattice spacing $a$, at subleading order, the two methods become equivalent if one sets $\tau =n_c/(3-15b_1)$ where $b_1$ is the Symanzik coefficient multiplying the rectangular term of the smoothing action. It is, thus, interesting to extend the study of Ref.~\cite{Bonati:2014tqa} and explore whether similar matching can also be derived for APE, stout and HYP smearings. Hence, we carry out such investigation and support it with the appropriate numerical results.

The second motivation of this paper is to attempt a systematic investigation of different topological charge definitions. We
have computed them on selected ensembles generated by the European Twisted Mass Collaboration
(ETMC)\footnote{In 2019, the name of the collaboration has been changed to Extended Twisted Mass Collaboration.}. The included definitions are:\footnote{For references on each of the following definitions, we refer to Sec.~\ref{sec:Field_Theoretic_Definition}.}
\begin{itemize}
\item index of the overlap Dirac operator on HYP-smeared and non-HYP-smeared configurations,
\item Wilson-Dirac operator spectral flow (SF),
\item spectral projector definition,
\item field theoretic (gluonic) definition, with gauge fields smoothed using:
\begin{itemize}
 \item {\it gradient flow} (GF)  with different smoothing actions and at different flow times. Namely, we smooth the gauge fields using the Wilson plaquette, Symanzik tree-level and Iwasaki actions at flow times $t_0$, $2t_0$ and $3t_0$; $t_0$ is defined in \sec{sec:Gradient_Flow}. 
 \item {\it cooling} (cool) with the three different smoothing actions and cooling steps matched to GF time for $t_0$, $2t_0$ and $3t_0$,
 \item {\it stout smearing} with three different values of the stout parameter $\rho_{\rm st}$ and smearing steps matched to GF time at flow times $t_0$, $2t_0$ and $3t_0$,
 \item {\it APE smearing} with three different values of the parameter $\alpha_{\rm APE}$ and smearing steps matched to GF time for $t_0$, $2t_0$ and $3t_0$,
 \item {\it HYP smearing} for a given set of parameters $\alpha_{\rm HYP1}$, $\alpha_{\rm HYP2}$, $\alpha_{\rm HYP3}$, and smearing steps numerically matched to GF time at flow times $t_0$, $2t_0$ and $3t_0$.
\end{itemize}
\end{itemize}

The outline of the paper is as follows: \sec{sec:Lattice_Setup} describes our lattice setup as well as the relevant details regarding the production of the $N_f=2$ configurations. \sec{sec:definitions_of_the_topological_charge} introduces the topological charge definitions that we are using and includes the derivation of matching conditions between different smoothing schemes. In \sec{sec:results}, we discuss and compare different definitions of the topological charge, we analyze the approach to the continuum limit and we show results for the topological susceptibility. Finally, in \sec{sec:conclusions} we summarize and conclude.
\section{Lattice setup}
\label{sec:Lattice_Setup}
Motivated by the desire to cover as many definitions of the topological charge as possible, including the costly overlap definition, we performed our comparison of different topological charge definitions on small volume ensembles (to keep the computational cost affordable and at the same time incorporate all definitions) generated by ETMC with $N_f=2$
\cite{Boucaud:2007uk,Boucaud:2008xu,Baron:2009wt}
dynamical twisted mass fermions. The action in the gauge sector is
\begin{eqnarray}
 S_G[U] = \frac{\beta}{3}\sum_x\Big( & b_0 \sum_{\substack{
      \mu,\nu=1\\1\leq\mu<\nu}}^4 \textrm{Re\,Tr} \big( 1 - P^{1\times
1}_{x;\mu,\nu}
\big) \nonumber \\
& +  b_1 \sum_{\substack{
      \mu,\nu=1\\\mu \ne \nu}}^4 \textrm{Re\,Tr}\big( 1 - P^{1 \times 2}_{x; \mu, \nu} \big)
\Big)\,,
\label{eq:gauge_action}
\end{eqnarray}
with $\beta=6/g_0^2$, $g_0$ being the bare coupling and $P^{1\times1}$, $P^{1\times 2}$ the plaquette and
rectangular Wilson loops respectively. The configurations were generated with the tree-level Symanzik improved action
\cite{Weisz:1982zw}, i.e. $b_1=-\frac{1}{12}$, $b_0=1-8b_1$. Note that for smoothing the gauge fields using the gradient flow and cooling, in addition to the tree-level Symanzik improved action, we use the Wilson plaquette action which corresponds to $b_1=0$ and $b_0=1$ as well as the Iwasaki improved action with $b_1=-0.331$ and $b_0=3.648$.

The fermionic action for the light quarks is the Wilson twisted mass
action \cite{Frezzotti:2000nk,Frezzotti:2003ni,Frezzotti:2004wz,Shindler:2007vp}, given in the so-called
twisted basis by
\begin{equation}
 S_l[\psi, \bar{\psi}, U] = a^4 \sum_x \bar{\chi}_l(x) \big( D_W + m_0 + i \mu_l \gamma_5 \tau_3
\big)
\chi_l(x)\,,
 \label{eq:tm_light}
\end{equation}
where $\tau^3$ acts in flavour space and $\chi_l=(\chi_u,\,\chi_d)$ is a
two-component vector in flavour space, related to the one in the physical basis ($\psi$) by a chiral rotation, $\psi = \exp(i\omega\gamma_5\tau_3/2)\chi$, with $\omega$ being the twist angle ($\omega=\pi/2$ at maximal twist).
The bare untwisted and twisted quark masses are, respectively, $m_0$ and $\mu_l$, while the
multiplicatively renormalized light quark mass is $\mu_R=Z_P^{-1}\mu_l$.
$D_W$ is given by:
\begin{equation}
 D_W = \frac{1}{2} \big( \gamma_{\mu} (\nabla_{\mu} + \nabla^*_{\mu}) - a \nabla^*_{\mu} \nabla_{\mu}
\big)\,,
\label{eq:massless_Wilson_Dirac}
\end{equation}
where $\nabla_{\mu}$ and $\nabla^*_{\mu}$ represent the forward and backward covariant
derivatives, respectively.

\begin{table*}[t!]
  \centering
  \begin{tabular}[]{cccccccccc}
    Ensemble & $\beta$ & $a$ [fm] & $r_0/a$ & $Z_P/Z_S$ & lattice & $a\mu_l$ & 
    $\kappa_c$  & $L$ [fm]& $m_\pi L$ \\
\hline
   b$40.16 $  & 3.90 & 0.085 & 5.35(4) & 0.639(3) & $16^3\times32$ & 0.004   & 0.160856   & 1.4 & 2.5 \\
  c$30.20$ & 4.05 & 0.067 &  6.71(4) &  0.682(2) &  $20^3\times40$  & 0.003 & 0.157010 &  1.3 & 2.4\\
  d$20.24$ & 4.20 & 0.054 &  8.36(6) & 0.713(3) & $24^3\times48$  & 0.002 & 0.154073 &  1.3 & 2.4\\
  e$17.32$ & 4.35 & 0.046 &  9.81(13) & 0.740(3) & $32^3\times64$  & 0.00175 & 0.151740 & 1.5 & 2.4\\
  \end{tabular}
  \caption{Parameters of the employed ETMC $N_f=2$ gauge field configuration ensembles
\cite{Boucaud:2007uk,Boucaud:2008xu,Baron:2009wt}. The columns contain: the inverse bare coupling
$\beta$, the approximate values of the lattice spacing $a$
\cite{Blossier:2010cr,Baron:2011sf,Jansen:2011vv},
$r_0/a$
\cite{Blossier:2010cr,Jansen:2011vv}, the scheme- and
scale-independent renormalization constants ratio $Z_P/Z_S$ \cite{Constantinou:2010gr,Alexandrou:2012mt,Cichy:2012is}, the lattice size $(L/a)^3\times(T/a)$, the bare twisted light quark mass $a\mu_l$,
the critical value of the hopping parameter (where the
PCAC mass vanishes), physical extent $L$ of the lattice in fm and the product $m_\pi L$.}
  \label{setupNf2}
\end{table*}

Twisted mass fermions are providing an automatically $\mathcal{O}(a)$-improvement if the twist angle is set to $\pi/2$
(maximal twist). This can be achieved by non-perturbative tuning of the hopping parameter $\kappa = (8+2
a m_0)^{-1}$ to its critical value, i.e. such that the PCAC quark mass vanishes
\cite{Frezzotti:2003ni,Chiarappa:2006ae,Farchioni:2004ma,Farchioni:2004fs,Frezzotti:2005gi,
Jansen:2005kk}.

The details of the gauge field configuration ensembles that were used in this work are shown in
Tab.~\ref{setupNf2}. Most of our investigations are performed using the ensemble b40.16.
However, we also investigate how the correlation between different topological charge definitions changes when approaching the continuum limit, thus using also ensembles c30.20, d20.24 and e17.32. For all of these ensembles the pion mass is close to 340 MeV

\section{Definitions of the topological charge}
\label{sec:definitions_of_the_topological_charge}
In this section, we introduce the definitions of the topological charge that we use below for numerical studies. We attempted to include the most commonly used, theoretically sound definitions of the topological charge. The relevant characteristics of each definition are summarized in Tab.~\ref{tab:defs}.

\begin{table*}[h]
  \centering
\begin{footnotesize}
  \begin{tabular}[]{lllll}
    nr & full name & smearing type & short name & type\\
\hline
1 & index of overlap Dirac operator $s=0.4$ & -- & index nonSmear $s=0.4$ & F\\
2 & index of overlap Dirac operator $s=0.0$ & -- & index nonSmear $s=0$ & F\\
3 & index of overlap Dirac operator $s=0.0$ & HYP1 & index HYP1 $s=0$ & F\\
4 & Wilson-Dirac op. spectral flow $s=0.0$ & HYP1 & SF HYP1 $s=0.0$ & F\\  
5 & Wilson-Dirac op. spectral flow $s=0.75$ & HYP1 & SF HYP1 $s=0.75$ & F\\ 
6 & Wilson-Dirac op. spectral flow $s=0.0$ & HYP5 & SF HYP5 $s=0.0$ & F\\
7 & Wilson-Dirac op. spectral flow $s=0.5$ & HYP5 & SF HYP5 $s=0.5$ & F\\  
8 & spectral projectors $M^2=0.00003555$ & -- & spec. proj. $M^2=0.0000355$ & F\\
9 & spectral projectors $M^2=0.0004$ & -- & spec. proj. $M^2=0.0004$ & F\\
10 & spectral projectors $M^2=0.0010$ & -- & spec. proj. $M^2=0.0010$ & F\\
11 & spectral projectors $M^2=0.0015$ & -- & spec. proj. $M^2=0.0015$ & F\\
12 & field theoretic (clover) & -- & cFT nonSmear & G\\
13 & field theoretic (plaquette) & GF (Wplaq,$t_0$) & pFT GF Wplaq $t_0$ & G\\
14 & field theoretic (plaquette) & GF (Wplaq,$2t_0$) & pFT GF Wplaq $2t_0$ & G\\
15 & field theoretic (plaquette) & GF (Wplaq,$3t_0$) & pFT GF Wplaq $3t_0$ & G\\
16 & field theoretic (clover) & GF (Wplaq,$t_0$) & cFT GF Wplaq $t_0$ & G\\
17 & field theoretic (clover) & GF (Wplaq,$2t_0$) & cFT GF Wplaq $2t_0$ & G\\
18 & field theoretic (clover) & GF (Wplaq,$3t_0$) & cFT GF Wplaq $3t_0$ & G\\
19 & field theoretic (improved) & GF (Wplaq,$t_0$) & iFT GF Wplaq $t_0$ & G\\
20 & field theoretic (improved) & GF (Wplaq,$2t_0$) & iFT GF Wplaq $2t_0$ & G\\
21 & field theoretic (improved) & GF (Wplaq,$3t_0$) & iFT GF Wplaq $3t_0$ & G\\
22 & field theoretic (clover) & GF (tlSym,$t_0$) & cFT GF tlSym $t_0$ & G\\
23 & field theoretic (clover) & GF (tlSym,$2t_0$) & cFT GF tlSym $2t_0$ & G\\
24 & field theoretic (clover) & GF (tlSym,$3t_0$) & cFT GF tlSym $3t_0$ & G\\
25 & field theoretic (clover) & GF (Iwa,$t_0$) & cFT GF Iwa $t_0$ & G\\
26 & field theoretic (clover) & GF (Iwa,$2t_0$) & cFT GF Iwa $2t_0$ & G\\
27 & field theoretic (clover) & GF (Iwa,$3t_0$) & cFT GF Iwa $3t_0$ & G\\
28 & field theoretic (clover) & cool (Wplaq,$t_0$) & cFT cool (GF Wplaq $t_0$) & G\\
29 & field theoretic (clover) & cool (Wplaq,$3t_0$) & cFT cool (GF Wplaq $3t_0$) & G\\
30 & field theoretic (clover) & cool (tlSym,$t_0$) & cFT cool (GF tlSym $t_0$) & G\\
31 & field theoretic (clover) & cool (tlSym,$3t_0$) & cFT cool (GF tlSym $3t_0$) & G\\
32 & field theoretic (clover) & cool (Iwa,$t_0$) & cFT cool (GF Iwa $t_0$) & G\\
33 & field theoretic (clover) & cool (Iwa,$3t_0$) & cFT cool (GF Iwa $3t_0$) & G\\
34 & field theoretic (clover) & stout (0.01,$t_0$) & cFT stout 0.01 (GF Wplaq $t_0$) & G\\
35 & field theoretic (clover) & stout (0.01,$3t_0$) & cFT stout 0.01 (GF Wplaq $3t_0$) & G\\
36 & field theoretic (clover) & stout (0.1,$t_0$) & cFT stout 0.1 (GF Wplaq $t_0$) & G\\
37 & field theoretic (clover) & stout (0.1,$3t_0$) & cFT stout 0.1 (GF Wplaq $3t_0$) & G\\
38 & field theoretic (clover) & APE (0.4,$t_0$) & cFT APE 0.4 (GF Wplaq $t_0$) & G\\
39 & field theoretic (clover) & APE (0.4,$3t_0$) & cFT APE 0.4 (GF Wplaq $3t_0$) & G\\
40 & field theoretic (clover) & APE (0.5,$t_0$) & cFT APE 0.5 (GF Wplaq $t_0$) & G\\
41 & field theoretic (clover) & APE (0.5,$3t_0$) & cFT APE 0.5 (GF Wplaq $3t_0$) & G\\
42 & field theoretic (clover) & APE (0.6,$t_0$) & cFT APE 0.6 (GF Wplaq $t_0$) & G\\
43 & field theoretic (clover) & APE (0.6,$3t_0$) & cFT APE 0.6 (GF Wplaq $3t_0$) & G\\
44 & field theoretic (clover) & HYP ($t_0$) & cFT HYP (GF Wplaq $t_0$) & G\\
45 & field theoretic (clover) & HYP ($3t_0$) & cFT HYP (GF Wplaq $3t_0$) & G\\
\end{tabular}
\end{footnotesize}
  \caption{\small The relevant characteristics of each topological charge definition. For each definition, we
give a number, full name, type of smearing of gauge fields (-- = no smearing, HYP$n$ = $n$ iterations of
HYP smearing, GF (action,$t$) = gradient flow with a given smoothing action (Wplaq = Wilson plaquette, tlSym = tree-level Symanzik improved, Iwa = Iwasaki) and at flow time $t$,
cool (action,$t$) = cooling (smoothing action as for GF) and a number of steps corresponding to GF at flow time $t$,
stout ($\rho_{\rm st}$,$t$) = stout smearing with a given $\rho_{\rm st}$ parameter and a number of steps corresponding to GF at flow time $t$, 
APE ($\alpha_{\rm APE}$,$t$) = APE smearing with a given $\alpha_{\rm APE}$ parameter  and a number of steps corresponding to GF at flow time $t$,
HYP ($t$) = HYP smearing with a number of steps corresponding to GF at flow time $t$), short name (used in plots) and definition type (G=gluonic,
F=fermionic).}
  \label{tab:defs}
\end{table*}

\subsection{Index of the overlap Dirac operator}
\label{sec:index}
For many years, it was considered impossible that chiral symmetry can be realized on the lattice without violating certain essential properties, like locality, translational invariance and the absence of doublers.
This feature of lattice Dirac operators was formulated in terms of a no-go theorem, the Nielsen-Ninomiya theorem \cite{Nielsen:1981hk}.
Only after several years, it was realized that this theorem can be overcome by allowing a modified definition of chiral symmetry on the lattice.
It was shown by L\"uscher \cite{Luscher:1998pqa} that if the lattice Dirac operator satisfies the so-called Ginsparg-Wilson relation \cite{Ginsparg:1981bj}, there is a corresponding exact symmetry and this symmetry becomes just the standard chiral symmetry in the continuum limit.
Thus, any Dirac operator that satisfies the Ginsparg-Wilson relation is chirally symmetric.
One of such operators is the overlap Dirac operator, introduced by Neuberger \cite{Neuberger:1997fp,Neuberger:1998wv}.

The overlap operator, as a chirally symmetric Dirac operator, can have exact zero modes \cite{Niedermayer:1998bi}.
The famous Atiyah-Singer index theorem \cite{Atiyah:1971rm} relates in a simple way the number of these zero modes to the topological charge $Q$ of a given gauge field configuration:
\begin{equation}
Q = n_{-} - n_{+}\,,
\label{eq:Atiya_Singer}  
\end{equation}
where $n_\pm$ denotes the number of zero modes with positive/negative chirality.
This remarkable result, thus, links a property of gauge fields to a fermionic observable.
By construction, it gives integer values of $Q$.
Note, however, that the definition of the overlap operator is not unique -- it depends on the details of the construction of the operator.
In common notation, the massless overlap operator is 
\begin{eqnarray}
\label{eq:index}
D=\frac{1}{a}\left( 1-\frac{A}{\sqrt{A^\dagger A}} \right)\,,\qquad A=1+s-aD_W\,,
\end{eqnarray}
with $D_W$ being the standard Wilson-Dirac operator, given by Eq.~(\ref{eq:massless_Wilson_Dirac}).
The $s$ parameter, appearing in the kernel operator $A$, can be tuned to optimize locality properties of the overlap operator $D$ \cite{Hernandez:1998et,Durr:2005an,Cichy:2012vg}.
It effectively introduces a dependence of the index obtained on a given configuration on the used value of $s$.
This dependence vanishes towards the continuum limit, but at practically used lattice spacings, $Q$ evaluated from the zero modes of the overlap operator shows a dependence on the value of the parameter $s$.
In a sense, this reflects the general property that topology is uniquely defined only for continuum gauge fields.
In Sec.~\ref{sec:results}, we will comment more on the dependence of $Q$ on $s$ by explicitly comparing results obtained for different values of the latter.

The overlap index definition of the topological charge is theoretically clean and very appealing, because it provides integer values of $Q$, while for most other definitions discussed in this paper, the $Q$ values at non-zero lattice spacing are driven away from integers by cut-off effects, ultraviolet fluctuations and/or stochastic noise. However, it has a severe practical drawback -- the cost of using the overlap operator is around one to two orders of magnitude larger than the one of e.g. variants of Wilson fermions \cite{Chiarappa:2006hz}.

\subsection{Wilson-Dirac operator spectral flow}
\label{sec:SF}
Closely related to the overlap index is the index derived from the spectral flow of the hermitian Wilson-Dirac operator \cite{Itoh:1987iy,Narayanan:1994gw}. Its definition is derived from the fact that the continuum hermitian Euclidean Dirac operator $H(m_0) = \gamma_5 D(m_0)$ with bare mass $m_0\neq 0$ has a gap, i.e. it has no eigenvalues in the region $(-|m_0|, |m_0|)$. As a consequence, eigenvalues crossing zero in the spectral flow of $H(m_0)$ can only occur at $m_0=0$, i.e.~they correspond to the zero modes of $D$, and, hence, the net number of crossings is related to the topological charge of the background gauge field \cite{Atiyah:1971rm}.

 On the lattice, zero crossings in the spectral flow of the hermitian Wilson-Dirac operator $H_W(m_0) = \gamma_5 (D_W + m_0)$ can occur for any value of $m_0$ in the region $-8 \leq m_0 \leq 0$ \cite{Narayanan:1993ss,Edwards:1998sh} and counting the net number of crossings in the region $-(1+s) \leq m_0 \leq 0$ enables one to associate an index to the Wilson-Dirac operator as a function of $s$. The interpretation from the overlap
formalism is essential to make this connection \cite{Narayanan:1994gw}. In fact, the correspondence between the index of the overlap operator
and the index from the spectral flow is exact, so the parameter $s$ in Eq.~(\ref{eq:index}) is the same as the one used here, and all the good properties of the overlap index carry over to the index from the spectral flow.

To be more specific, we consider the hermitian Wilson-Dirac operator
\begin{equation}
H_W(m_0) = \gamma_5 (D_W + m_0) \,
\end{equation}
and its eigenvalues $\lambda_k^{H_W}(m_0)$. Their dependence 
 on $m_0$ defines the spectral flow. Since the  Wilson-Dirac operator $D_W$ is non-normal, the eigensystems of $H_W(m_0)$ and $D_W+m_0$ are related in a non-trivial way, except for the modes of $H_W(m_0)$ which are zero for a particular value of $m_0$,
\begin{equation}
H_W(m_0) \psi = 0 \qquad \Longleftrightarrow \qquad D_W \psi = -m_0 \psi \,.
\end{equation}
It follows from this equation that the real modes $\lambda_k^W \in \mathbb{R}$ of $D_W$ correspond to zero modes of $H_W(m_0=-\lambda_k^W)$, while the chirality of the modes is given through first order perturbation theory by the derivative of the spectral flow at $m_0 =-\lambda_k^W$ \cite{Itoh:1987iy},
\begin{equation}
\left. \frac{d \lambda_k^{H_W}}{dm_0} \right|_{m_0=-\lambda_k^W} = \langle k |\gamma_5| k\rangle \, .
\end{equation}
Finally, summing up the chiralities of the real modes $\lambda_k^W \in \mathbb{R}$ of $D_W$ up to $1+s$ yields an index of the Wilson-Dirac operator, and hence the topological charge from the spectral flow,
\begin{equation}
Q = \sum_{\lambda_k^W \in \mathbb{R}} \textrm{sign}(\langle k |\gamma_5| k\rangle) \, ,
\end{equation}
where the sum is over $\lambda_k^W < 1+s$ only, i.e.~it excludes the real doubler modes.

Smoothing the gauge fields in the covariant derivatives of $D_W$ reduces the non-normality of $D_W$, and hence improves the chirality of the real modes \cite{Durr:2005an}. In addition, it also improves the separation of the physical modes from
the doubler modes and in this way reduces the ambiguity of the charge definition due to the choice of the parameter $s$. Interestingly, this ambiguity can be quantified in the context of Wilson Random Matrix Theory \cite{Akemann:2010em, Kieburg:2011uf, Damgaard:2011eg, Deuzeman:2011dh, Kieburg:2013xta}.

\subsection{Spectral projectors}
\label{sec:specproj}
Another fermionic definition of the topological charge was introduced by Giusti and L\"uscher \cite{Giusti:2008vb,Luscher:2010ik}.
One introduces the projector $\PM$ to the subspace of eigenmodes of the squared Hermitian Dirac operator $D^\dagger D$  with eigenvalues below $M^2$.
The projectors $\PM$ can be evaluated stochastically and for chirally
symmetric fermions, the topological charge can be defined in terms of it as
$Q=\Tr\,\{\gamma_5\PM\}$.
This definition is then equivalent to the index definition, apart from the fact that the counting of modes proceeds stochastically, instead of determining it from zero modes.
For non-chirally symmetric fermions, the chirality of modes is no longer $\pm1$, but it can be schematically written as $\pm1+\mathcal{O}(a)$.
Thus, the above definition of $Q$ still holds, but it gives in general non-integer values, contaminated by cut-off effects and by noise from the stochastic evaluation.
In practice, the spectral projector computation of the topological charge proceeds in the following way for Wilson-type fermions \cite{Luscher:2010ik}.
One introduces in the theory a set $\eta_1$, $\ldots$, $\eta_{N_{src}}$ of $N_{src}$ pseudofermion fields with the action $S_\eta=\sum_{j=1}^{N_{src}} (\eta_j,\eta_j)$, where the bracket denotes the scalar product. The fields are generated randomly, thus their gauge ensemble average is distributed according to the introduced action.
Then, one defines the observable
\begin{equation}
\label{eq:C}
{\cal C}=\frac{1}{N_{src}}\sum_{j=1}^{N_{src}}
\left(\PM\eta_j,\dirac{5}\PM\eta_j\right)\,.
\end{equation}
which plays the role of the topological charge.
To compute the topological susceptibility from this definition, one needs a correction to account for a finite number of stochastic noise samples $N_{src}$ and the ratio of renormalization constants $Z_P/Z_S$.
This is done using other observables
\begin{equation}
{\cal B}=\frac{1}{N_{src}}\sum_{j=1}^{N_{src}}
\left(\PM\gamma_5\PM\eta_j,\PM\gamma_5\PM\eta_j\right)\,,
\label{eq:B}
\end{equation}
\begin{equation}
{\cal A} =\frac{1}{N_{src}}\sum_{j=1}^{N_{src}}
\left(\PM^2\eta_j,\PM^2\eta_j\right)\,.
\label{eq:A}
\end{equation}
Then, the topological susceptibility, $\chi$, is defined as
\begin{equation}
\label{eq:chi}
\chi=\frac{1}{V}\frac{\langle{\cal A}\rangle^2}{\langle{\cal B}\rangle^2}
  \left( {\langle{\cal C}^2\rangle-\frac{\langle{\cal B}\rangle}{N_{src}}} \right) \,.
\end{equation}
If the ratio $Z_P/Z_S$ is known from another computation, one can replace $\langle{\cal A}\rangle^2/\langle{\cal B}\rangle^2$ in the above equation with $Z_S^2/Z_P^2$. We can, therefore, define as a proxy of the topological charge the quantity
\begin{eqnarray}
  Q_{\rm eff} = \frac{Z_S}{Z_P} {\cal C} \,,
\end{eqnarray}
with $Q = \lim_{N_{src} \to \infty} Q_{\rm eff}$. It can be shown that the spectral projector definition is manifestly ultraviolet finite \cite{Giusti:2008vb,Luscher:2004fu,Cichy:2013lat,Cichy:2014yca} and hence theoretically very appealing, especially for the computations of the topological susceptibility, as done in Refs.~\cite{Luscher:2010ik,Cichy:2011an,Cichy:2013lat,Cichy:2013rra,Bruno:2014ova,Cichy:2015jra}.
However, if the aim is to e.g. separate topological sectors, i.e. to choose configurations from a given sector, then the stochastic noise present in the spectral projector evaluated observables strongly contaminates the results, if using a relatively small $N_{src} \approx 6$, while for large $N_{src} \to \infty$, the method becomes expensive.
As we show in the results section, the stochastic ingredient also makes the correlation with respect to other definitions only moderate and much smaller than e.g. the correlation between the field theoretic definitions (evaluated with different kinds of smearing).

The obtained result also depends on the spectral thre\-shold $M$ chosen for the projector $\PM$. As stated in Ref.~\cite{Giusti:2008vb}, $M$ can be chosen arbitrarily, but it is wise to avoid its large values in lattice units (that enhance cut-off effects) and also values close to the quark mass. We will check a few values of $M$ and investigate the implications of choosing different values.

Recently, the spectral projector formalism was extended to staggered fermions \cite{Bonanno:2019xhg}.
In this case, the bare topological charge is also multiplicatively renormalizable, but the renormalization constants are different, which is due to a different pattern of chiral symmetry breaking.

\subsection{Field theoretic definition}
\label{sec:Field_Theoretic_Definition}

The topological charge of a gauge field can be naturally defined as the four-dimensional integral over
space-time of the topological charge density. In the continuum, this reads
\bea
Q=\int d^4 x \,q(x)\, ,
\label{eq:topological_charge_continuum}
\eea
where $q(x)$ denotes the topological charge density defined as
\bea
q(x) = \frac{1}{32 \pi^2} \epsilon_{\mu \nu \rho \sigma} {\rm Tr} \left\{ F_{\mu \nu}  F_{\rho \sigma} \right\}\, .
\label{eq:topological_charge_density_def}
\eea
On the lattice, one has to choose a valid discretization $q_L(x)$ of $q(x)$ in order to evaluate \eq{eq:topological_charge_continuum}, which now takes the form of the sum
\bea
Q=a^4 \sum_{x} q_{L}(x)\, .
\label{eq:discretisation_sum}
\eea
In practice, any discretization which gives the right continuum limit can be used for the evaluation of \eq{eq:discretisation_sum}, but depending on the discretization $q_L(x)$,  lattice artifacts affecting the total topological charge $Q$ can vary. This means that using such an operator, on smoothed configurations as we explain in \sec{sec:Smoothing_Procedures}, we do not expect to obtain an exact integer value for the total topological charge $Q$, but rather that the obtained value of $Q$ would be approaching an integer as we tend to the continuum limit i.e. $Q={\rm integer} \pm O(a^2)$. In addition, we expect that the total topological charge for some definitions of $q_{L} (x)$ converges faster and closer to an integer than that obtained by others. One can build such operators from closed path-ordered products of links which lead to the field strength tensor $F_{\mu \nu}$ if we perturbatively expand them in $a$. Namely, by using a number of different Wilson loop shapes and sizes, we cancel, step by step, the leading lattice artifacts contributions. Examples of such operators are demonstrated in the next paragraph.

The simplest lattice discretization of $q_{L}$ is based on the simple plaquette and can be noted as 
\begin{eqnarray}
  q_L^{\rm plaq}(x) = \frac{1}{32 \pi^2} \epsilon_{\mu \nu \rho \sigma} {\rm Tr} \left( C^{\rm plaq}_{\mu \nu} C^{\rm plaq}_{\rho \sigma}  \right)\,,
\end{eqnarray}
with 
\begin{eqnarray}
 C^{\rm plaq}_{\mu \nu}(x)=  {\rm Im} \left( \parbox{0.8cm}{\rotatebox{0}{\includegraphics[height=0.8cm]{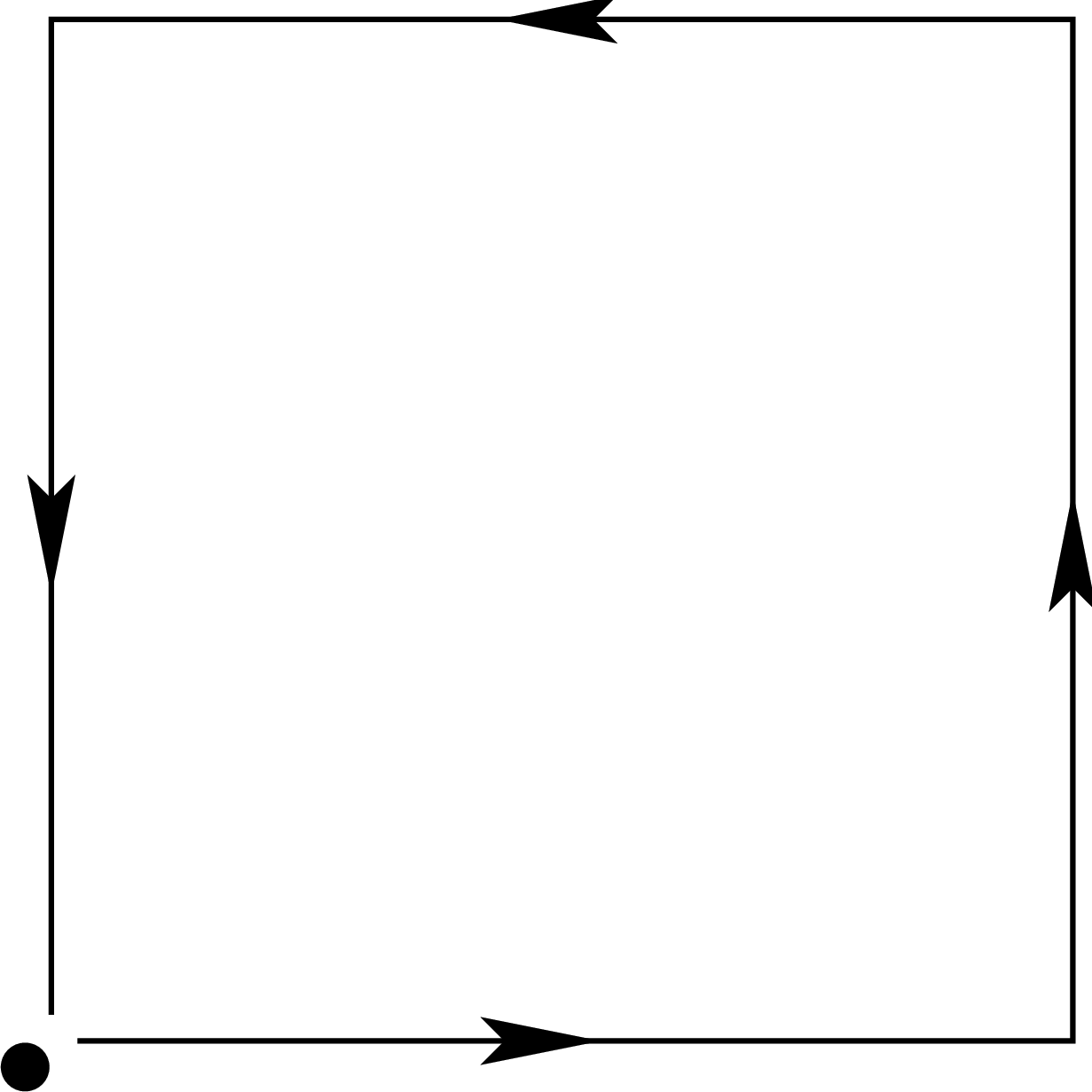}\put(-0.4,0.1){\tiny ${\hat \mu}$}\put(-0.7,0.4){\tiny ${\hat \nu}$}}} \right)\,, 
\end{eqnarray}
where the square pictorializes the path ordered product of the links lying along plaquette sides in the directions ${\hat \mu}$ and ${\hat \nu}$. This definition of $q_{L}(x)$ has a low computational cost and leads to lattice artifacts of order ${\cal O}(a^2)$. Furthermore, it has been used in several determinations of the topological susceptibility and investigations of the instanton properties~\cite{Hart:2001pj,Hart:2004ij}.

Without question, the most commonly used definition of $q_{L}$ is the symmetrized clover leaf noted as
\begin{eqnarray}
  q_L^{\rm clov}(x) = \frac{1}{32 \pi^2} \epsilon_{\mu \nu \rho \sigma} {\rm Tr} \left( C^{\rm clov}_{\mu \nu} C^{\rm clov}_{\rho \sigma}  \right)\,,
\label{eq:clover_discretisation}
\end{eqnarray}
with 
\begin{eqnarray}
 C^{\rm clov}_{\mu \nu}(x)=  \frac{1}{4}{\rm Im} \left( \parbox{1.6cm}{\rotatebox{0}{\includegraphics[height=1.6cm]{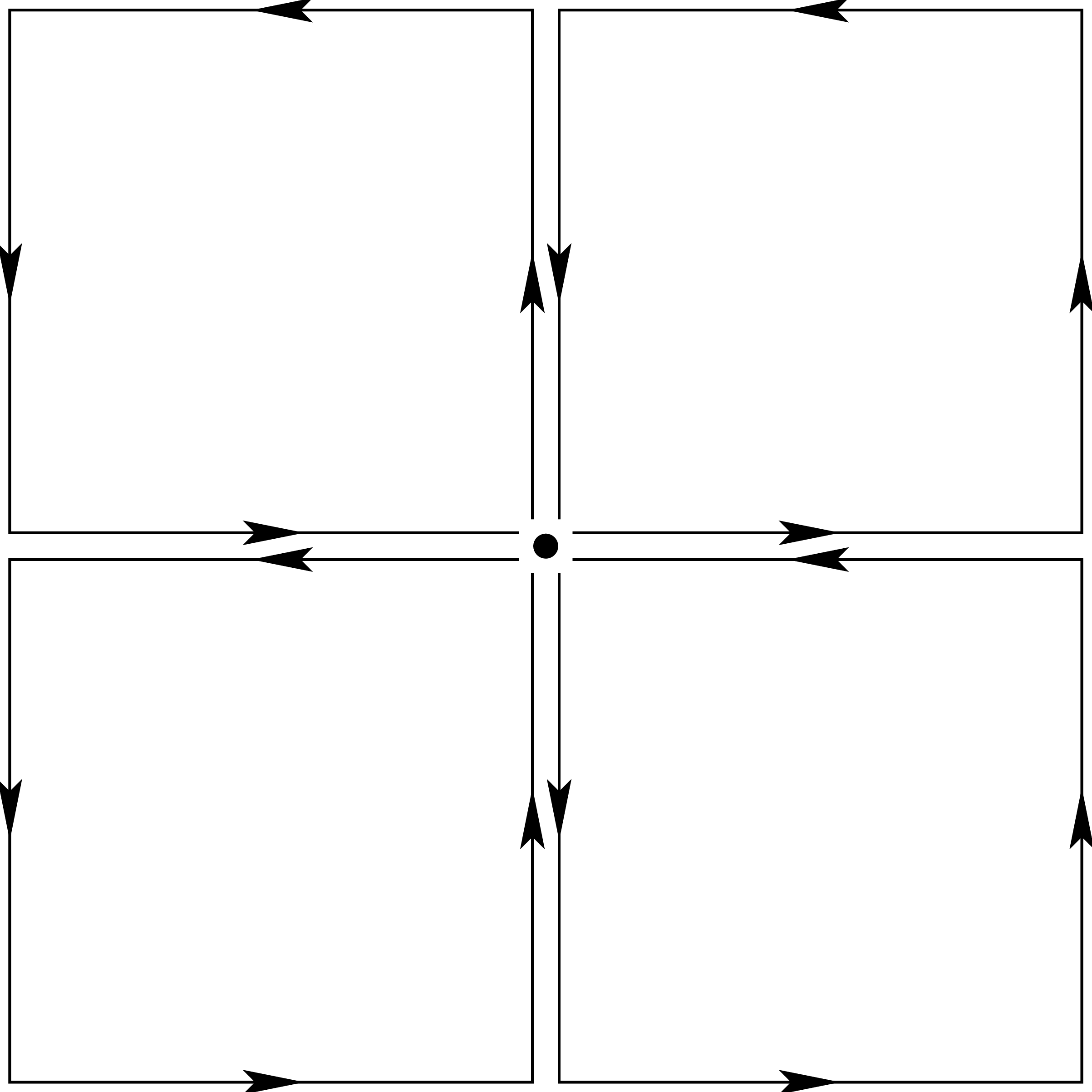}\put(-0.4,0.9){\tiny ${\hat \mu}$}\put(-0.7,1.2){\tiny ${\hat \nu}$}}} \right)\,. 
\end{eqnarray}
Like in the plaquette definition, clover includes lattice artifacts of order ${\cal O}\left( a^2 \right)$. This can be viewed easily by perturbatively expanding $C^{\rm plaq}_{\mu \nu}(x)$ and $ C^{\rm clov}_{\mu \nu}(x)$ and obtaining $1+a^4 F_{\mu \nu}(x)+ {\cal O}(a^6)$. Nevertheless, one can also construct improved definitions of topological charge density operators by including additional Wilson loop shapes in the definition of $q_{L}\left( x \right)$ and then perturbatively canceling the terms which contribute to higher powers of $a$. Such a definition is the Symanzik tree-level improved expressed as
\begin{eqnarray}
q^{\rm imp}_{L}(x)=b_0 q_L^{\rm clov}(x) + b_1 q_L^{\rm rect} (x)\, ,
\end{eqnarray} 
with 
\begin{eqnarray}
q_L^{\rm rect}(x) = \frac{2}{32 \pi^2} \epsilon_{\mu \nu \rho \sigma} {\rm Tr} \left( C^{\rm rect}_{\mu \nu} C^{\rm rect}_{\rho \sigma}  \right)\, ,
\end{eqnarray} 
and
\begin{eqnarray}
C^{\rm rect}_{\mu \nu}(x) = \frac{1}{8} {\rm Im} \left( \parbox{2.4cm}{\rotatebox{0}{\includegraphics[height=1.2cm]{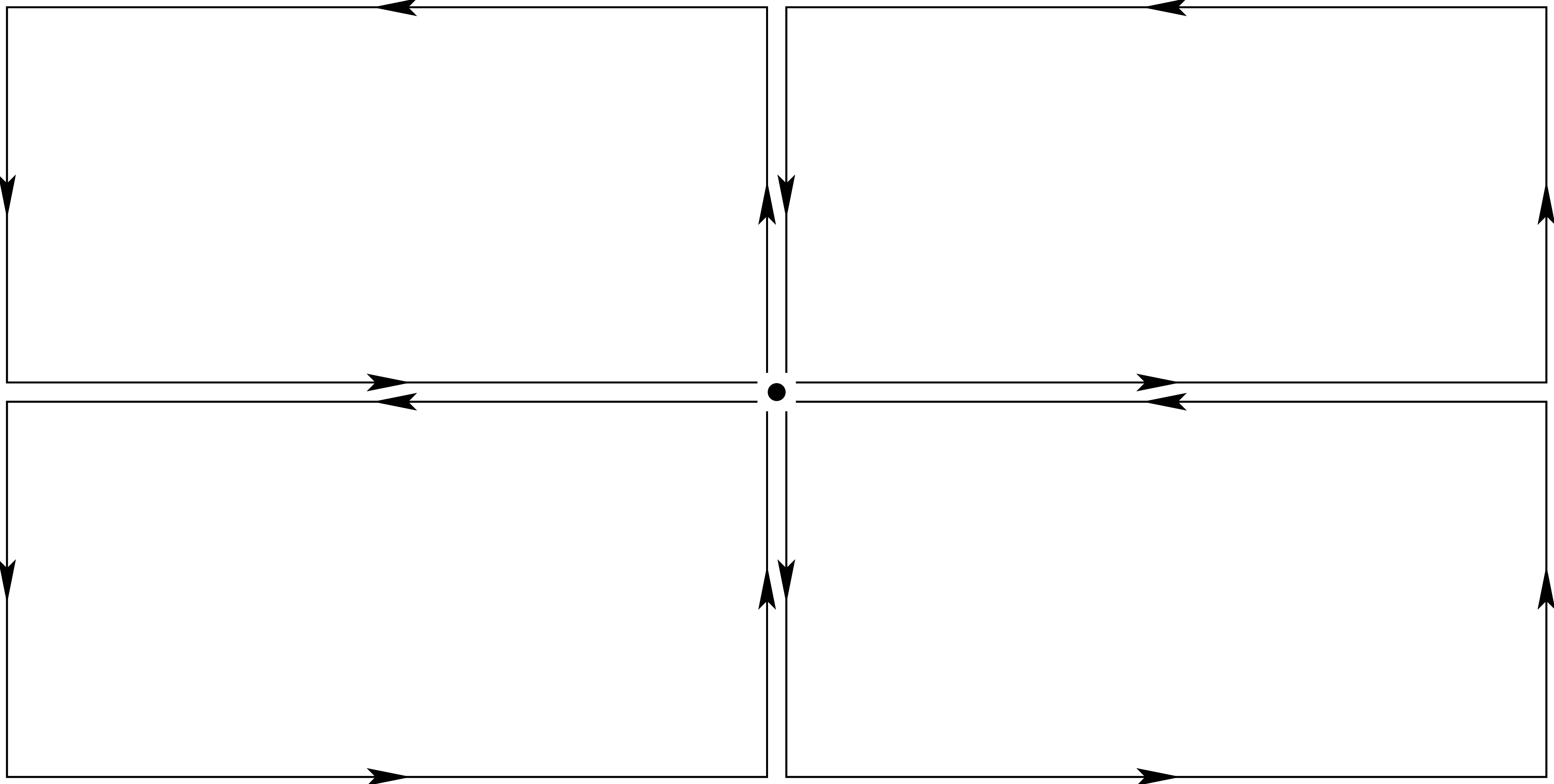}\put(-0.8,0.7){\tiny ${\hat \mu}$}\put(-1.15,0.9){\tiny ${\hat \nu}$}}} + \parbox{1.2cm}{\rotatebox{0}{\includegraphics[height=2.4cm]{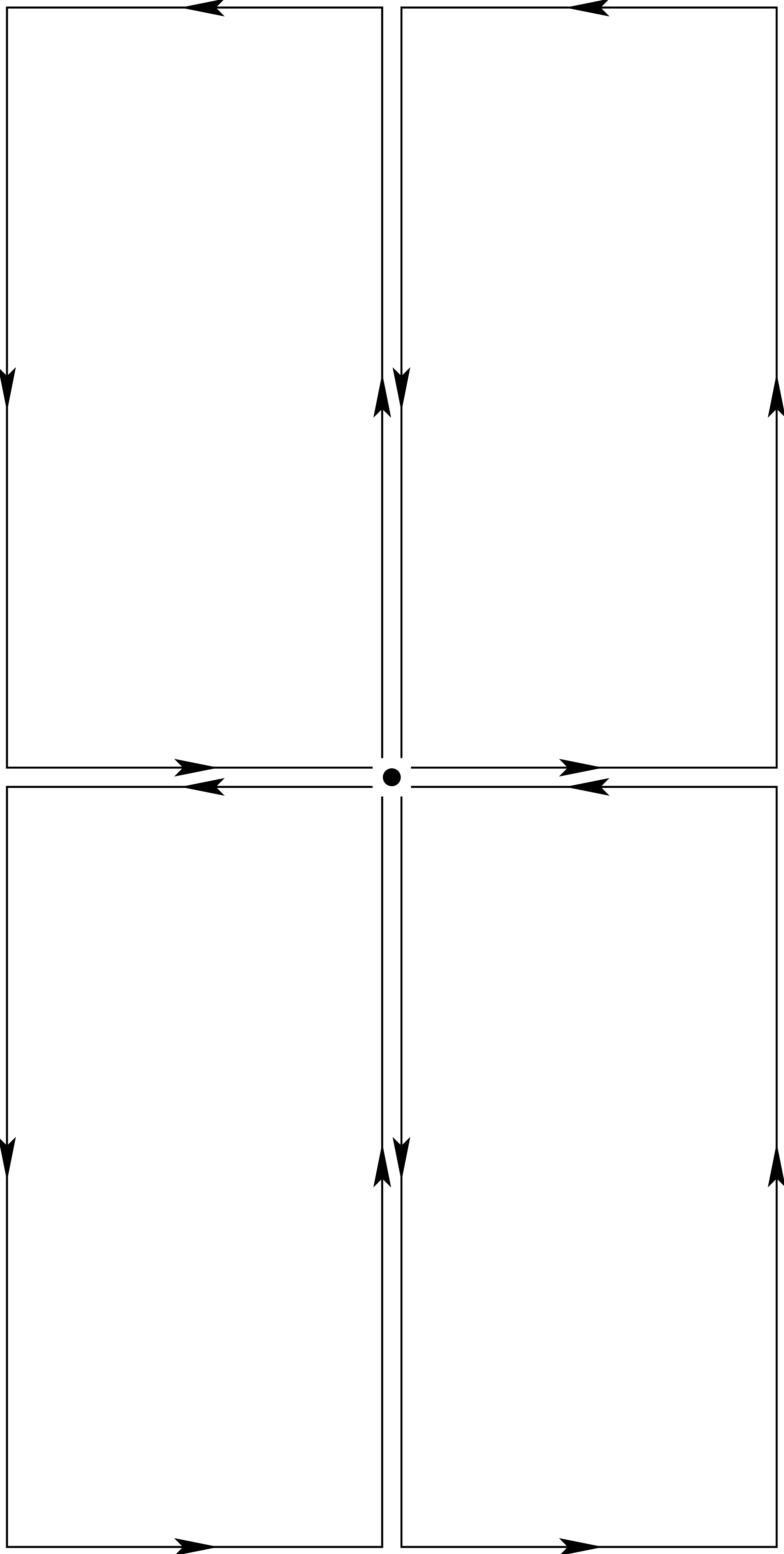}\put(-0.4,1.3){\tiny ${\hat \mu}$}\put(-0.55,1.7){\tiny ${\hat \nu}$}}} \ \right) .
\end{eqnarray}
$q^{\rm rect}_{L}(x)$ is the clover-like operator where instead of squares we make use of horizontally and vertically oriented rectangular Wilson loops of size $2 \times 1$. We remove the discretization errors at tree-level using the Symanzik tree-level coefficients $b_1$ and $b_0$ as these were previously used in \eq{eq:gauge_action}. Thus, this definition of the topological charge density, by a semiclassical inspection, converges as ${\cal O}(a^4)$ in the continuum limit\footnote{Other improved discretizations also exist, see e.g.\ Ref.~\cite{BilsonThompson:2002jk}.}. Hence, a way to obtain topological quantities with small lattice artifact contributions is by using improved topological density operators.\footnote{\label{fn}An approach to reduce further the lattice artifacts and improve the convergence in the continuum limit is to shift the spikes of the topological charge distribution obtained with a given definition of $q_{\rm L}$ around exact integers. This has been introduced in Ref.~\cite{DelDebbio:2002xa} and used extensively since then. However, since this is a rather arbitrary  redefinition of the topological charge, we omit its discussion in this manuscript.}

Ultraviolet fluctuations of the gauge fields entering in the definition of the topological charge density lead to contamination of the topological charge. Hence, we employ methods to suppress these UV fluctuations. Such techniques include the gradient flow, the extensively used cooling and several smearing schemes such as APE, HYP and stout. We examine all the above smoothers and investigate their analytic as well as numerical relations.

It is worth mentioning that one can recover the correct definition of topological quantities from unsmoothed configurations by subtracting additive and multiplicative renormalization constants at zero smoothing step. However, the extraction of these renormalization constants may be subject to more systematic effects. Of course, this method is proven useful for particular studies such as the investigation of the $\theta$-dependence of quantities extracted on configurations produced with the topological density weighted by an imaginary theta term $i \theta$ being included in the QCD action~\cite{Bonati:2015sqt}. Although this method of defining the topological charge is out of the scope of this work, the reader can find more information in Refs.~\cite{DElia:2003zne} and \cite{campostrini1988topological}.

\subsection{Smoothing procedures}
\label{sec:Smoothing_Procedures}
Smoothing a gauge link $U_{\mu}(x)$ can be accomplished by its replacement by some other link that minimizes a local gauge action. To this purpose, it makes more sense to rewrite the lattice gauge action as
\begin{eqnarray}
S_G & = & \frac{\beta}{3} {\rm Re} \Tr \{ {\it X}^\dagger_{\mu}(x) {\it U}_{\mu}(x) \}  \nonumber \\ & & \ \ \ \ \ \ \ \ \ \ \ \ + \{ {\rm terms \  independent \ of} \ {\it U}_{\mu}(x) \}\,,
\label{eq:Local_Gauge_Action}
\end{eqnarray}
where $X_{\mu}(x)$ is the sum of all the path ordered products of link matrices, called the ``staples'', which interact with the link $U_{\mu}(x)$. If we consider the Wilson gauge action, the main components in $X_{\mu}(x)$ are the staples extending over $1 \times 1$ squares (in lattice units). We can, therefore, write  $X_{\mu}(x)$ as
\begin{eqnarray}
  X_{\mu}(x) &\!\!=\!\! & \sum_{\nu\ge 0, \nu\neq\mu}\Big[ U_{\nu}(x)U_{\mu}(x+a\hat{\nu})U_{\nu}^{\dag}(x+a\hat{\mu}) \nonumber \\
    & + & \!U_{\nu}^{\dag}(x-a\hat{\nu})U_{\mu}(x-a\hat{\nu})U_{\nu}(x-a\hat{\nu}+a\hat{\mu})\Big].
\label{eq:Wilson_Staples}
\end{eqnarray}
According to the above equation, for a given link $U_{\mu}(x)$, the total number of plaquette staples interacting with it is 6. There are several ways to iteratively minimize the local action. These include procedures such as the gradient flow, cooling, APE and stout smearing, which make use of the original staples $X_{\mu}(x)$ to minimize the local action; this provides the opportunity to perturbatively relate these smoothers and obtain a more concrete understanding on the numerical equivalence among them.  Furthermore, other more sophisticated smearing procedures such as HYP smearing, which makes use of a more complicated construction of staples, also exist. However, the latter, although it leads to numerical equivalence with the other smoothing techniques, prohibits us from relating it perturbatively at tree-level order with other smoothing techniques. In the next subsections, we give a brief overview of these most commonly used smoothing techniques used for the calculation of the topological charge.   

\subsubsection{Gradient Flow}
\label{sec:Gradient_Flow}
Modern non-perturbative studies of QCD have been employing the gradient flow, which has been proven to be a perturbatively and numerically well-defined smoothing procedure. It has good, perturbatively proven renormalization properties and the fields which have been smoothed via gradient flow do not need to be renormalized. From a historic point of view, the gradient flow is related to the streamline idea of Refs.~{\cite{Shuryak:1987iz,Shuryak:1987ja,Shuryak:1987tr}} and its lattice counterpart was previously introduced in the context of Morse theory \cite{Atiyah:1982fa}. 

The gradient flow is defined as the solution of the evolution equations~\cite{Narayanan:2006rf,Luscher:2010iy,Luscher:2011bx,Lohmayer:2011si,Luscher:2013vga} \bea
\dot{V}_{\mu} \left( x, \tau \right) & = & -g_0^2 \left[ \partial_{x, \mu} S_G(V(\tau)) \right] V_{\mu} \left(x, \tau \right) \,, \nonumber \\
V_{\mu} \left( x, 0 \right) & = & U_{\mu} \left( x \right)\,,
\label{eq:Definition_Gradient_Flow}
\eea
where $\tau$ is the dimensionless gradient flow time. In the above equation, the link derivative is defined as
\bea
\partial_{x, \mu} S_G(U) & = & i \sum_{a} T^a \frac{\rm d}{{\rm d} s} S_G \left( e^{is Y^a} U  \right) \Bigg|_{s=0} \, \nonumber \\
                     & \equiv & i \sum_{a} T^a \partial^{(a)}_{x,\mu} S_G(U)\,,
\eea
with
\begin{equation}
Y^a(y,\nu)=\left\{\begin{array}{ll} T^a & \mathrm{if}\ (y,\nu)=(x,\mu) \, , \\ 0 & 
\mathrm{if}\ (y,\nu) \neq(x,\mu)\,, \end{array} \right. \ 
\end{equation}
and $T^a$ ($a=1,\cdots, 8$) the Hermitian generators of the $SU(3)$ group. If we now set $\Omega_{\mu}(x) = U_{\mu}(x) X^{\dagger}_{\mu} (x)$, we obtain
\begin{eqnarray}
g_0^2 \partial_{x,\mu} S_G(U) (x) = & & \frac{1}{2} \left( \Omega_{\mu} (x)  - \Omega^{\dagger}_{\mu} (x)  \right) \nonumber  \\ & & - \frac{1}{6} {\rm Tr} \left( \Omega_{\mu} (x) - \Omega^{\dagger}_{\mu} (x)  \right).
\label{eq:g02_partial_S}
\end{eqnarray}
The last equation provides all we need in order to smooth the gauge fields according to the Eqs.~(\ref{eq:Definition_Gradient_Flow}). Evolving the gauge fields via the gradient flow requires the numerical integration of Eqs.~(\ref{eq:Definition_Gradient_Flow}) manifested by an integration step $\epsilon$. This is performed using the third order Runge-Kutta scheme, as explained in Ref.~\cite{Luscher:2013vga}. We set $\epsilon=0.01$ for the integration step, since this has been shown to be a safe option~\cite{Bonati:2014tqa}. For the exponentiation of the Lie-algebra fields required for the integration, we apply the algorithm described in Ref.~\cite{Morningstar:2003gk}. 

An important use of the gradient flow is the determination of a reference scale $t_0$, defined as the gradient flow time $t=a^2 \tau$ in physical units for which
\begin{eqnarray}
t^2 \langle E(t) \rangle |_{t=t_0} = 0.3 \,,
\label{eq:reference_scale}
\end{eqnarray}  
where $E(t)$ is the action density
\begin{eqnarray}
  E(t)=-\frac{1}{2V} \sum_{x} {\rm Tr} \left\{ F_{\mu \nu}(x,t)  F_{\mu \nu}(x,t)   \right\}\,.
\end{eqnarray} 
To evaluate numerically $E(t)$, we use the clover discretization similarly to \eq{eq:clover_discretisation}.

Having defined the reference scale $t_0$, a question is raised: for what value of $t$ shall we read an observable? It has been argued that cut-off effects in some observables can be reduced
by chosing larger flow times as reference length scales \cite{Borsanyi:2012zs,Sommer:2014mea}. We therefore chose to commit a numerical comparison of the topological charge obtained with gradient flow with that extracted using different smoothers for three different gradient flow times, namely, $t_0$, $2 t_0$ and $3 t_0$.

\subsubsection{Cooling}
\label{Cooling}
The smoothing technique of cooling \cite{Berg:1981nw,Iwasaki:1983bv,Itoh:1984pr,Teper:1985rb} was one of the first methods used to remove ultraviolet fluctuations from gauge fields.
Cooling is applied to a link variable $U_{\mu}(x) \in SU(3)$ by updating it, from an old value $U^{\rm old}_{\mu}(x)$ to $U^{\rm new}_{\mu}(x)$, according to the probability density
\be
P(U) \propto {\rm exp}\left\{ -\lim_{\beta \to \infty} \beta \frac{1}{3} {\rm Re} \Tr {\it X_{\mu}}^\dagger(x) {\it U_{\mu}}(x)  \right \}\,.
\label{eq:Probability_Density_Cooling}
\ee
The basic step of the cooling algorithm is to replace the given link $U^{\rm old}_{\mu}(x)$ by an $SU(3)$ group element, which minimizes locally the action, while all the other links remain untouched. This is done by choosing a matrix $U^{\rm new}_{\mu}(x) \in SU(3)$ that maximizes 
\bea
{\rm Re} \Tr \{ {\it U}^{\rm new}_{\mu}(x) {\it X}^{\dagger}_{\mu}(x)  \}\,.
\eea
In the case of an $SU(2)$ gauge theory, the maximization is achieved by
\be
U^{\rm new}_{\mu}(x) = \frac{X_{\mu} (x)}{ \sqrt{{\deter} X_{\mu} (x)}}\,.
\label{eq:case_of_su2}
\ee  
For $SU(3)$, the maximization can be implemented using the Cabibbo-Marinari algorithm~\cite{Cabibbo:1982zn} according to which one has to iterate the maximization over all the $SU(2)$ subgroups embedded into $SU(3)$.

We iterate this procedure so that all the links on all sites are updated. Such a  sweep over the whole lattice is called one cooling step $n_c=1$. 
Traditionally, during such a sweep the link variables, which have already been updated, are subsequently used for the update of the links still retaining their old value. Nevertheless, one can also consider to use the updated links only after the whole lattice is covered, increasing the CPU time by a factor of two. 

\subsubsection{APE (Array Processor Experiment) smearing}
\label{sec:description_APE}
An alternative way to smooth the gauge fields is to apply APE smearing \cite{Albanese:1987ds} on the gauge configurations. According to this smoothing procedure, we create fat links by adding to the original links the neighbouring staples weighted by a relative strength $\alpha_{\rm APE}$, which represents the smearing fraction and can be tuned according to its use. This operation breaks unitarity of the resulting ``fat'' links and shifts them away from ${\rm det}\,U =1$, thus, we should project back to $SU(3)$. The above operation is noted as 
\begin{eqnarray}
U_{\mu}^{\left( n_{\rm APE} +1 \right)} \left( x  \right) =  {Proj}_{SU(3)} &[ \left( 1 - \alpha_{\rm APE}   \right) U_{\mu}^{\left( n_{\rm APE}  \right)} \left(x \right)  \nonumber \\  & + \frac{\alpha_{\rm APE}}{6} X_{\mu}^{\left( n_{\rm APE}  \right)} (x)]\,. 
\label{eq:APE_step}
\end{eqnarray}

The APE smearing scheme can be iterated $n_{\rm APE}$ times to produce smeared links. In addition to the simple APE smearing, variations which make use of ``chair'' and ``diagonal'' staples have been proposed from time to time. For the purposes of this investigation, we have considered just the simple APE smearing for different values of the $\alpha_{\rm APE}$ parameter:
\begin{eqnarray}
  \alpha_{\rm APE} = 0.1, \; 0.2, \; 0.3, \; 0.4, \; 0.5 \;\; {\rm and} \;\; 0.6 \,.
\end{eqnarray}

One can project back to $SU(3)$ by maximizing the expression ${\rm Re} \Tr \{ {\tilde U}^{\left( n_{\rm APE} +1 \right)}_{\mu}(x) {\it X}^{\dagger}_{\mu}(x) \}$ with ${\tilde U}^{\left( n_{\rm APE} +1 \right)}_{\mu}(x)$ being the unprojected smeared link. Nonetheless, one can project onto $SU(3)$ using other iterative procedures which suggest that there is not a unique way to do so. This subtlety, however, becomes irrelevant in analytic smearing schemes such as the stout. 

\subsubsection{Stout Smearing}
\label{sec:Stout_Smearing}
A method which allows analytical derivation of smooth\-ed configurations in $SU(3)$ is the so--called stout smearing proposed in Ref.~\cite{Morningstar:2003gk}. This smoothing scheme works in the following way. Once again, let $X_{\mu} (x)$ denote the weighted sum of the perpendicular staples which begin
at lattice site $x$ and terminate at neighboring site $x+a{\hat \mu}$. Now, we give a weight $\rho_{\rm st}$ 
to the staples according to
\begin{eqnarray}
C_{\mu} (x) = \rho_{\rm st} X_{\mu} (x)\,.
\label{eq:cmu}
\end{eqnarray} 
The weight $\rho_{\rm st}$ is a tunable real parameter. Then, the matrix $Q_{\mu} (x)$, defined in $SU(3)$ by
\bea
Q_{\mu} (x) & = & \frac{i}{2} \left( \Xi^{\dagger}_{\mu}(x) - \Xi_{\mu}(x)  \right) \nonumber \\  & & \quad \quad \quad -  \frac{i}{6} {\rm Tr} \left( \Xi^{\dagger}_{\mu}(x) - \Xi_{\mu}(x)  \right)\,
\label{eq:Ximu}
\eea
is Hermitian and traceless, where
\bea
\Xi_{\mu}(x) = C_{\mu} (x) U^{\dagger}_{\mu}(x)\,.
\eea  
Thus, we define an iterative, analytic link smearing algorithm in which the links $U^{(n_{\rm st})}_{\mu}(x)$ at stout smearing step $n_{\rm st}$ are mapped into links $U^{(n_{\rm st}+1)}_{\mu}(x)$ at stout smearing step  $n_{\rm st}+1$ using
\bea
U^{(n_{\rm st}+1)}_{\mu}(x) = {\rm exp} \left( i Q^{n_{\rm st}}_{\mu} (x)   \right) U^{(n_{\rm st})}_{\mu}(x)\,.
\label{eq:exponentiation}
\eea 
This step can be iterated $n_{\rm st}$ times to finally produce link variables in $SU(3)$ which we call stout links. The structure of the stout smearing procedure resembles the exponentiation
steps of the gradient flow and, as a matter of fact, the gradient flow using the Wilson gauge action can be considered as a continuous generalization of stout smearing  \cite{Bergner:2014ska}, using gauge paths more complicated than just staples. Hence, for a small enough lattice spacing, one would expect the two smoothing schemes to provide extremely similar results. The level of this similarity is investigated in this work.

\subsubsection{HYP (Hypercubic) smearing}
\label{sec:description_HYP}
We turn now to the discussion of the HYP (Hypercubic) smearing which has been introduced in Ref.~\cite{Hasenfratz:2001hp}. The smeared links of the HYP smoothing procedure are constructed in three steps. These steps are described in the next bullet points.
\begin{itemize}

\item[3.] The final step of the HYP smearing consists of applying an APE smearing routine in which the staples are constructed by decorated links which have undergone HYP smearing levels 1 and 2:
\begin{eqnarray}
U_{\mu}^{\left( n_{\rm HYP} +1 \right)} \left( x  \right) = {Proj}_{SU(3)} & [ \left( 1 - \alpha_3\right) U_{\mu}^{\left( n_{\rm HYP} \right)} \left(x  \right)   \nonumber \\ & +  \frac{\alpha_3}{6} {\tilde X}_{\mu}(x) ]\,,
\end{eqnarray}
with
\begin{eqnarray}
 {\tilde X}_{\mu}(x)  = \!\!\!\!\!\! \sum_{\nu\ge 0, \nu\neq\mu}\!\!\!\!\!\Big[ {\tilde U}_{\nu ; \mu}(x){\tilde U}_{\mu; \nu}(x+a\hat{\nu}) {\tilde U}_{\nu; \mu}^{\dag}(x+a\hat{\mu}) \nonumber \\
+ {\tilde U}_{\nu ; \mu}^{\dag}(x-a\hat{\nu}){\tilde U}_{\mu; \nu}(x-a\hat{\nu}) {\tilde U}_{\nu ; \mu}(x-a\hat{\nu}+a\hat{\mu})\Big]\,.
\label{eq:HYP_staple}
\end{eqnarray}
The link $U_{\mu}^{\left( n_{\rm HYP} \right)} \left(x \right)$ is the original link in ${\hat \mu}$ direction which has been smoothed $n_{\rm HYP}$ times. ${\tilde U}_{\mu ; \nu}(x)$ are fat links along the direction ${\hat \mu}$ resulting from the second step of the HYP smearing procedure with the staples extending over direction ${\hat \nu}$ being not smeared. The parameter $\alpha_3$ is tunable and real.

\item[2.] The second step of HYP smearing creates the decorated links ${\tilde U}_{\mu ; \nu}(x)$ by applying a modified APE smearing procedure according to
\begin{eqnarray}
{\tilde U}_{\mu;\nu} \left( x  \right) & = & {Proj}_{SU(3)} [ ( 1  - \alpha_2) U_{\mu}^{\left( n_{\rm HYP} \right)} \left(x  \right)   \nonumber \\  & & \quad \quad \quad \quad \quad \quad +   \frac{\alpha_2}{4} {\overline X}_{\mu; \nu} (x) ]\,,
\end{eqnarray}
with
\begin{eqnarray}
&&{\overline X}_{\mu;\nu}(x) = \\
&& \sum_{\rho \ge 0, \rho \neq \nu, \mu}\!\!\!\!\!\!\!\Big[ {\overline U}_{\rho ; \nu, \mu}(x){\overline U}_{\mu; \rho, \nu}(x+a\hat{\rho}) {\overline U}_{\rho; \nu, \mu}^{\dag}(x+a\hat{\mu}) \nonumber \\
&&+ {\overline U}_{\rho ; \nu, \mu}^{\dag}(x-a\hat{\rho}){\overline U}_{\mu; \rho, \nu}(x-a\hat{\rho}) {\overline U}_{\rho ; \mu, \nu}(x-a\hat{\rho}+a\hat{\mu})\Big]\,.\nonumber
\end{eqnarray}
Once more, the link $U_{\mu}^{\left( n_{\rm HYP} \right)} \left(x \right)$ is the link in ${\hat \mu}$ direction which has been smoothed $n_{\rm HYP}$ times. Now, ${\overline U}_{\mu ; \rho, \nu}(x)$ are fat links along direction ${\hat \mu}$ resulting from the first step of the HYP smearing procedure with staples extending in the directions ${\hat \rho}$, ${\hat \nu}$ being non smeared. The parameter $\alpha_2$ is again tunable and real.

\item[1.]  The decorated links ${\overline U}_{\rho ; \nu, \mu}(x)$ are built from the links which have been smeared $n_{\rm HYP}$ using the modified APE smearing step:
\begin{eqnarray}
{\overline U}_{\mu;\nu, \rho} \left( x  \right) & = & {Proj}_{SU(3)} [ \left( 1 - \alpha_1   \right) U_{\mu}^{\left( n_{\rm HYP} \right)} \left(x  \right)   \nonumber \\ & &  \quad \quad \quad \quad \quad \quad + \frac{\alpha_2}{2} {\breve X}_{\mu ; \nu \rho} (x) ]\,,
\end{eqnarray}
with
{\small
\begin{eqnarray}
  && {\breve X}_{\mu;\nu, \rho}(x) = \\
  && \!\!\sum_{\sigma \ge 0, \sigma \neq \rho, \nu, \mu}\!\!\!\!\!\!\!\!\!\!\Big[ {U}^{\left( n_{\rm HYP} \right)}_{\sigma ; \rho, \nu, \mu}(x){U}^{\left( n_{\rm HYP} \right)}_{\mu; \sigma, \rho, \nu}(x+a\hat{\sigma}) {U^{\dagger}}_{\sigma; \rho, \nu, \mu}^{\left( n_{\rm HYP} \right)}(x+a\hat{\mu}) \nonumber \\
&& +{U^{\dagger}}_{\sigma ; \rho, \nu, \mu}^{\left( n_{\rm HYP} \right)}(x-a\hat{\sigma}){U}_{\mu; \sigma, \rho, \nu}^{\left( n_{\rm HYP} \right)}(x-a\hat{\sigma}) {U}_{\sigma ; \rho, \mu, \nu}^{\left( n_{\rm HYP} \right)}(x-a\hat{\sigma}+a\hat{\mu})\Big]\,. \nonumber
\end{eqnarray}}
In the above expression, only the two staples orthogonal to directions ${\hat \mu}$, ${\hat \nu}$, ${\hat \rho}$ are included in the smearing procedure.
\end{itemize}
For the purpose of our work, we choose the values \cite{Hasenfratz:2001hp}
\begin{eqnarray}
\alpha_1 = 0.75, \; \alpha_2 = 0.6, \; \alpha_3 = 0.3\, .
\end{eqnarray}

\subsubsection{Perturbative relation between smoothing techniques}
\label{sec:perturbative_procedure}
The gradient flow, cooling as well as APE, stout and HYP smearing schemes can be used to remove the ultraviolet fluctuations and should lead to the same topological properties,
provided that we are close enough to the continuum limit.  Assuming that $a$ is small enough so that we are in the perturbative regime, we can carry out a comparison between the different smoothing procedures in order to obtain an analytic relation among the associated smoothing scales involved, following Refs. \cite{Bonati:2014tqa,Alexandrou:2015yba}. Since the gradient flow has the advantage of being the only smoothing scheme with good, perturbatively proven, renormalizability properties, we first relate all other smoothing schemes with the gradient flow and, subsequently, with each other.

\paragraph{Gradient flow.}
\label{sec:perturbative_wilson_flow}
The perturbative investigation of the relation between the gradient flow using the Wilson action and cooling has been studied in Ref.~\cite{Bonati:2014tqa}. The authors demonstrated that the two smoothing schemes alter the links of a gauge configuration by the same amount if one rescales the flow time and the number of cooling steps according to
\begin{eqnarray}
\tau \simeq \frac{n_c}{3}\,.
\label{eq:perturbative_expression}
\end{eqnarray}
In the following few lines, we sketch the extraction of the derivative evolution of a gauge link in order to compare it to other smoothing schemes. 

In the perturbative regime, a link variable which has been smoothed via the Wilson flow\footnote{For convenience, here and in the following we refer to the gradient flow using the Wilson gauge action in short as the 'Wilson flow'. This is not to be confused with the spectral flow of the Hermitian Wilson Dirac operator which in the past sometimes has also been called 'Wilson flow'.} for a finite flow time $\tau$ can be expanded as
\bea
  U_{\mu} (x, \tau) \simeq \idnty + i \sum_{a} u^a_{\mu}(x, \tau) T^a\,,
\label{eq:Link_Variable_Expansion}
\eea
with $u^a_{\mu}(x, \tau) \in \mathbb{R}$ assumed to be infinitesimal. Using Eq.~(\ref{eq:Wilson_Staples}), the plaquette staples are written as
\bea
X_{\mu} (x, \tau)   \simeq   6 \cdot \idnty + i \sum_{a} w^a_{\mu}(x, \tau) T^a \,,
\label{eq:equation_xmu}
\eea
where  $w^a_{\mu}(x, \tau)$ is an infinitesimal quantity. The leading coefficient with the value 6 appearing in the above equation is just the number of plaquettes interacting with the link on which the Wilson flow evolution is applied. We can, therefore, write $\Omega_{\mu}(x, \tau)$ as 
\bea
\Omega_{\mu}(x, \tau) \simeq 6  \cdot \idnty  + \sum_{a} \left[ 6 u^a_{\mu}(x, \tau ) - w_{\mu}^a(x, \tau) \right] T^a\,.
\eea
Hence, \eq{eq:g02_partial_S} becomes
\bea
g_0^2 \partial_{x,\mu} S_G(U) \simeq i \sum_{a} \left[ 6 u^a_{\mu}(x,\tau) - w_{\mu}^a(x,\tau)  \right] T^a\,.
\eea
Using the above expression, the evolution of the gradient flow by an infinitesimally small flow time $\epsilon$ can be approximated as
\bea
u_{\mu}^a (x,\tau+\epsilon) \simeq u_{\mu}^{a} (x,\tau) - \epsilon \left[  6 u^a_{\mu}(x,\tau) \!-\! w_{\mu}^a(x,\tau)   \right]\,.
\label{WilsonUpdate}
\eea
Since the gradient flow is the only smoothing scheme with a concrete theoretical foundation and good renormalizability properties, we will attempt to relate it to other smoothing schemes. Hence, through the resulting matching formulae of any smoothing technique with the gradient flow, we will be able to relate all the different smoothing schemes with each other.

\paragraph{Cooling.}
\label{sec:prrturbative_cooling}
In the cooling procedure, the link $U_{\mu}(x,n_c)$ is substituted with the projection of $X_{\mu}(x,n_c)$ over the gauge group. Namely, for the case of the $SU(2)$ gauge theory, this projection is manifested by \eq{eq:case_of_su2} where we substitute $X_{\mu}(x,n_c)$ by \eq{eq:equation_xmu}. In the perturbative approximation, this leads to
\bea
U_{\mu}^{\left( n_c+1  \right)}(x) \simeq \idnty + i \sum_{a} \frac{w_{\mu}^a(x, n_c)}{6} T^a\,.
\eea
The above update corresponds to the substitution
\bea
u^a_{\mu} (x, n_c+1) = \frac{w_{\mu}^a(x,n_c)}{6}\,.
\label{CoolingUpdate}
\eea
Comparing Eqs.~(\ref{WilsonUpdate}) and~(\ref{CoolingUpdate}), one sees that the flow would evolve the same as cooling if one chooses a step of $\epsilon = 1/ {6}$. In addition, during a whole cooling step, the link variables which have already been updated are subsequently used for the update of the remaining links that await update; this corresponds to a speed-up of a factor of two. Therefore, the predicted perturbative relation between the flow time $\tau$ and the number of cooling steps $n_c$ so that both smoothers have the same effect on the gauge field is given by \eq{eq:perturbative_expression}. This has been shown analytically and demonstrated numerically in Ref.~\cite{Bonati:2014tqa}. Moreover, the authors in Ref.~\cite{Alexandrou:2015yba} have generalized this equivalence for the case of the gradient flow and cooling employing smoothing actions which in addition to the square term multiplied by a factor of $b_0$, also included a rectangular term multiplied by $b_1=(1-b_0)/8$. This equivalence in manifested by the formula
\bea
\tau \simeq \frac{n_c}{3-15 b_1}\,.
\label{eq:perturbative_expression_rectangles}
\eea
In Section ~\ref{sec:numerical_equivalence}, we investigate and confirm that the equivalence for the Wilson smoothing action is manifested by \eq{eq:perturbative_expression}.

\paragraph{APE smearing.}
\label{sec:perturbative_APE}
We now move to the case of the APE smearing and relate it perturbatively to the Wilson flow and consequently to cooling and stout smearing. Once more, we express the gauge link $U_{\mu} (x, n_{\rm APE})$ in terms of elements of its Lie algebra
\bea
  U_{\mu}^{\left( n_{\rm APE} \right)} (x) \simeq \idnty + i \sum_{a} u^a_{\mu}(x, n_{\rm APE}) T^a\,.
\label{eq:Link_Variable_Expansion}
\eea
Subsequently, we apply this expansion to \eq{eq:APE_step} and obtain the evolution equation
{
\begin{eqnarray}
&& u_{\mu}^a (x,n_{\rm APE}  +  1)  \simeq  u_{\mu}^{a} (x,n_{\rm APE}) \nonumber \\  && \quad -   \frac{\alpha_{\rm APE}}{6} \left[  6 u^a_{\mu}(x,n_{\rm APE}) - w_{\mu}^a(x,n_{\rm APE})   \right].
\label{eq:APE_Update}
\end{eqnarray}}
Comparing  Eqs.~(\ref{WilsonUpdate}) and~(\ref{eq:APE_Update}), we observe that the Wilson flow would evolve the gauge links the same as APE smearing if one chooses a flow step of $\epsilon = {\alpha_{\rm APE}}/{6}$. Hence, the perturbative relation between the Wilson flow time $\tau$ and the number of APE smearing steps $n_{\rm APE}$ so that both smoothers have the same effect on the gauge field is
\bea
\tau = \frac{\alpha_{\rm APE}}{6}  n_{\rm APE} \,.
\label{eq:ape_ground_expression}
\eea
The above perturbative matching relation is investigated numerically in Section~\ref{sec:numerical_equivalence}. 

\paragraph{Stout smearing.}
\label{sec:perturbative_Stout}
Let us now turn to the stout smearing smoothing procedure and check whether we could demonstrate analytic equivalence with the Wilson flow and consequently with cooling.  According to equation \eq{eq:cmu} $C_{\mu} (x, n_{\rm st})$ can be written as
\bea
C_{\mu} (x, n_{\rm st})   \simeq   6 \rho_{\rm st} \cdot \idnty  + i \sum_{a} \rho_{\rm st} w^a_{\mu}(x, n_{\rm st}) T^a \,.
\label{eq:equation_cmu}
\eea
Hence, $\Xi_{\mu} (x, n_{\rm st})$ is written as
\begin{eqnarray}
\Xi_{\mu}(x, n_{\rm st}) & \simeq & 6 \rho_{\rm st} \cdot \idnty  \nonumber \\ & - &  i \rho_{\rm st} \sum_{a} \left[ 6 u^a_{\mu}(x,n_{\rm st}) - w_{\mu}^a(x,n_{\rm st}) \right] T^a\,
\end{eqnarray}
The above leads to 
\bea
Q_{\mu}(x, n_{\rm st}) \!\simeq\! - \rho_{\rm st}  \sum_{a} \!\left[ 6 u^a_{\mu}(x, n_{\rm st}) \!-\! w_{\mu}^a(x, n_{\rm st})  \right] T^a\,.
\eea
If we now apply the exponantiation and multiplication according to \eq{eq:exponentiation}, we obtain that
\begin{eqnarray}
U^{(n_{\rm st}+1)}_{\mu}(x) & \simeq  & \idnty + i \sum_{a} [  u_{\mu}^{a} (x,n_{\rm st}) \nonumber \\  & - & \rho_{\rm st} \left[  6 u^a_{\mu}(x,n_{\rm st}) - w_{\mu}^a(x,n_{\rm st}) ]    \right] T^a\,,
\end{eqnarray}
and in terms of $u_{\mu}^{a} (x,n_{\rm st})$ that
\bea
u_{\mu}^a (x,n_{\rm st}+1) & \simeq & u_{\mu}^{a} (x,n_{\rm st}) \nonumber \\  & - & \rho_{\rm st} \left[  6 u^a_{\mu}(x,n_{\rm st}) - w_{\mu}^a(x,n_{\rm st})   \right]\,.
\label{StoutUpdate}
\eea
Comparing  Eqs.~(\ref{WilsonUpdate}) and~(\ref{StoutUpdate}), we observe that the Wilson flow would evolve the same as stout smearing if one chooses a step of $\epsilon = \rho_{\rm st}$. Therefore, the predicted perturbative relation between the flow time $\tau$ and the number of stout smearing steps $n_{\rm st}$ so that both smoothers have the same effect on the gauge field is
\bea
\tau = \rho_{\rm st}  n_{\rm st} \,.
\label{eq:stout_ground_expression}
\eea
The above perturbative correspondence is also studied numerically in Section~\ref{sec:numerical_equivalence}.
The result that we found is in accordance with previous investigations on the matching between stout and APE smearing \cite{Durr:2007cy}.

\paragraph{HYP smearing.}
\label{sec:perturbative_APE}
Let us now consider the HYP smearing procedure. We have already mentioned that the construction of the decorated staples in the case of HYP smearing prohibits the extraction of a tree-level perturbative relation with the other four smearing procedures. Indubitably, HYP smearing is a valid numerical scheme for smoothing gauge links and removing the ultraviolet fluctuations. Hence, the average action density decreases as we iterate the smoothing procedure. We can, therefore, obtain a numerical equivalence between HYP and another smoother. We attempt to relate the Wilson flow with HYP smearing by calculating the function $\tau(n_{\rm HYP})$ and interpolating it with an ansatz. The function $\tau(n_{\rm HYP})$, once more, is defined as the Wilson flow time $\tau$ for which the average action density changes by the same amount as when $n_{\rm HYP}$ smearing steps are performed. The perturbative nature of the equivalence and the fact that we need at least three full cooling sweeps to relate the ordinary staple of \eq{eq:Wilson_Staples} with all the components of the HYP staple in \eq{eq:HYP_staple}, as well as the dependence on three HYP parameters, suggests that the ansatz could be a polynomial such as 
\begin{eqnarray}
\tau(n_{\rm HYP}) = A n_{\rm HYP} + B n^2_{\rm HYP} + C n^3_{\rm HYP}\,.
\label{eq:HYP_ansatz}
\end{eqnarray}
Since the above equation involves the numerical determination of the associated coefficients, we proceed with this in Section~\ref{sec:numerical_equivalence}.

\paragraph{Fixing the smoothing scale.}
\label{sec:fixing_smoothing_scale}

As the continuum limit is approached, one has to tune the smoothing scale, i.e. gradient flow time, cooling and smearing steps, so that the physics under investigation does not change. When applying a smoothing procedure, the ultraviolet properties of the theory are modified up to some length scale $\lambda_{S}$ because the ultraviolet fluctuations at smaller length scales are suppressed. In order to have a well defined smoothing procedure towards the continuum limit, one has to make sure that changing the ultraviolet part of the theory leaves the continuum results unaltered, thus, the underlying physics should not depend on $\lambda_{S}$. This can be successfully applied by fixing the length scale $\lambda_{S}$, which depends on the smoothing parameters. For APE, HYP and stout smearing schemes as well as for cooling, this scale was chosen in the past using different kind of arguments. Nevertheless, for the case of gradient flow, this length scale is quantified. Namely, it has been demonstrated that we can renormalize composite operators at fixed physical length scale related to the flow time by
\begin{eqnarray}
\lambda_{S} = \sqrt{8 t} = a \sqrt{8 \tau}\,.
\end{eqnarray}  
The above equation enables us to translate the length scale $\lambda_{S}$ to the number of smearing and cooling steps. For cooling, as was shown in Refs.~\cite{Bonati:2014tqa,Alexandrou:2015yba}, we can define the length scale as a function of $n_c$ according to the formula
\begin{eqnarray}
\lambda_S = a \sqrt{\frac{8 n_c}{3} }\,.
\label{eq:scale_cooling}
\end{eqnarray}
For APE smearing, we can write $\lambda_S$ as a function of $n_{\rm APE}$ as
\begin{eqnarray}
 \lambda_S = a \sqrt{\frac{4 \alpha_{\rm APE} n_{\rm APE}}{3} }.
\label{eq:scale_ape}
\end{eqnarray}
In the case of stout smearing, $\lambda_S$ takes the form
\begin{eqnarray}
\lambda_S = a \sqrt{{8 \rho_{\rm st}} n_{\rm st} }\,.
\label{eq:scale_stout}
\end{eqnarray}
Finally, for HYP smearing, we can use the numerically extracted $\tau_{\rm HYP} (n_{\rm HYP})$ in \eq{eq:HYP_ansatz} to define $\lambda_S$ as a function of $n_{\rm HYP}$ according to
\begin{eqnarray}
\lambda_S = a \sqrt{{8  \tau_{\rm HYP} (n_{\rm HYP}) }}\,.
\label{eq:scale_hyp}
\end{eqnarray}
Using the above four equations, we can extract a matching relation between two different smoothing schemes with the corresponding matching coefficients given in \tbl{tab:rescaling}.  

To explain how we can use the length scale in order to obtain a continuum observable, let us consider the calculation of the continuum limit of the topological susceptibility in quenched QCD. One calculates the topological charge using a given smoothing scheme at a fixed value (in physical units) of $\lambda_S =\sqrt{8 t} = \mathcal{O}(0.1 {\rm fm})$. The value of $\lambda_S$ should be chosen such that it is not too small so that ultraviolet contamination is adequately suppressed, as well as not too large so that the underlying topological structure of the gauge fields is preserved. In most cases, $\lambda_S$ corresponds to a plateau for the topological susceptibility reflecting the scale invariance of the observable. We, therefore, extract the topological susceptibility at fixed $\lambda_S$ for a sequence of lattice spacings and then extrapolate it to the continuum limit. 

Of course, the above scenario cannot hold if we consider unquenched QCD, where the physical reference scale depends on the pion mass. In such case, one needs to keep fixed a reference scale such as $t_0$ using \eq{eq:reference_scale}. We can generalize the above procedure for any different smoothing scheme with an effective smoothing flow time defined as $a^2 \tau(n)$ where $\tau(n)$ corresponds to the matching condition between gradient flow and a given smoothing scheme with smooting scale $n$.

In our investigation, we evaluated $t_0$ for the ensemble b40.16 and found that it is equal to $t_0=a^2 \tau_0 = a^2 \times 2.5$; in other words the dimensionless flow time $\tau_{0}=2.5$. For each individual smoother, we can extract a smoothing scale which matches this flow time. 

\begin{table}[ht]
\begin{center}
\begin{tabular}{c|cccc}
 & $\tau$ & $n_c$ & $n_{\rm APE}$ & $n_{\rm st}$ \\ \hline
$\tau$ &  1 & $\frac{1}{3}$ & $\frac{\alpha_{\rm APE}}{6}$ & $\rho_{\rm st}$ \\ 
$n_c$ &  3  & 1 & $\frac{\alpha_{\rm APE}}{2}$ & $3\rho_{\rm st}$ \\
$n_{\rm APE}$ &  $\frac{6}{\alpha_{\rm APE}}$  & $\frac{2}{\alpha_{\rm APE}}$ & 1 & $\frac{6\rho_{\rm st}}{\alpha_{\rm APE}}$ \\
$n_{\rm st}$ &  $\frac{1}{\rho_{\rm st}}$  & $\frac{1}{3 \rho_{\rm st}}$  & $\frac{\alpha_{\rm APE}}{6 \rho_{\rm st}}$ & 1 \\
\end{tabular}
\caption{\label{tab:rescaling} The matching prefactors between the smoothing schemes of the Wilson flow with time $\tau$, cooling at level $n_c$, APE smearing with level $n_{\rm APE}$ and finally stout smearing with level $n_{\rm st}$. The leftmost column corresponds to the left hand side of the matching equation while the uppermost row to the scale of the right hand side i.e. $n_{\rm APE} \simeq \frac{2}{\alpha_{\rm APE}} n_{c}$.}
\end{center}
\end{table}

\section{Results}
\label{sec:results}

\subsection{Numerical equivalence between different smoothers}
\label{sec:numerical_equivalence}

\begin{figure*}[!h]
 \rotatebox{0}{\includegraphics[width=16cm]{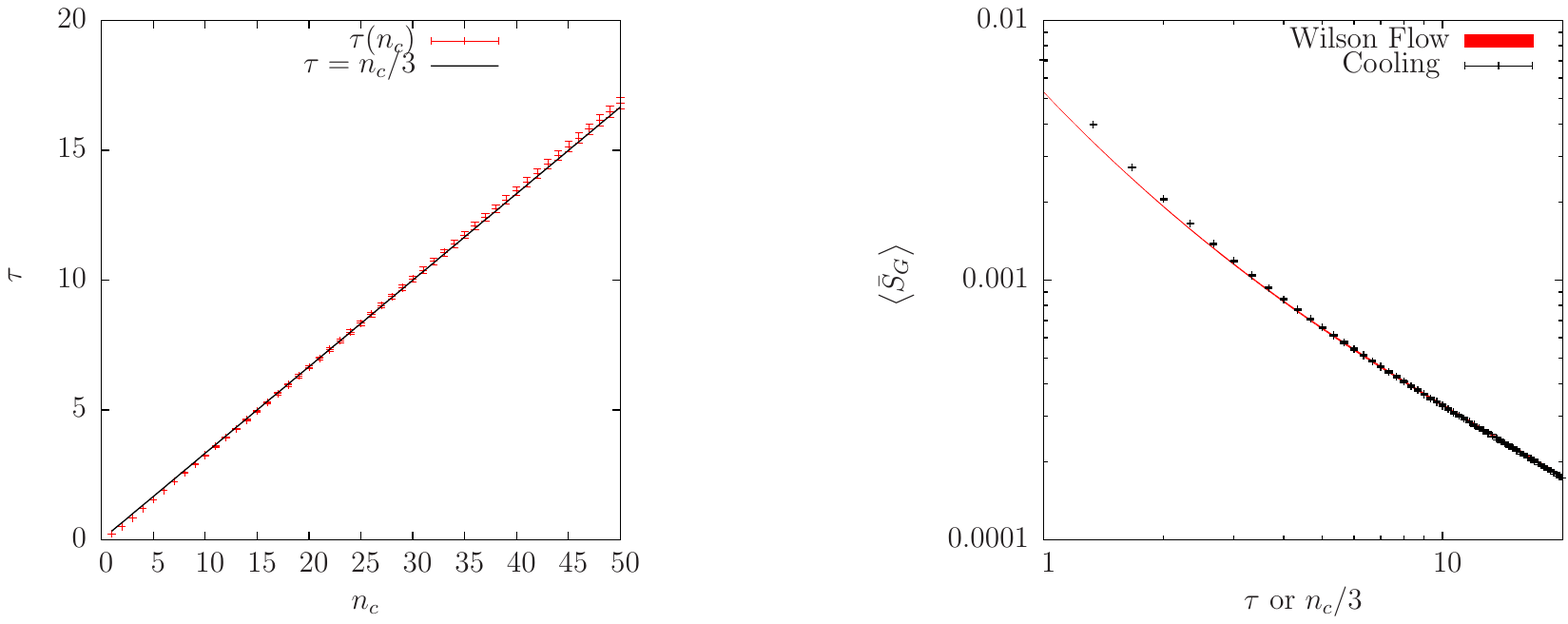}}
 \vspace{-4cm}
\caption{\label{fig:equivalence_cooling} Left Panel: The behavior of $\tau(n_c)$ as a function of $n_c$ for the Wilson smoothing action. The line corresponds to $\tau = n_c/3$. Right Panel: The average action density as a function of the Wilson flow time $\tau$ or the corresponding cooling step $n_c/3$.}
\end{figure*}

We start the presentation of our results by investigating the perturbative matching between the different smoothing schemes which can be used to remove the ultraviolet divergences. We do so by exploring the relation between the average action extracted via both smoothers, looking at the correlation coefficient as well as comparing the topological charge and topological susceptibilities obtained using the two different smoothing schemes. In this section we investigate how the average action density reduces as a function of the two smoothing scales and in the next sections of this paper we investigate the correlation coefficient, the topological charge as well as the topological susceptibility.
\paragraph{Cooling vs. Wilson flow.}
First, we consider the numerical results which have been shown in Refs.~\cite{Bonati:2014tqa,Alexandrou:2015yba}. We confirm the equivalence realized by \eq{eq:perturbative_expression} by investigating how the average action density reduces as a function of the two smoothing scales $\tau$ and $n_c$. In the left panel of \fig{fig:equivalence_cooling} we show the function $\tau(n_c)$ defined as the Wilson flow time $\tau$ for which the average action density 
\begin{eqnarray}
\left\langle \bar{S}_{G} \right\rangle & = & 1 - \left\langle \frac{\sum_{x} \sum_{\substack{
      \mu,\nu=1\\1\leq\mu<\nu}}^4 {\rm Re} \Tr U^{1\times1}_{x,\mu,\nu} (\tau)}{6 Va^{-4}N}   \right\rangle
\label{eq:action_density_definition}
\end{eqnarray}
changes by the same amount as when $n_c$ cooling steps are performed. After a few cooling steps, the data appear to lie on the line $\tau=n_c/3$. Furthermore, in the right panel of \fig{fig:equivalence_cooling}, we present the average action density as a function of $\tau$ or the perturbatively determined values of cooling step $n_c/3$. We observe that after approximately 20 cooling steps, the two sets of data coincide. This confirms that the relation $\tau=n_c/3$ leads to equivalent results for the average action density between the Wilson flow and cooling for small values of $\tau$ and $n_c$. 

\begin{figure*}[!h]
\rotatebox{0}{\includegraphics[width=16cm]{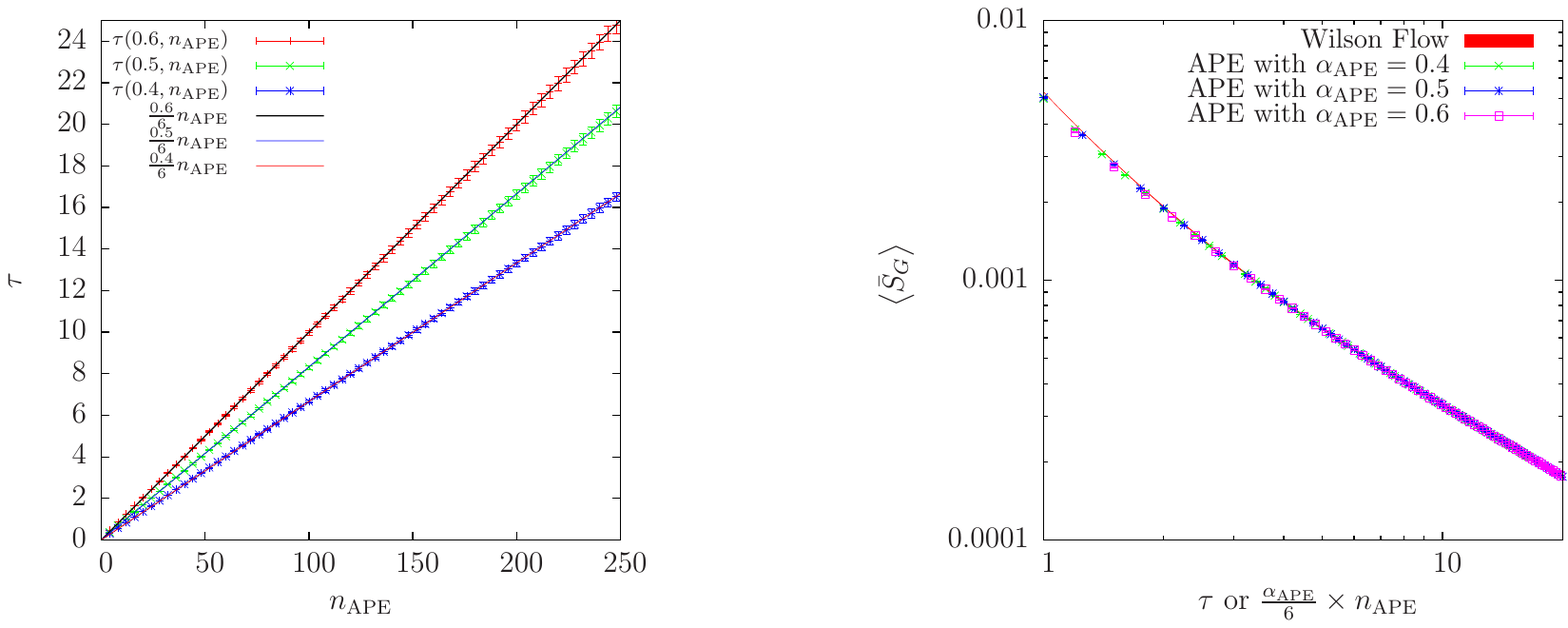}}
 \vspace{-4cm}
\caption{\label{fig:equivalence_ape} Left Panel: The behavior of $\tau(\alpha_{\rm APE}, n_{\rm APE})$ as a function of $n_{\rm APE}$ for $\alpha_{\rm APE}=0.4$, $0.5$, $0.6$. The red, blue and black lines correspond to $\tau = (0.4/6) n_{\rm APE}$, $(0.5/6) n_{\rm APE}$ and $(0.6/6) n_{\rm APE}$ respectively. Right Panel: The average action density as a function of the Wilson flow time $\tau$ or the corresponding rescaled APE smearing step $\frac{\alpha_{\rm APE}}{6} n_{\rm APE}$.}
\end{figure*}

\begin{figure*}[!h]
\vspace{-1cm}
\rotatebox{0}{\includegraphics[width=16cm]{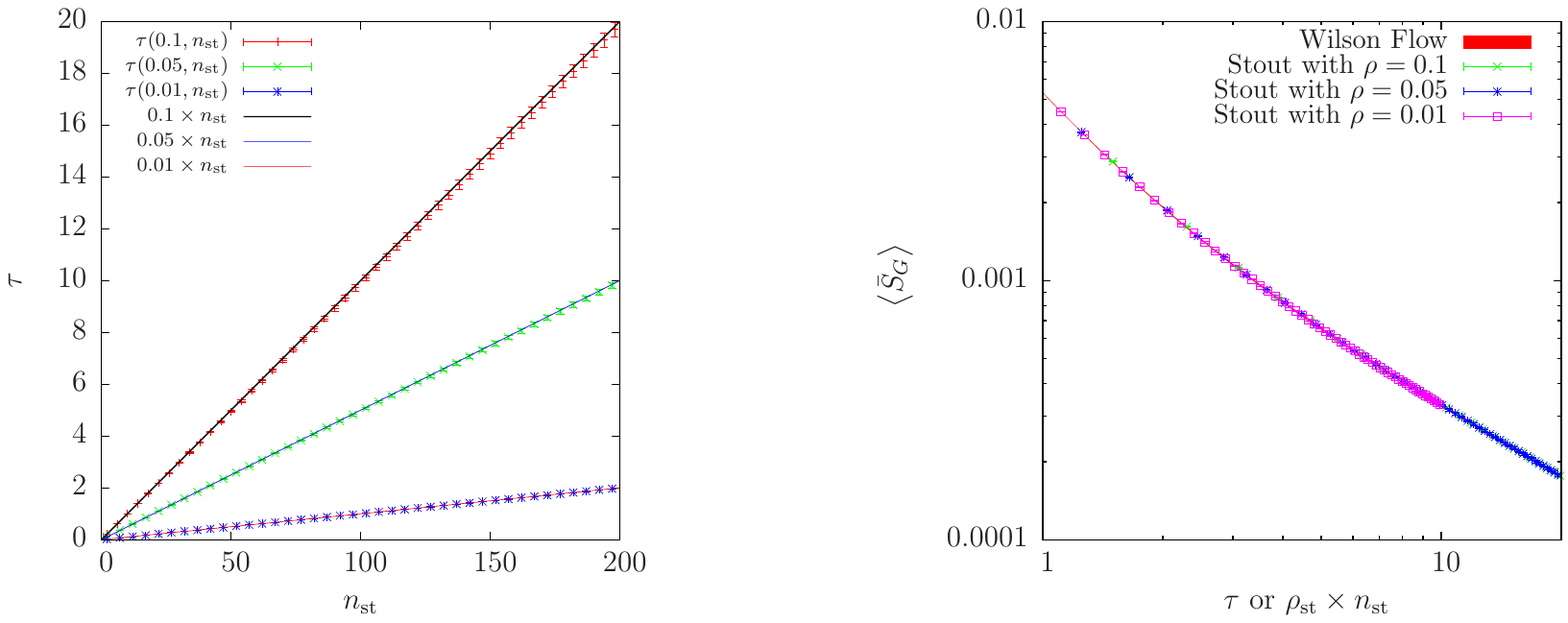}}
 \vspace{-4cm}
\caption{\label{fig:equivalence_stout} Left Panel: The behavior of $\tau(\rho_{st}, n_{\rm st})$ as a function of $n_{\rm st}$ for $\rho=0.01$, $0.05$, $0.1$. The red, blue and black lines corresponds to $\tau = 0.01 n_{\rm st}$, $0.05 n_{\rm st}$ and $0.1 n_{\rm st}$, respectively. Right Panel: The average action density as a function of the Wilson flow time $\tau$ or the corresponding rescaled stout smearing step $\rho_{\rm st} \times n_{\rm st}$.}
\end{figure*}

\begin{figure*}[!h]
\vspace{-1cm}
\rotatebox{0}{\includegraphics[width=16cm]{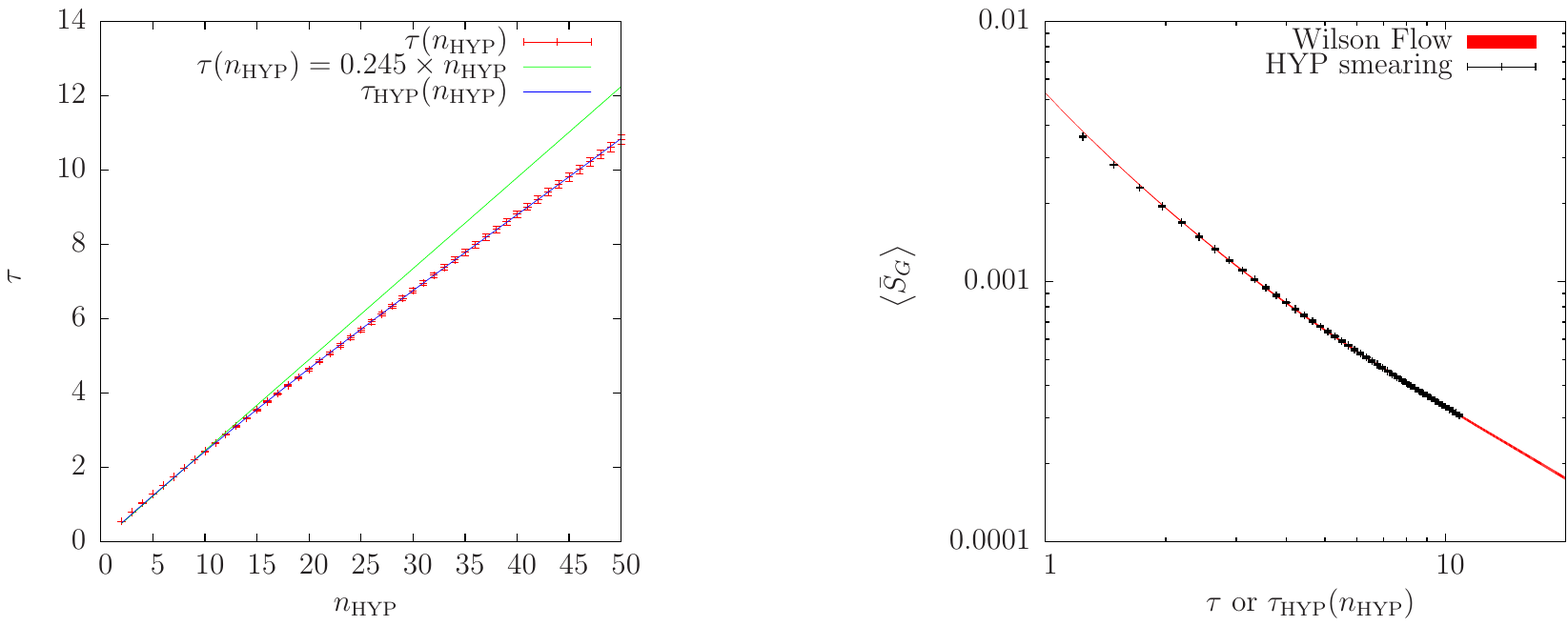}}
 \vspace{-4cm}
\caption{\label{fig:equivalence_hyp} Left Panel: The behavior of $\tau(n_{\rm HYP})$ as a function of $n_{\rm HYP}$. The green line corresponds to a linear approximation which is valid up $t_0$ while the blue line corresponds to the numerical fit $\tau(n_{\rm HYP})$. Right Panel: The average action density as a function of the Wilson flow time $\tau$ and the corresponding numerical matching $\tau({n_{\rm HYP}})$.}
\end{figure*}

\paragraph{APE smearing vs. Wilson flow.}
We move now to the investigation of the numerical equivalence between the APE smearing and the Wilson flow. To test the formula of~\eq{eq:ape_ground_expression}, we smoothed the gauge configurations via APE smearing for three different values of $\alpha_{\rm APE}$, namely $\alpha_{\rm APE}=0.4$, $0.5$ and $0.6$. Subsequently, we calculated the function $\tau(\alpha_{\rm APE}, n_{\rm APE})$ defined as the Wilson flow time $\tau$ for which the average action density reduces by the same amount as when $n_{\rm APE}$ smearing steps, for a fixed value of $\alpha_{\rm APE}$, are performed. In the left panel of \fig{fig:equivalence_ape}, we demonstrate $\tau(\alpha_{\rm APE}, n_{\rm APE})$ for the three different values of $\alpha_{\rm APE}$. The data points for the three values of $\alpha_{\rm APE}$ appear to agree with the lines $\tau=(\alpha_{\rm APE}/6) \times n_{\rm APE}$ providing strong evidence that \eq{eq:ape_ground_expression} provides the right rescaling for which Wilson flow and APE smearing become numerically equivalent. Furthermore, in the right panel of \fig{fig:equivalence_ape}, we provide the average action density as a function of $\tau$ and $(\alpha_{\rm APE}/6) \times n_{\rm APE}$ demonstrating that the four sets of data perfectly agree with each other. 
\paragraph{Stout smearing vs. Wilson flow.}
We now move to the numerical correspondence between the two smoothing schemes of stout smearing and the Wilson flow. In order to test this equivalence, we smoothed the gauge configurations using stout smearing and three different values of the $\rho_{\rm st}$, namely $\rho_{\rm st}=0.01$, $0.05$ and $0.1$. Now we define the function  $\tau(\rho_{\rm st}, n_{\rm st})$, like in the previous cases, as the Wilson flow time $\tau$ for which the average action density alters by the same amount as when $n_{\rm st}$ stout smearing steps with a given $\rho_{\rm st}$ are performed. In the left panel of \fig{fig:equivalence_stout}, we present the function $\tau(\rho_{\rm st}, n_{\rm st})$ for the three different values of the parameter $\rho_{\rm st}$. The data points for the three values of $\rho_{\rm st}$ appear to agree with the lines $\tau=\rho_{\rm st} \times n_{\rm st}$ providing strong evidence that \eq{eq:stout_ground_expression} provides the right rescaling for which Wilson flow and stout smearing become numerically equivalent. This agreement sets in for $\tau \simeq 5$, $2$, $1$ for $\rho_{\rm st} = 0.1, \ 0.05, \ 0.01$, respectively, demonstrating that the smaller the value of $\rho_{\rm st}$ the closer we approach the Wilson flow. Furthermore, in the right panel of \fig{fig:equivalence_stout}, we provide the average action density as a function of $\tau$ or $\rho_{\rm st} \times n_{\rm st}$ for the Wilson flow or stout smearing, respectively, demonstrating that the four sets of data perfectly agree with each other. Similar comparisons of the topological charge and susceptibility are provided in Section~\ref{sec:comparison_topological_charge}. 
\paragraph{HYP smearing vs. Wilson flow.}
As we have already mentioned in Section~\ref{sec:perturbative_APE}, the peculiar construction of the HYP smearing staples prohibits the extraction of a linear perturbative rescaling between the Wilson flow time $\tau$ and the number of HYP smearing steps $n_{\rm HYP}$. Thus, instead, we attempted a numerical fit using a parametrization of $\tau(n_{\rm HYP})$ in $n_{\rm HYP}$ according to Eq.~(\ref{eq:HYP_ansatz}). In the left panel of \fig{fig:equivalence_hyp}, we provide the function $\tau(n_{\rm HYP})$. Obviously, the sketched behaviour deviates from a linear response (green line) such as those observed for cooling, stout and APE smearing. This suggests that it is impossible to extract a tree-level perturbative expression which relates this smoother with the others. We fit the data using \eq{eq:HYP_ansatz} and extract the coefficients $A=0.25447(32)$, $B=-0.001312(90)$ as well as $C=1.217(91)\times 10^{-5}$. Of course, these numbers depend on the parameters $\alpha_{\rm HYP1, \ 2, \ 3}$. In the right panel of \fig{fig:equivalence_hyp}, we show the average action density for the HYP smearing as a function of the rescaling equation of \eq{eq:HYP_ansatz} as well as the average action density for the Wilson flow. Clearly, the two lines coincide, demonstrating the realization of a numerical equivalence through Eq.~(\ref{eq:HYP_ansatz}). 

\subsection{Field theoretic topological charges on a single configuration}
\label{sec:comparison_topological_charge}

\begin{figure}[h!]
 \vspace{0cm}
\centerline{\rotatebox{0}{\includegraphics[width=10cm]{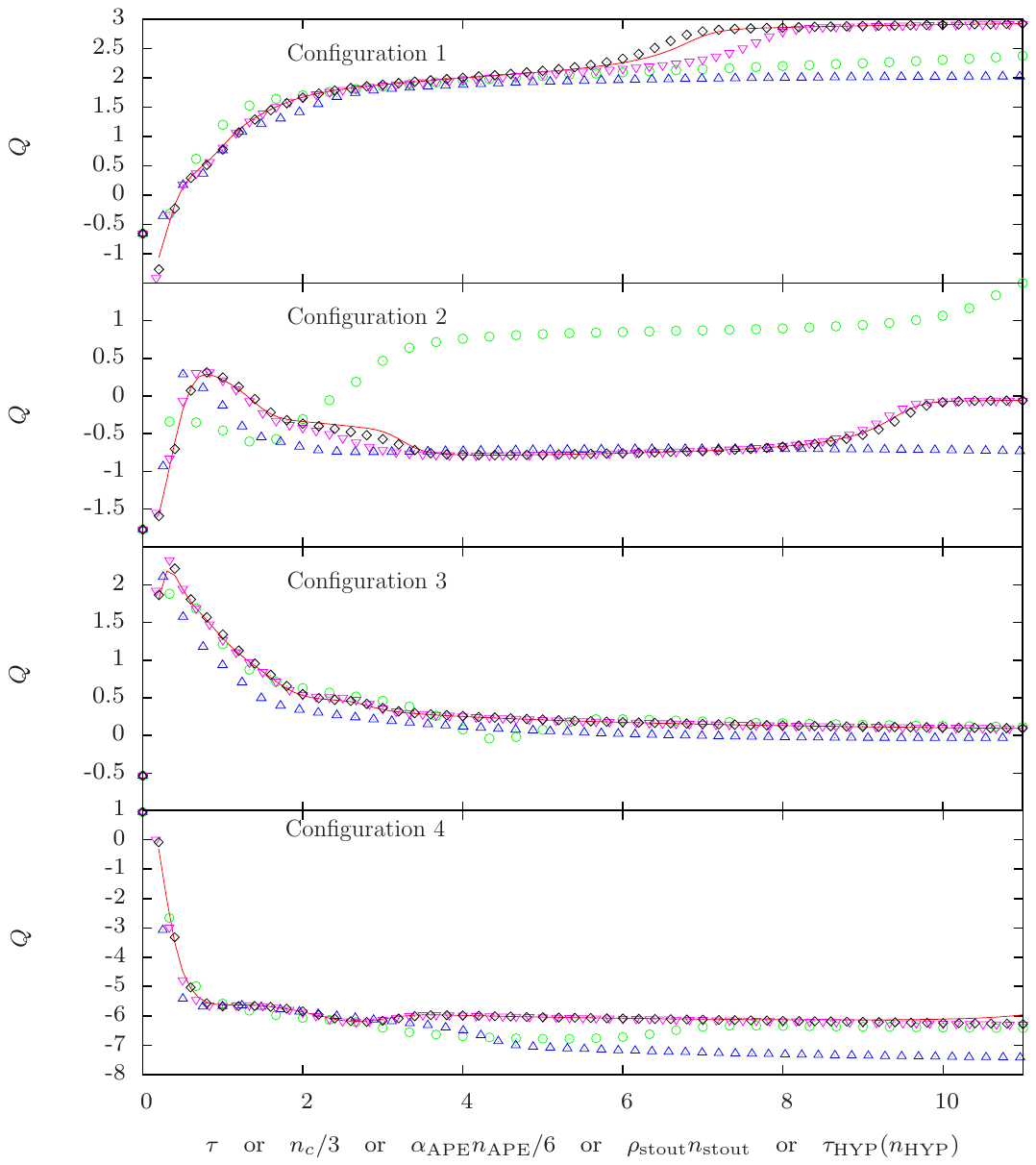}}}
 \vspace{-2cm}
\caption{\label{fig:comparison_topological_charge} For four different gauge field configurations, we show the clover definition of the topological charge as a function of the gradient flow time $\tau$ for Wilson flow (line in red), $n_c/3$ for cooling (${\Large\color{green} {\circ}}$), $0.5 \times n_{\rm APE}/6$ for APE smearing ($\color{magenta} \triangledown$), $0.05 \times n_{\rm stout}$ for stout smearing ($\color{black} \diamond$) and  $\tau_{\rm HYP} (n_{\rm HYP}))$ for HYP smearing ($\color{blue} \triangle$). For this ensemble $t_0\simeq 2.5a^2$.}
\end{figure}

The behaviour of the topological charge $Q$ for single configurations as a function of the gradient flow time $\tau$ and for matched smoothing scales for cooling, APE, stout as well as HYP smearing has been investigated. In \fig{fig:comparison_topological_charge}, we present the clover definition of the topological charge as a function of $\tau$, $n_c/3$, $\alpha_{\rm APE} n_{\rm APE}/6$ for $\alpha_{\rm APE}=0.5$, $\rho_{\rm st} n_{\rm st}$  for $\rho_{\rm st}=0.05$ and $\tau_{\rm HYP}(n_{\rm HYP})$, for four randomly chosen configurations in the ensemble b40.16; each panel corresponds to each configuration.

Strikingly, the topological charge obtained with APE smearing as well as stout smearing appears to exhibit significant agreement with that extracted via the Wilson flow.  This, of course, occurs if $n_{\rm APE}$ and $n_{\rm st}$ are rescaled according to  \eq{eq:ape_ground_expression} and \eq{eq:stout_ground_expression}, respectively. Namely, the approximate plateaus observed in \fig{fig:comparison_topological_charge} for the three smoothing schemes appear to coincide. An interesting phenomenon is the fine structure occuring when a small instanton or anti-instanton (dislocations) start to drop off the lattice (in case one considers the semiclasical instanton picture). For instance, in the uppermost panel of \fig{fig:comparison_topological_charge}, and between $\tau=6-8$, the approximate plateau shifts from $Q\simeq 2$ to $Q \simeq 3$. During this transition, we observe that the topological charge $Q$ between the two smoothing schemes and the Wilson flow diverge.  Nonetheless, this disagreement appears to vanish as we choose smaller values of $\rho_{\rm st}$ and $\alpha_{\rm APE}$. In addition, we observe that $Q$ obtained via stout smearing is closer to Wilson flow than APE. The above suggest that the effect of APE and stout smearing on the gauge fields resemble, to a high extent, the Wilson flow. In fact, one does not expect two smoothing procedures to provide equal topological charges since different smoothers carry different lattice artifacts and do not need to agree at non-zero values of the lattice spacing.  
Indubitably, the topological charges will become closer as the lattice spacing decreases. Thus, as one approaches the continuum limit, any two
different procedures converge. Nevertheless, for APE and more strikingly for stout smearing and at finite lattice spacing, the topological charge is, in good approximation, equal to the value extracted via the Wilson flow. We note that the topological charge itself is not the main quantity of interest -- the physically relevant observable is the topological susceptibility, which measures the fluctuations of the topological charge.

Turning now to cooling, we demonstrate that the topological charge for a given configuration yields not necessarily the same values as the gradient flow even if we rescale $n_c$ according to \eq{eq:perturbative_expression}. Of course, this is not a new observation. Similar comparison which reveals such a possible difference has been published in \cite{Alexandrou:2015yba}. Once again, we emphasize that the topological charge for both smoothing procedures will  become equivalent if one decreases enough the lattice spacing. As we already know from Ref.~\cite{Alexandrou:2015yba}, both smoothers yield approximately the same topological susceptibility although the topological charge is not necessarily the same; this is demonstrated also in \sec{sec:topological_susceptibility} of this manuscript.

Finally, we discuss the comparison of the results on $Q$ obtained with HYP smearing and the Wilson flow. Similarly with cooling, if we rescale $n_{\rm HYP}$ with the numerically-extracted formula of \eq{eq:HYP_ansatz}, the topological charge $Q$ exhibits approximate equivalence with $Q$ resulted by the Wilson flow for some short range of low values of $\tau$; however, for this range of $\tau$ the value of $Q$ we get is highly dominated by the UV noise. The structure of HYP smearing includes staples which extend beyond the nearest neighbouring links to the original link. This may lead to a supposition that the topological charge obtained via HYP would differ enough from that obtained by the other four smoothers. Interestingly, this does not occur in a noteworthy manner. Of course, once more, the fluctuations of the topological charge are just a measure of the topological susceptibility. Hence, it would be interesting to investigate the response of HYP in this physical quantity; we discuss this topic in \sec{sec:topological_susceptibility}.

\subsection{Monte Carlo histories and distribution histograms}
\label{sec:histories_histograms}

In this subsection, we show some typical features of the topological charge evaluated with one representative fermionic definition\footnote{Mind that here we do not match the smoothing scales, but choose typical smoothing levels applied for the overlap definition and the gluonic definition. The smoothing in both cases serves very different purposes. For the gluonic definition, smoothing is mandatory, since the topological charge from unsmeared configurations is dominated by ultraviolet noise and a meaningful extraction of the charge is not possible. In the case of the overlap index definition, the topological charge can be meaningfully extracted even without any smoothing and the latter is applied only to reduce the computational time.} (index of the overlap operator evaluated on configurations with 1 step of HYP smearing applied) and one representative gluonic definition (with the Wilson flow at flow time $t_0$).
This serves two purposes.
First, we show that correlations between the two classes of definitions are apparent even from a visual inspection of Monte Carlo histories.
Second, we want to investigate whether the distribution of the topological charge is approximately Gaussian and, for the gluonic case, whether clustering of the values around integers occurs.

\begin{figure}[h!]
\begin{center}
\rotatebox{0}{\hspace{-0.725cm} \includegraphics[width=10cm]{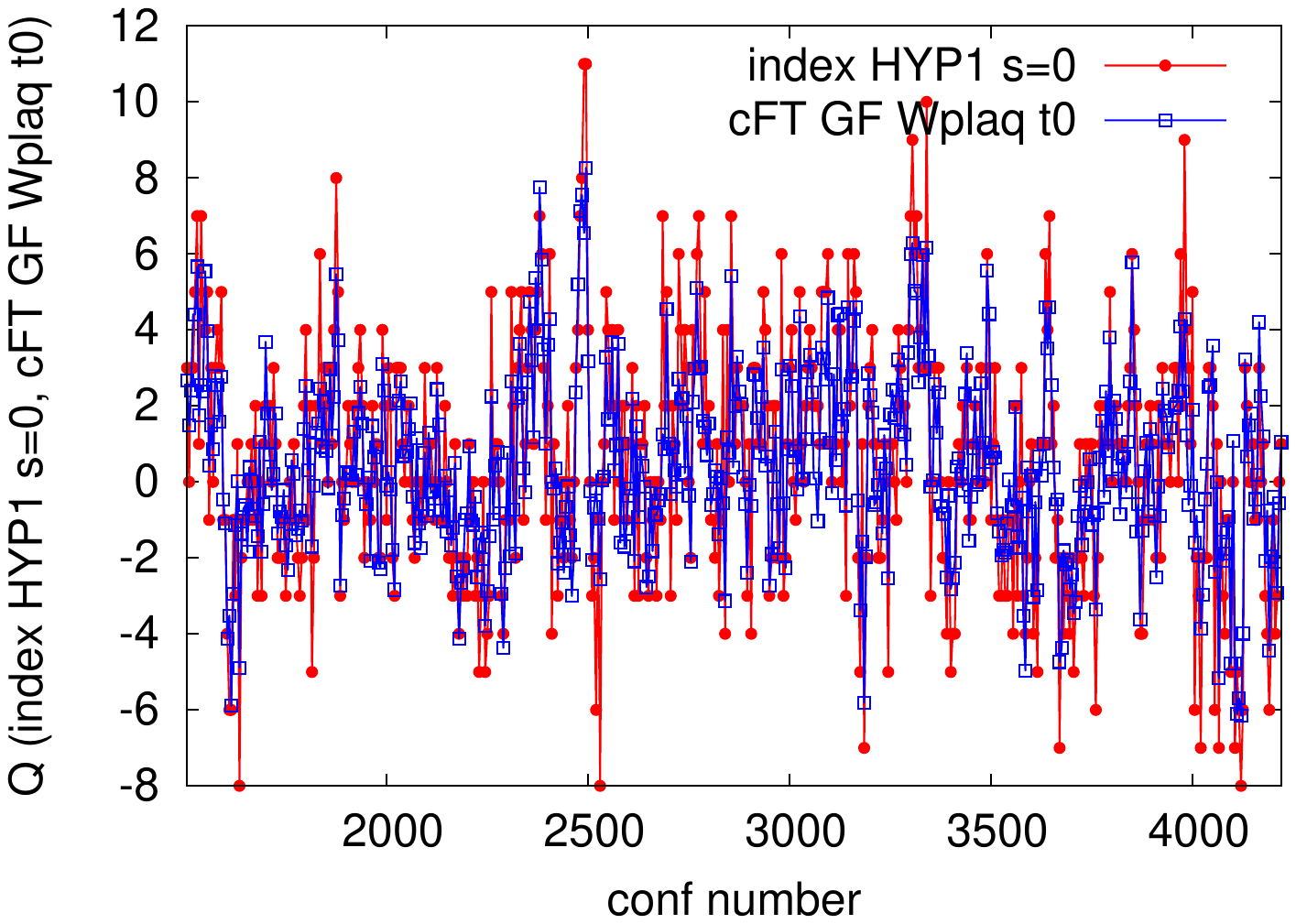}}
\end{center}
\caption{\label{fig:MC}  Monte Carlo histories for one representative fermionic definition (index of the overlap operator evaluated on configurations with 1 step of HYP smearing applied) and one representative gluonic definition (Wilson flow at flow time $t_0$).}
\end{figure}

\begin{figure}[h!]
\begin{center}
\rotatebox{0}{\hspace{-0.725cm} \includegraphics[width=10cm]{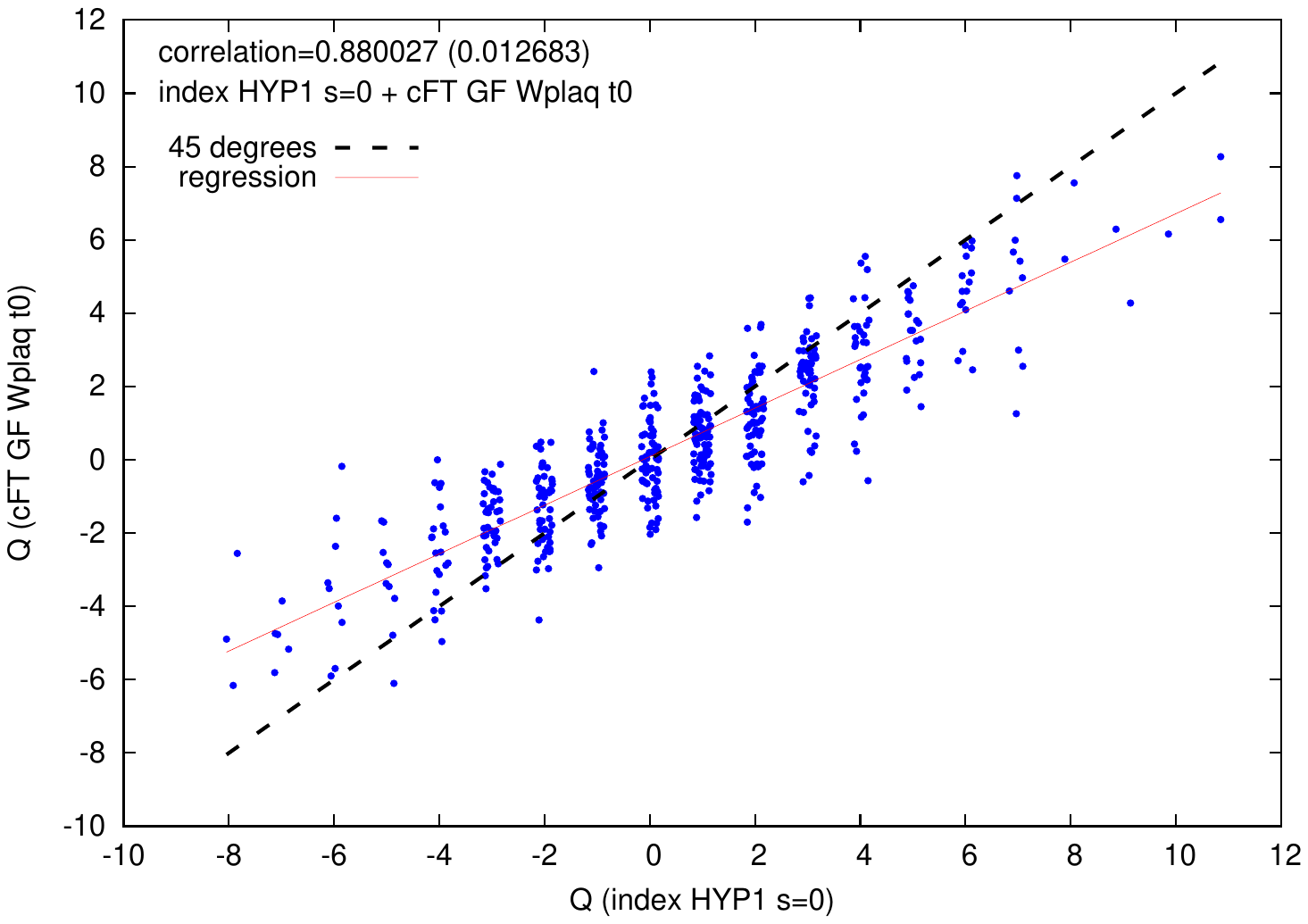}}
\end{center}
\caption{\label{fig:scatter} 
Scatterplot of the topological charge for one representative fermionic definition (index of the overlap operator evaluated on configurations with 1 step of HYP smearing applied) and one representative gluonic definition (Wilson flow at flow time $t_0$).
For better visibility, the integer values of the index were randomly shifted by a small non-integer value.
}
\end{figure}

The Monte Carlo histories are shown in Fig.~\ref{fig:MC}. All relevant topological sectors seem to be scanned correctly and there are no excessive autocorrelations.
For the latter, we used the bootstrap procedure with blocking and we find that for different definitions, measurements with a step of 5 configurations (10 Monte Carlo trajectories, {i.e.\ after each saved trajectory, one was unsaved in the generation}) yield an integrated autocorrelation time $\tau_{\rm int}\in[0.5,\,2]$ for the topological charge {in units of measured configurations}. The lowest autocorrelation is obtained obviously for the field theoretic definition without smearing, since one then basically observes uncorrelated ultraviolet fluctuations. All meaningful definitions yield compatible autocorrelation times with $\tau_{\rm int}\approx1.7(3)$. 
The correlation between the values of $Q$ from the two definitions in Fig.~\ref{fig:MC} is obvious even without computing the correlation coefficient (which is 88\%; see the next subsection for a systematic analysis of correlations between different definitions).
This is also illustrated in a scatterplot (Fig.~\ref{fig:scatter}).
Although the correlation is evident in this plot, it demonstrates that the value of the topological susceptibility is larger for the index definition, indicating that cut-off effects affect the two definitions in a somewhat different manner.

\begin{figure}[h!]
\begin{center}
\vspace{-0.5cm}
\rotatebox{0}{\hspace{-0.725cm} \includegraphics[width=7.8cm]{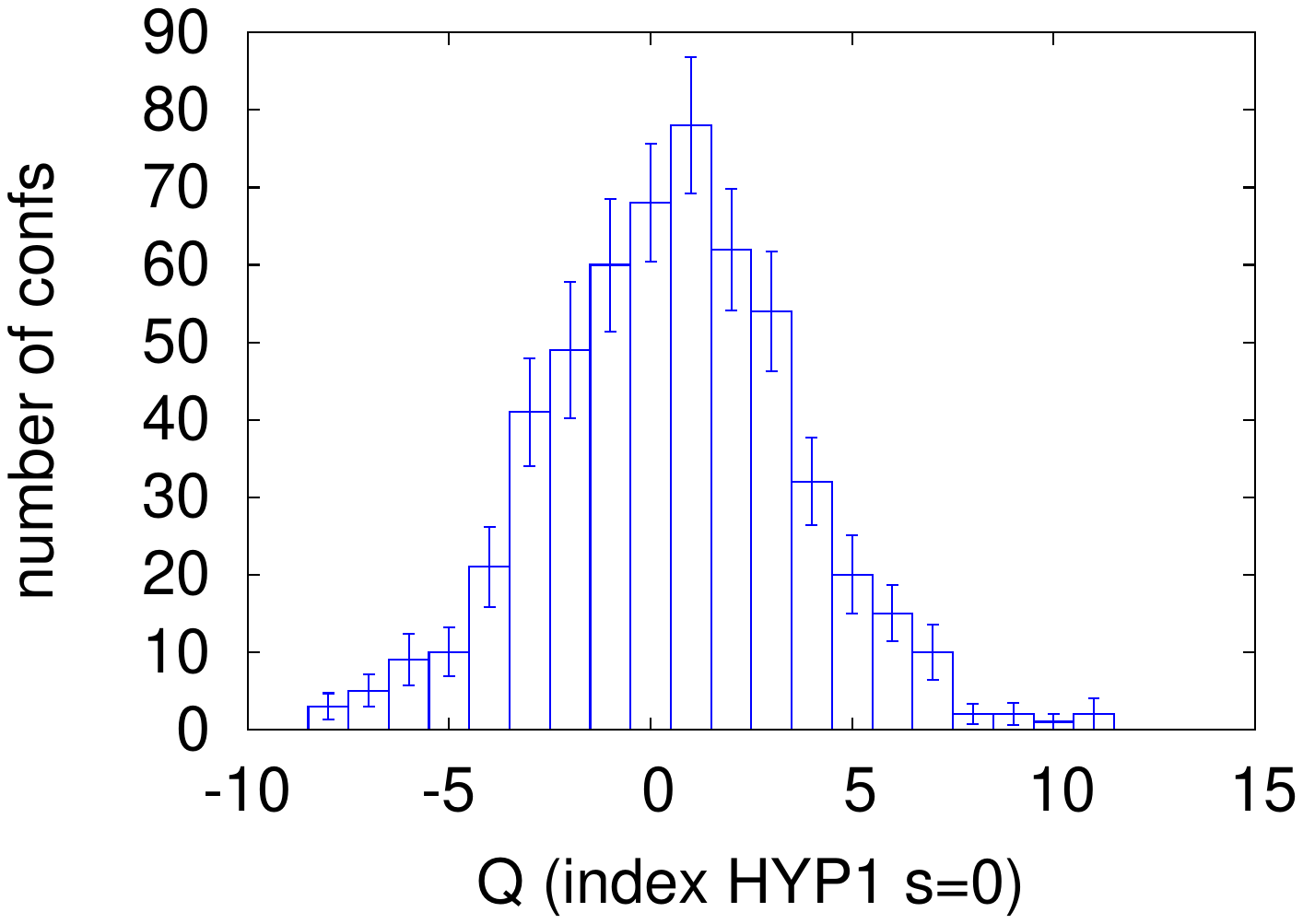}}
\vspace{-0.8cm}
\rotatebox{0}{\hspace{-0.725cm} \includegraphics[width=7.8cm]{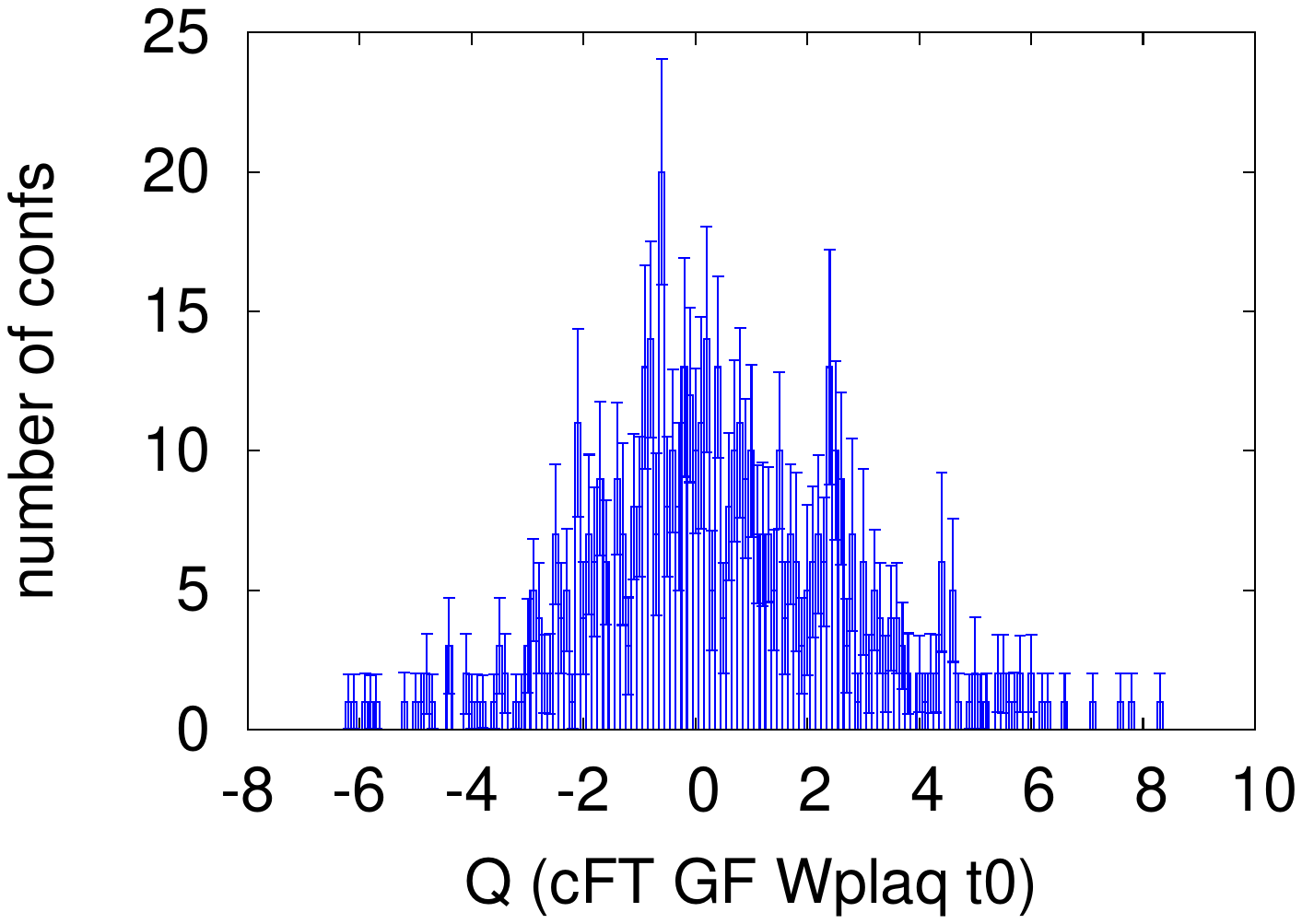}}
\vspace{-0.8cm}
\rotatebox{0}{\hspace{-0.725cm} \includegraphics[width=7.8cm]{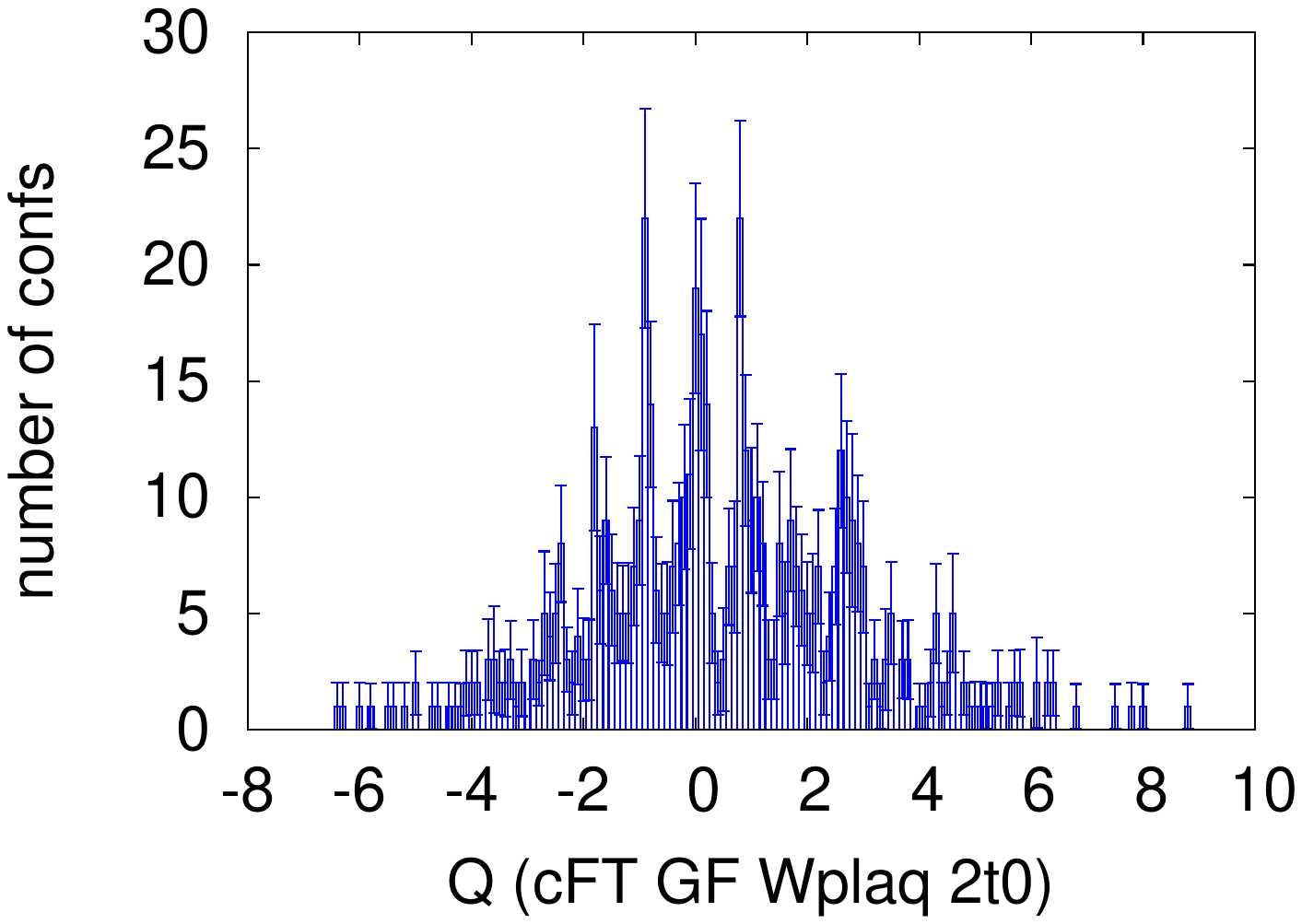}}
\vspace{-0.8cm}
\rotatebox{0}{\hspace{-0.725cm} \includegraphics[width=7.8cm]{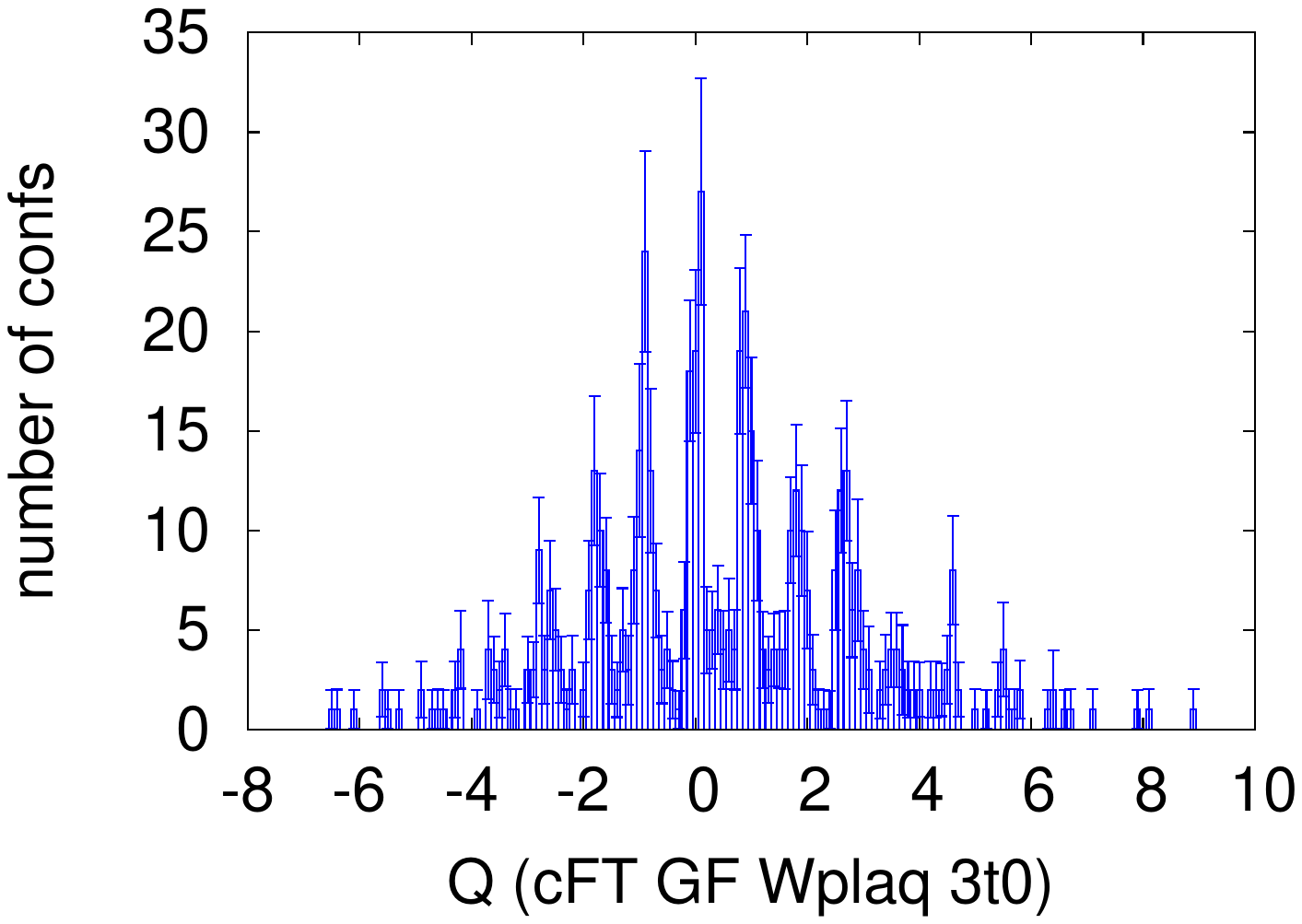}}
\end{center}
\caption{\label{fig:b40-hist} Histograms of the topological charge for ensemble
b40.16. The employed definitions are: index of the overlap operator evaluated on configurations with 1 step of HYP smearing applied (top left) and the Wilson flow at flow times $t_0$ (top right), $2t_0$ (bottom left) and $3t_0$ (bottom right).} 
\end{figure}

Next, we show typical histograms (Fig.~\ref{fig:b40-hist}) obtained with a fermionic definition (again, index HYP1 $s=0$) and a gluonic one (Wilson flow at flow times $t_0$, $2t_0$ and $3t_0$).
For the index definition, we obtain a distribution that is compatible with a Gaussian\footnote{It is worth to mention that the distribution is not expected to be ideally Gaussian. For an investigation of non-Gaussianities in the quenched case, see Ref.~\cite{DElia:2003zne,Panagopoulos:2011rb,Ce:2015qha,Bonati:2015sqt}. However, the detection of such non-Gaussianities requires very large statistics, at least an order of magnitude larger than in our present work.}.
For the field theoretic definition, we used an interval width of 0.1 to detect clustering around values close to an integer.

When the flow time is relatively small, of the order of $t_0$, basically no clustering is observed. However, when increasing the flow time, at $2t_0$ and $3t_0$, the filtering out of the ultraviolet noise is enough to discern peaks at positions close to 0.9, 1.8, 2.7 etc. (for the ensemble b40.16; for other lattice setups or other discretizations of the field stength tensor, the values can be different, but they will also be multiples of some number relatively close to 1).
These non-integer positions of the peaks are sometimes ``corrected'' as mentioned in footnote \ref{fn}, but this procedure is artificial.
In particular, it is not needed to obtain the correct value of the topological susceptibility in the continuum limit.
What is, however, relevant for a correct continuum limit is that the smearing procedure defines a proper smoothing scale, as discussed in the previous section.
Such a scale is naturally defined in the gradient flow procedure and in the other smoothing schemes via the matching to gradient flow.
If one applies e.g. APE smearing without proper matching to GF, one can not define a consistent procedure of extrapolating to the continuum limit.
The traditional method of looking for a plateau in the smearing history is not enough, as it does not define a valid smoothing scale.
However, if APE smearing (or any other non-GF type of smoothing) is matched to GF, such a smoothing scale is well-defined and one expects the proper continuum limit for the topological susceptibility.

\subsection{Correlations between different definitions}
For a complete comparison of as many definitions of the topological charge as possible, we concentrated on our ensemble b40.16, i.e. one with the smallest lattice volume and hence the smallest cost of the computations.
For this ensemble, we took into account all of the definitions listed in Tab.~\ref{tab:defs}.
We focus on the correlations between different definitions, expressed by the standard correlation coefficient, normalized to be in the interval $[-1,1]$.
We used the bootstrap procedure (with blocking) to compute the error and the influence of autocorrelations.

To understand better the relations between all definitions, we discuss below correlations between different groups of definitions.
We start with a general comparison, including the typical representatives of each family from  Tab.~\ref{tab:defs}.
Then, we concentrate on the fermionic definitions. The comparison between the most abundant family of field theoretic definitions with several types of smearing that can be applied to the gauge fields to filter out the ultraviolet noise as well as different smoothing actions for the gradient flow is provided in the two sections in the Appendix.

\subsubsection{Main comparison}
\label{sec:main}
In this subsection, we choose the following definitions as typical representatives:
\begin{itemize}
\item index of the overlap Dirac operator applied to non-smeared and smeared gauge fields (with one iteration of HYP smearing) (definitions 1, 3 from Tab.~\ref{tab:defs}),
\item spectral flow of the Wilson-Dirac operator computed on gauge fields with one or five iterations of HYP smearing (4, 6),
\item spectral projectors with two values of the threshold parameter $M$ (9, 10),
\item field theoretic without smearing (12),
\item field theoretic with GF at flow time $t_0$, three types of smoothing action (16, 22, 25),
\item field theoretic with cooling matched to GF at flow time $t_0$, three types of smoothing action (28, 30, 32),
\item field theoretic with stout smearing matched to GF at flow time $t_0$ (34),
\item field theoretic with APE smearing matched to GF at flow time $t_0$ (40),
\item field theoretic with HYP smearing matched to GF at flow time $t_0$ (44).
\end{itemize}

\begin{figure}[h!]
\rotatebox{0}{\hspace{-0.725cm} \includegraphics[width=10cm]{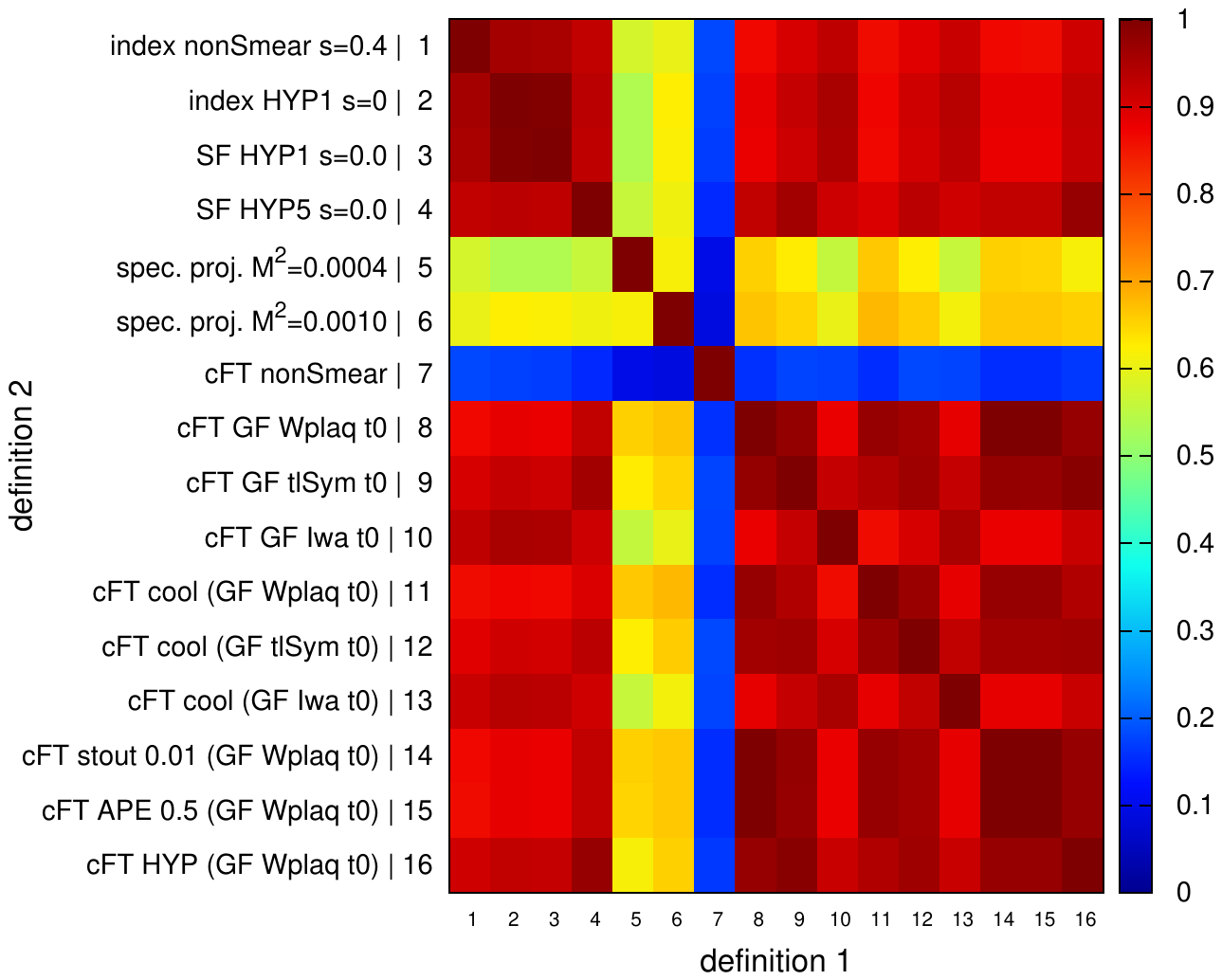}}
\caption{\label{fig:b40-corr0} 
Main comparison of selected definitions of the topological charge. The correlation between different definitions is colour-coded (note the scale is different than in Figs.~\ref{fig:b40-corr1},~\ref{fig:b40-corr2},~\ref{fig:b40-corr3},~\ref{fig:b40-corr4}).}
\end{figure}

\renewcommand{\tabcolsep}{0.06cm}
\begin{table*}[h!]
  \centering
\begin{scriptsize}
  \begin{tabular}[]{c|cccccccccccccccc}
& 1 & 2 & 3 & 4 & 5 & 6 & 7 & 8 & 9 & 10 & 11 & 12 & 13 & 14 & 15 & 16 \\
\hline
1 & 1 & 0.96(0) & 0.95(0) & 0.92(1) & 0.58(4) & 0.60(3) & 0.18(6) & 0.86(1) & 0.90(1) & 0.93(0) & 0.86(1) & 0.89(1) & 0.91(0) & 0.86(1) & 0.86(1) & 0.91(1) \\
2 & 0.96(0) & 1 & 0.99(0) & 0.93(0) & 0.54(4) & 0.62(3) & 0.17(4) & 0.88(1) & 0.92(0) & 0.95(0) & 0.87(1) & 0.91(1) & 0.94(0) & 0.88(1) & 0.88(1) & 0.92(0) \\
3 & 0.95(0) & 0.99(0) & 1 & 0.93(0) & 0.54(4) & 0.62(3) & 0.17(4) & 0.88(1) & 0.91(0) & 0.95(0) & 0.86(1) & 0.90(0) & 0.93(0) & 0.88(1) & 0.88(1) & 0.92(0) \\
4 & 0.92(1) & 0.93(0) & 0.93(0) & 1 & 0.56(4) & 0.61(3) & 0.15(4) & 0.92(0) & 0.96(0) & 0.91(0) & 0.90(1) & 0.93(0) & 0.91(0) & 0.92(0) & 0.92(0) & 0.97(0) \\
5 & 0.58(4) & 0.54(4) & 0.54(4) & 0.56(4) & 1 & 0.62(4) & 0.10(3) & 0.66(3) & 0.63(3) & 0.56(4) & 0.66(3) & 0.62(3) & 0.56(4) & 0.65(3) & 0.65(3) & 0.62(3) \\
6 & 0.60(3) & 0.62(3) & 0.62(3) & 0.61(3) & 0.62(4) & 1 & 0.09(4) & 0.67(3) & 0.65(3) & 0.60(4) & 0.68(3) & 0.66(3) & 0.61(4) & 0.66(3) & 0.66(3) & 0.65(3) \\
7 & 0.18(6) & 0.17(4) & 0.17(4) & 0.15(4) & 0.10(3) & 0.09(4) & 1 & 0.16(4) & 0.18(4) & 0.17(4) & 0.15(4) & 0.18(4) & 0.18(4) & 0.16(4) & 0.16(4) & 0.17(4) \\
8 & 0.86(1) & 0.88(1) & 0.88(1) & 0.92(0) & 0.66(3) & 0.67(3) & 0.16(4) & 1 & 0.97(0) & 0.88(1) & 0.97(0) & 0.96(0) & 0.88(1) & 1.00(0) & 1.00(0) & 0.97(0) \\
9 & 0.90(1) & 0.92(0) & 0.91(0) & 0.96(0) & 0.63(3) & 0.65(3) & 0.18(4) & 0.97(0) & 1 & 0.92(0) & 0.94(0) & 0.96(0) & 0.92(0) & 0.97(0) & 0.97(0) & 0.99(0) \\
10 & 0.93(0) & 0.95(0) & 0.95(0) & 0.91(0) & 0.56(4) & 0.60(4) & 0.17(4) & 0.88(1) & 0.92(0) & 1 & 0.86(1) & 0.90(1) & 0.95(0) & 0.88(1) & 0.88(1) & 0.91(0) \\
11 & 0.86(1) & 0.87(1) & 0.86(1) & 0.90(1) & 0.66(3) & 0.68(3) & 0.15(4) & 0.97(0) & 0.94(0) & 0.86(1) & 1 & 0.97(0) & 0.88(1) & 0.97(0) & 0.97(0) & 0.94(0) \\
12 & 0.89(1) & 0.91(1) & 0.90(0) & 0.93(0) & 0.62(3) & 0.66(3) & 0.18(4) & 0.96(0) & 0.96(0) & 0.90(1) & 0.97(0) & 1 & 0.92(0) & 0.96(0) & 0.96(0) & 0.96(0) \\
13 & 0.91(0) & 0.94(0) & 0.93(0) & 0.91(0) & 0.56(4) & 0.61(4) & 0.18(4) & 0.88(1) & 0.92(0) & 0.95(0) & 0.88(1) & 0.92(0) & 1 & 0.88(1) & 0.88(1) & 0.91(0) \\
14 & 0.86(1) & 0.88(1) & 0.88(1) & 0.92(0) & 0.65(3) & 0.66(3) & 0.16(4) & 1.00(0) & 0.97(0) & 0.88(1) & 0.97(0) & 0.96(0) & 0.88(1) & 1 & 1.00(0) & 0.97(0) \\
15 & 0.86(1) & 0.88(1) & 0.88(1) & 0.92(0) & 0.65(3) & 0.66(3) & 0.16(4) & 1.00(0) & 0.97(0) & 0.88(1) & 0.97(0) & 0.96(0) & 0.88(1) & 1.00(0) & 1 & 0.97(0) \\
16 & 0.91(1) & 0.92(0) & 0.92(0) & 0.97(0) & 0.62(3) & 0.65(3) & 0.17(4) & 0.97(0) & 0.99(0) & 0.91(0) & 0.94(0) & 0.96(0) & 0.91(0) & 0.97(0) & 0.97(0) & 1 \\
  \end{tabular}
\end{scriptsize}
\caption{Main comparison of selected definitions of the topological charge. The numbers correspond to the numbering given in Fig.~\ref{fig:b40-corr0}. We give the correlation coefficient between different definitions and its error (0 means that the error is smaller than 0.005).}
  \label{tab:corr0}
\end{table*}

\noindent For the field theoretic definitions, we always use the clover discretization of the topological charge operator for this comparison.
The effects of using other discretizations will be considered in one of the further subsections.

Our results are summarized in Fig.~\ref{fig:b40-corr0} and Tab.~\ref{tab:corr0}.
In general, we observe very high correlations among different definitions of the topological charge, typically between 85\% and 100\% (the latter for equivalent definitions).

There are two exceptions to this feature.
As expected, the field theoretic definition applied to non-smeared configuration measures basically only ultraviolet noise.
The correlation coefficient with respect to other definitions is very small, although non-zero, which suggests that even on non-smeared configurations, some residual signal of the topological charge remains (the correlation coefficient as well as the topological susceptibility are non-zero with statistical significance; nevertheless, reliably extracting the susceptibility from non-smeared gluonic definition is not possible).
Nevertheless, smoothing of gauge fields is mandatory in the field theoretic definition to obtain a meaningful result.
The second exception is the spectral projector method, which yields a 55\%-65\% correlation with respect to other cases.
One reason for this is obviously the stochastic ingredient in the estimation of $Q$ with this method.
However, with 12 stochastic sources that were used, this stochastic ingredient is largely, although not completely, eliminated.
Apparently, there are other effects which result in the rather moderate correlation -- it is very likely that these are cut-off effects at the considered, relatively coarse, lattice spacing.
Also, one should keep in mind that the spectral projector observable ${\cal C}$ was never intended to be used as a topological charge observable -- it was rather introduced for computations of the topological susceptibility, for which the gauge ensemble average and the stochastic correction play an important role.

Within the group of highly correlated definitions, we observe that the fermionic definitions are slightly more correlated with themselves than with the gluonic ones.
Concerning the correlation of fermionic and gluonic definitions, it is interesting to note that the former are visibly better correlated with field theoretic ones with improved smoothing actions -- while the correlation with GF/cooling with the Wilson plaquette smoother is around 86\%-88\%, the one with the Iwasaki smoothing action is up to 95\%.
If, however, one considers the spectral flow (or index) computed on configurations with 5 steps of HYP smearing applied, this effect is alleviated and actually the Iwasaki smoother gives a consistent result with the Wilson plaquette one, while tree-level Symanzik improved is slightly more correlated.
We also observe that the correlation between fermionic definitions and GF (Wilson plaquette smoother), stout smearing and APE smearing (both matched to GF at flow time $t_0$) is basically the same.
Interesting is the case of the correlation of the index/SF with the gluonic definition on HYP-smeared configurations (matched to GF at flow time $t_0$, which for this ensemble implies 10 steps of HYP smearing).
It is systematically higher than the one to stout/APE and follows the pattern of the correlation between fermionic and gluonic with GF and the tree-level Symanzik improved smoothing action.
In particular, it yields a 97\% correlation with SF HYP5.
This suggests that HYP smearing has a somewhat similar effect both for fermionic and gluonic definitions.
We have also checked the correlation of SF HYP5 and field theoretic with different numbers of HYP iterations and indeed the best correlation is achieved when this number is between 5 and 10 (no statistically significant difference between the latter).
This suggests also that some matching between fermionic and gluonic definitions could be achieved.

In the next subsections, we analyze in detail the correlations between definitions inside some selected groups, with specific questions in mind, e.g.\ about the role of the used discretization of the topological charge operator or about the role of the used smoothing action.

\subsubsection{Comparison of fermionic definitions}
\label{sec:ferm}
\begin{figure}[h!]
\begin{center}
\rotatebox{0}{\hspace{-0.725cm} \includegraphics[width=10cm]{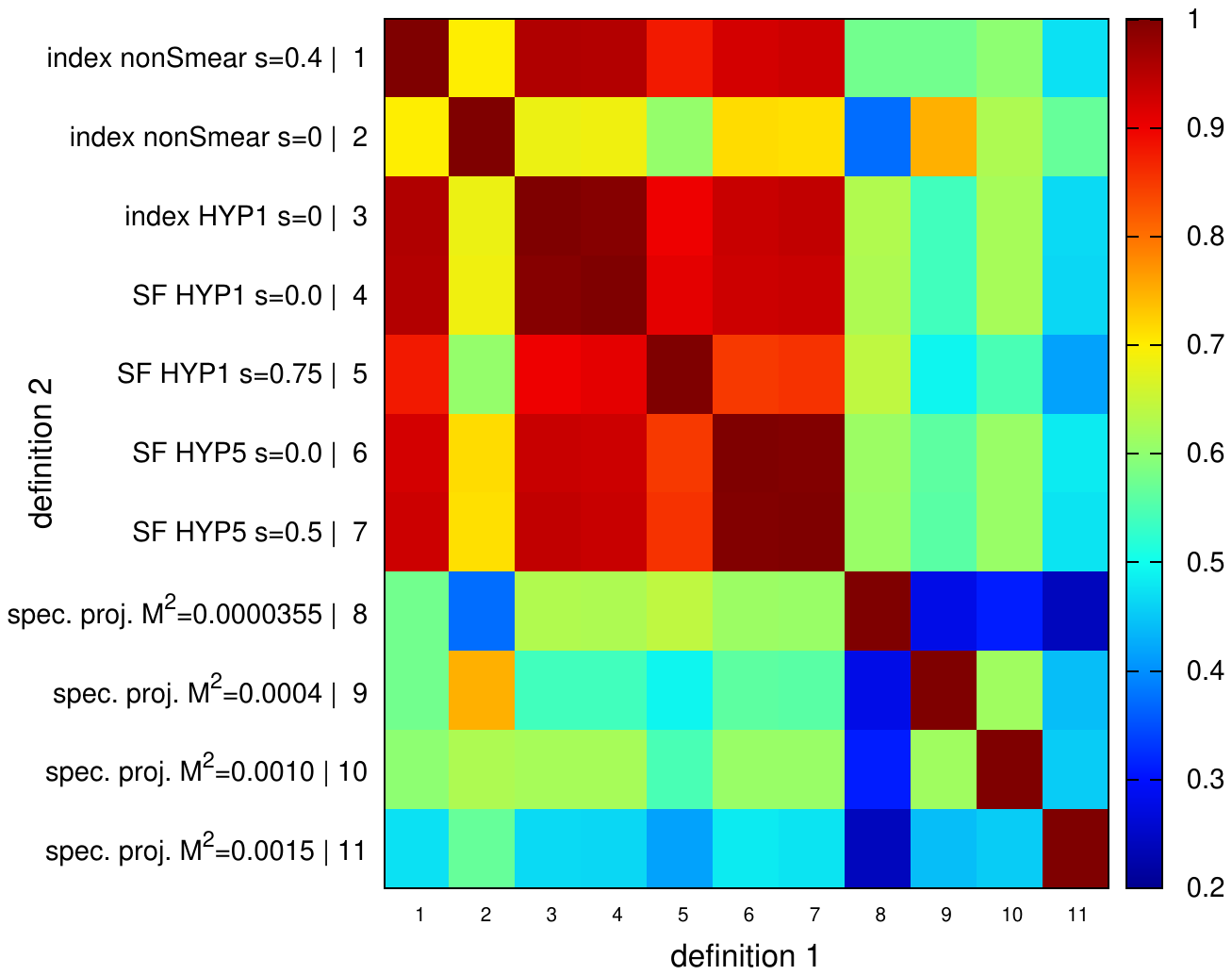}}
\end{center}
\caption{\label{fig:b40-corr1} 
Comparison of fermionic definitions of the topological charge. The correlation between different definitions is colour-coded {(note the scale is different than in Figs.~\ref{fig:b40-corr0},~\ref{fig:b40-corr2},~\ref{fig:b40-corr3},~\ref{fig:b40-corr4})}.} 
\end{figure}

In this subsection, we make a comprehensive comparison of all fermionic definitions (1-11 from Tab.~\ref{tab:defs}), see Fig.~\ref{fig:b40-corr1} and Tab.~\ref{tab:corr1} for a summary.
With respect to the main comparison of Sec.~\ref{sec:main}, we are able to conclude more about different parameter values that can be used in the definition of $Q$, i.e. the kernel parameter $s$ for overlap and spectral flow, the number of HYP smearing steps applied before the calculation of $Q$ and different values of the threshold parameter for spectral projectors.

We start by investigating the role of the $s$ parameter of the Wilson-Dirac kernel operator (see Sec.~\ref{sec:index}).
It is, in particular, responsible for the locality properties of the overlap Dirac operator \cite{Hernandez:1998et,Durr:2005an,Cichy:2012vg} and thus is expected to be important for measurements of $Q$.
For example, if the overlap operator is not local enough, then small topological objects may not be visible (they ``fall through the lattice'').
Indeed, one can notice sizable effects when $s$ is changed from 0.4 (value that guarantees optimal locality for the non-smeared gauge fields case \cite{Cichy:2012vg}) to 0 (very bad locality) -- the correlation is only around 70\%, which is much lower than the correlation with index/SF definitions with good locality properties (in particular, $s=0$ for the case of 1 iteration of HYP smearing with 96\% correlation).
Similarly, the violation of locality in the HYP1 case (change from the optimal value $s=0$ to $s=0.75$ (more than twice smaller value of the decay rate of the norm of the overlap Dirac operator)) also leads to decreasing correlations.
For the case HYP5, locality was not investigated in Ref.~\cite{Cichy:2012vg}.
However, from the practically identical results at $s=0$ and $s=0.5$, one can infer that locality is similar for both of them and guarantees high correlation with respect to the index extracted with the optimally local $s=0$ (HYP1) or $s=0.4$ (no smearing) values.

\begin{table*}[h!]
  \centering
\begin{scriptsize}
  \begin{tabular}[]{c|ccccccccccc}
& 1 & 2 & 3 & 4 & 5 & 6 & 7 & 8 & 9 & 10 & 11 \\
\hline
1 & 1 & 0.70(3) & 0.96(0) & 0.95(0) & 0.88(1) & 0.92(1) & 0.93(0) & 0.58(3) & 0.58(4) & 0.60(3) & 0.47(4) \\
2 & 0.70(3) & 1 & 0.68(3) & 0.68(3) & 0.60(4) & 0.71(2) & 0.71(3) & 0.37(5) & 0.75(3) & 0.63(4) & 0.57(3) \\
3 & 0.96(0) & 0.68(3) & 1 & 1.00(0) & 0.90(0) & 0.93(0) & 0.94(0) & 0.63(2) & 0.54(4) & 0.62(3) & 0.47(3) \\
4 & 0.95(0) & 0.68(3) & 1.00(0) & 1 & 0.91(0) & 0.93(0) & 0.93(0) & 0.63(2) & 0.54(4) & 0.62(3) & 0.47(3) \\
5 & 0.88(1) & 0.60(4) & 0.90(0) & 0.91(0) & 1 & 0.85(1) & 0.85(1) & 0.64(2) & 0.49(4) & 0.55(4) & 0.42(3) \\
6 & 0.92(1) & 0.71(2) & 0.93(0) & 0.93(0) & 0.85(1) & 1 & 0.99(0) & 0.61(3) & 0.56(4) & 0.61(3) & 0.48(3) \\
7 & 0.93(0) & 0.71(3) & 0.94(0) & 0.93(0) & 0.85(1) & 0.99(0) & 1 & 0.61(3) & 0.56(4) & 0.61(3) & 0.48(3) \\
8 & 0.58(3) & 0.37(5) & 0.63(2) & 0.63(2) & 0.64(2) & 0.61(3) & 0.61(3) & 1 & 0.28(4) & 0.31(4) & 0.24(4) \\
9 & 0.58(4) & 0.75(3) & 0.54(4) & 0.54(4) & 0.49(4) & 0.56(4) & 0.56(4) & 0.28(4) & 1 & 0.62(4) & 0.44(4) \\
10 & 0.60(3) & 0.63(4) & 0.62(3) & 0.62(3) & 0.55(4) & 0.61(3) & 0.61(3) & 0.31(4) & 0.62(4) & 1 & 0.45(3) \\
11 & 0.47(4) & 0.57(3) & 0.47(3) & 0.47(3) & 0.42(3) & 0.48(3) & 0.48(3) & 0.24(4) & 0.44(4) & 0.45(3) & 1 \\
\end{tabular}
\end{scriptsize}
\caption{Comparison of fermionic definitions of the topological charge. The numbers correspond to the numbering given in Fig.~\ref{fig:b40-corr1}. We give the correlation coefficient between different definitions and its error (0 means that the error is smaller than 0.005).}
  \label{tab:corr1}
\end{table*}

As stated in Sec.~\ref{sec:SF}, the index and spectral flow definitions (with the same value of the $s$ parameter) are exactly equivalent, i.e. should yield a 100\% correlation.
However, with the spectral flow at a coarse lattice spacing, it may be difficult to disentangle all the zero crossings that determine the value of $Q$.
Similarly, with the index of the overlap operator, numerical precision issues may appear when using too relaxed (to decrease the cost) tolerance criterion for the solver in the procedure of finding zero modes.
As a result, the obtained correlation was very close to, but not ideally 1, due to the occurrence of few cases where the value from overlap and from SF differed by $\pm1$.

We now move on to discuss the role of the $M$ parameter for spectral projectors.
In Refs.~\cite{Luscher:2010ik,Cichy:2013rra}, the renormalized $M$ parameter ($M_R$) was set to around 100 MeV ($\MSb$ scheme at the renormalization scale of 2 GeV).
This is expected to be a reasonable choice, since it avoids both the region close to the renormalized quark mass and the region where $aM\approx1$.
However, the above references considered relatively large lattices, while the small-volume lattice that we consider for this comparison can suffer from another effect that we discuss below.
Namely, with a value of $M_R\approx100$ MeV, the number of eigenmodes of the operator $D^\dagger D$ is relatively small (typically 5-10), while the number of zero modes can be as high as 15.
Hence, one can expect that the projector $\PM$ may not include (``count'') all the zero modes in some cases, leading to a too small value of the topological susceptibility.
To investigate how this feature can affect correlations, we performed the spectral projector calculations for four values of $M$.
Interestingly, we found that the correlations between topological charges extracted from different $M$ values are very small -- the ones with the smallest $M$ are only 25\%-30\% correlated with the ones at higher values of $M$. The correlations among higher values of $M$ ($M^2\geq0.0004$) are somewhat higher (45\%-60\%), but still only moderate.
This suggests that the stochastic noise is not entirely suppressed and that cut-off effects are possibly very different for different values of $M$.

Concerning correlations between spectral projectors and index-type definitions, they are typically between 50\% and 60\%.
However, there is no obvious tendency, like an improving correlation when increasing or decreasing $M$.

In the end, these results provide some warning about interpreting the spectral projector observable ${\cal C}$ as the topological charge.
Although the topological susceptibility is well-defined with spectral projectors, it requires to correct $\langle{\cal C}^2\rangle$ with $\langle{\cal B}\rangle/N$ when the number of stochastic sources is finite.
Together with the presented results for correlations (in particular the ones for index-type definitions, which \emph{a priori} could be expected to be highly correlated with ${\cal C}$), this makes the interpretation of ${\cal C}$ on a single gauge field configuration difficult.
It is, moreover, likely that the values of ${\cal C}$ extracted with different values of $M$ are affected by different cut-off effects.

\subsection{Correlation towards the continuum limit}
\begin{figure}[t!]
\begin{center}
\rotatebox{0}{\hspace{-0.725cm} \includegraphics[width=10cm]{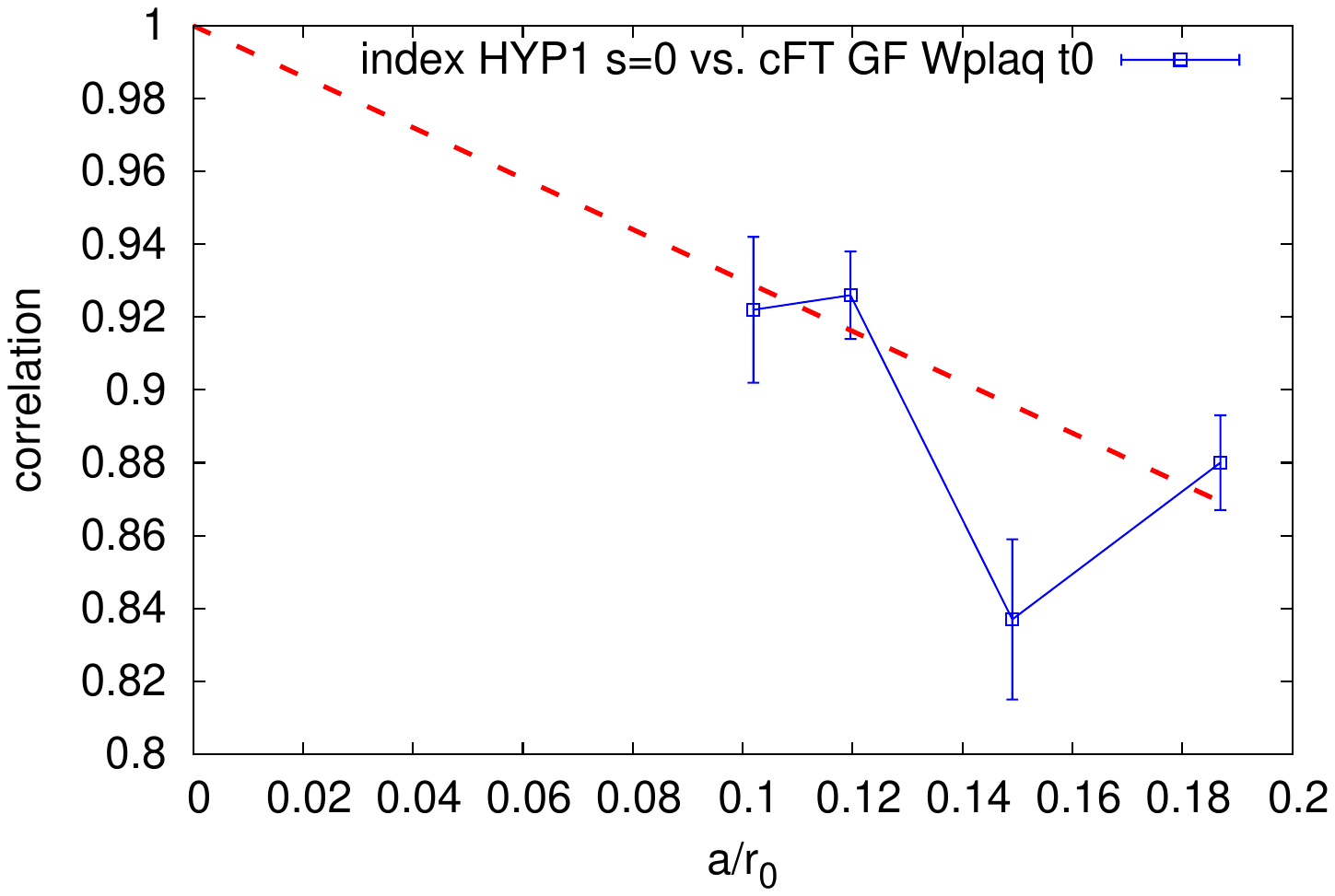}}
\end{center}
\caption{\label{fig:b40-cont} 
Increase of correlation towards the continuum limit between one representative fermionic definition (index of the overlap operator evaluated on configurations with 1 step of HYP smearing applied) and one representative gluonic definition (with gradient flow, Wilson plaquette smoothing action at flow time $t_0$).
The dashed red line is a guide to the eye, showing that it is plausible that the correlation will become 1 in the continuum limit.} 
\end{figure}

Our next aim is to investigate how the correlations behave towards the continuum limit.
The expectation is that very close to the continuum, all definitions agree -- hence an increase of correlation coefficients should be observed when decreasing the lattice spacing.
We chose one representative fermionic definition (index of the overlap operator evaluated on configurations with 1 step of HYP smearing applied) and one representative gluonic definition (with gradient flow, Wilson plaquette smoothing action at flow time $t_0$).
In Fig.~\ref{fig:b40-cont}, we show that the correlation coefficient indeed increases towards the continuum limit, from around 84\%-88\% for the two coarser lattice spacings to 92\%-93\% for the two finer.
It is therefore plausible that, as expected, the differences between results at a finite lattice spacing are cut-off effects.

\subsection{Topological susceptibility}
\label{sec:topological_susceptibility}
\begin{figure}[t!]
\begin{center}
\rotatebox{0}{\hspace{-0.725cm} \includegraphics[width=10cm]{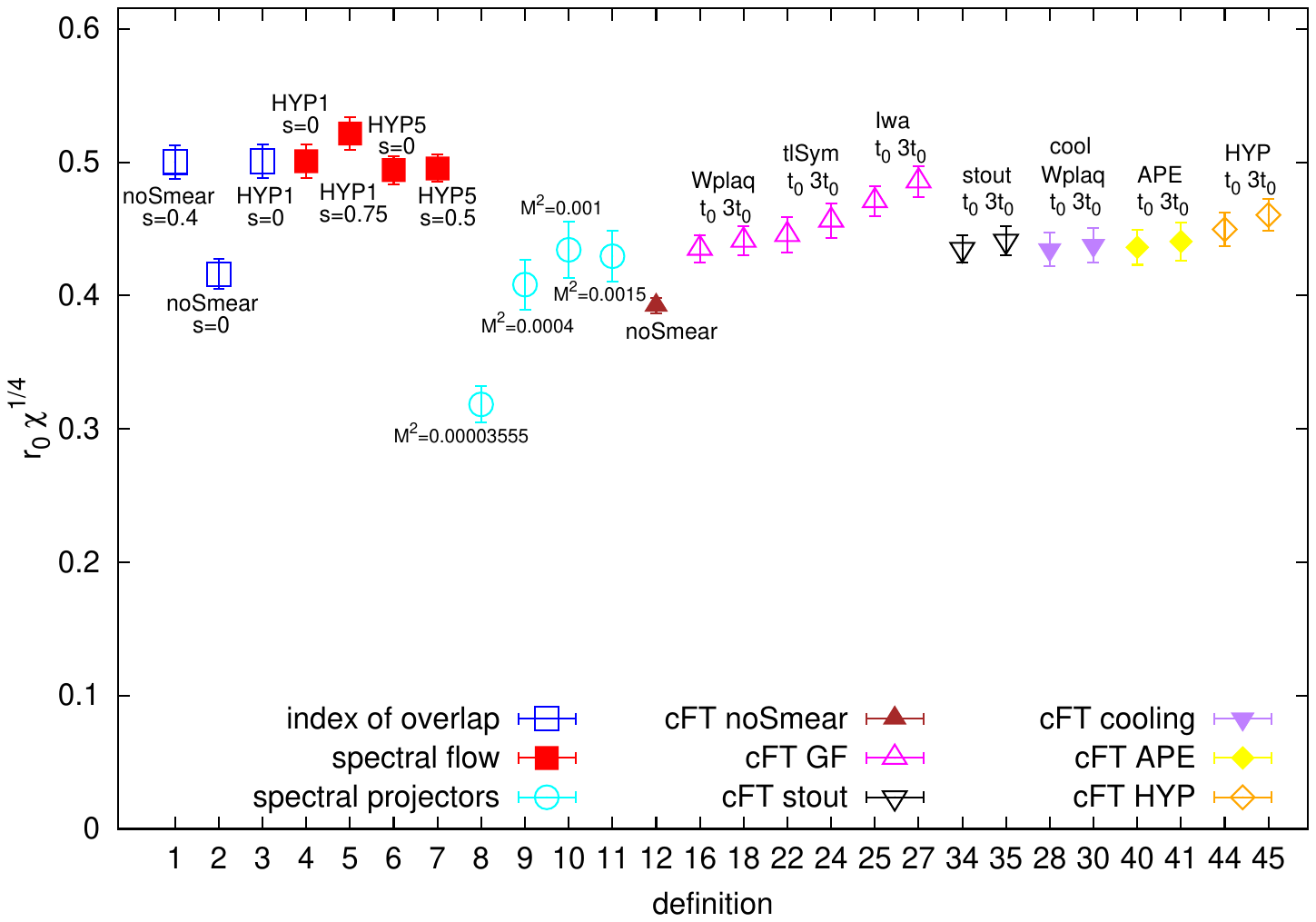}}
\end{center}
\caption{\label{fig:b40-chi} 
Topological susceptibility for ensemble b40.16 (with pion mass around 340 MeV), using representatives of different kinds of definitions. The definition numbers corresponds to the ones in Tab.~\ref{tab:defs}.} 
\end{figure}

We also compared the topological susceptibility computed with different methods. We defined the topological susceptibility as 
\begin{eqnarray}
\chi=\frac{\langle Q^2\rangle}{V} \,,
\end{eqnarray}
for all cases, except spectral projectors, where one needs a correction for a finite number of stochastic sources and renormalization with $(Z_S/Z_P)^2$, see Eq.\ (\ref{eq:chi}).

The comparison is shown in Fig.~\ref{fig:b40-chi}.
We find rather nice agreement between different definitions, with most of them giving a value in the range $r_0\chi^{1/4}\in[0.4,0.5]$.
It is interesting to observe that the outlying values concern definitions for which certain theoretical doubts appear about their validity.
In particular, the index definition for the non-smeared case and $s=0$ gives a 20\% smaller result than other index definitions.
This may be due to the fact that with decreased locality, some topological structures are not counted properly and hence $|Q|$ is too small on some configurations.
Similarly, the lowest value of $M^2$ for spectral projectors might be too small to count all zero modes.
However, these effects should go away in the continuum limit -- for the index definition, strict locality is then recovered and for spectral projectors, all relevant modes become actual zero modes and hence are counted with any value of $M^2$.
The situation is somewhat different with the third outlier, the value of $\chi$ from the field theoretic definition without any smearing, since by using it one is basically averaging over ultraviolet noise and it is not clear whether any physical signal for the topological susceptibility is left.

The remaining differences between valid definitions, e.g. field theoretic ones with different kinds of smearing, are most likely due to cut-off effects.
In particular, changing the smoothing action has a rather strong effect on the computed value of $\chi$.
It would certainly be desirable to perform the continuum limit extrapolation for the topological susceptibility from several definitions, but this is inconclusive with the current precision of our data (with typical error of $\chi$ at the level of 10\%).
Nevertheless, with current theoretical understanding, one can be rather certain that the continuum limit is correct for all the cases, excluding the ones for which obvious reservations can be made.

\begin{figure*}[h!]
\rotatebox{0}{\hspace{-0.725cm} \includegraphics[width=18.75cm]{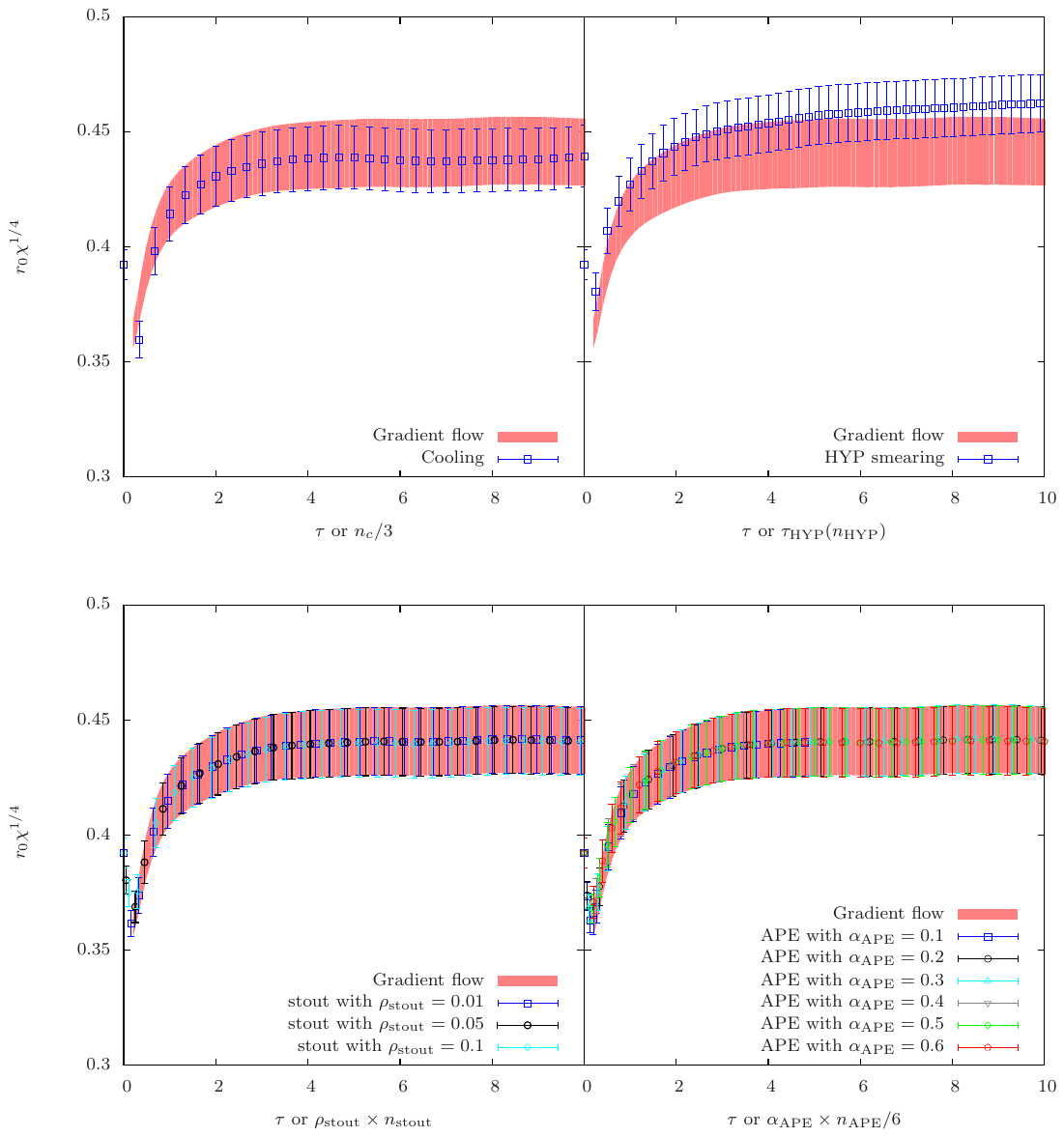}}
\vspace{-4.725cm}
\caption{\label{fig:equivalence_susc} A comparison of the topological suscpeptibility $r_0 \chi^{1/4}$ as a function of the gradient flow time $\tau$ for the Wilson flow, or, in the upper left panel, the rescaled cooling step $n_c/3$ for cooling, or, in the upper right panel, the rescaled HYP smearing $\tau_{\rm HYP}(n_{\rm HYP})$, or, in the lower left panel, the rescaled $n_{\rm stout}$ smearing $\rho_{\rm st} \times n_{\rm stout}$ for three values of $\rho_{\rm st}$, or, finally in the lower right panel, the rescaled APE smearing $\alpha_{\rm APE} \times n_{\rm APE} /6$ for six different values of the $\alpha_{\rm APE}$ parameter. For this ensemble $t_0\simeq 2.5a^2$.}
\end{figure*}

An interesting question regarding the field theoretic definition of $Q$ is how the resulting topological susceptibility behaves as a function of the smoothing scale. Furthermore, we would like to test how the matching between the different smoothers affects $\chi$. To this purpose, in \fig{fig:equivalence_susc}, we present the topological susceptibility extracted using the clover definition of the corresponding charge density for the Wilson flow as a function of the flow time and compare it to other smoothing procedures with smoothing scales adjusted according to Tab.~\ref{tab:rescaling}. 

In the upper left panel of \fig{fig:equivalence_susc}, we provide a comparison of $r_0 \chi^{1/4}$ extracted via the Wilson flow and cooling. By rescaling $n_c$ $\to$ $n_c/3$ according to \eq{eq:perturbative_expression} we observe that both smoothers give results which agree for the whole range of $\tau$. The above picture resembles similar comparisons demonstrated in Ref.~\cite{Alexandrou:2015yba}. This suggests that cooling, which is much faster compared to the Wilson flow, provides very similar topological susceptibility as long as the number of cooling steps is rescaled appropriately.

The upper rightmost panel of \fig{fig:equivalence_susc} \ reveals an interesting feature of the HYP smoothing technique. Namely, a comparison between HYP smearing with the number of smearing steps rescaled according to  \eq{eq:HYP_ansatz} with the Wilson flow shows an approximate agreement (within the statistical accuracy). However, the topological susceptibility via the Wilson flow appears to manifest a plateau, and thus scale invariance, starting at small values of $\tau \sim 3$. On the contrary, when smoothing with HYP smearing, the plateau sets in at larger values of $\tau$ suggesting that this picture could be the outcome of larger cut-off effects in the topological charge. Without question, the non-local character of the HYP smearing introduces different instantonic properties, which could potentially explain this behaviour; this should be investigated in more detail.  

In the lower left panel of \fig{fig:equivalence_susc}, we demonstrate a comparison of $r_0 \chi^{1/4}$ obtained via the Wilson flow as a function of the flow time with $r_0 \chi^{1/4}$ calculated via stout smearing as a function of the rescaled smearing steps of $\rho_{\rm st} n_{\rm st}$. We consider three values of $\rho_{\rm st}=0.1, \ 0.05$ and $0.01$. Amazingly, the four sketched bands appear to fall on top of each other, yielding a message that the level of similarity between stout smearing and Wilson flow is indeed very high. In fact, the perfect matching of topological susceptibilities in addition to the correlation coefficient of $1.00$ flags the exact numerical equivalence between the two smoothers. Once more, this result suggests that we could safely use stout smearing instead of the Wilson flow to measure topological observables and define physical reference scales according to \eq{eq:scale_stout}.

Finally, in the lower right panel of \fig{fig:equivalence_susc}, we provide a comparison of $r_0 \chi^{1/4}$ measured with the Wilson flow as a function of $\tau$ with the value of $r_0 \chi^{1/4}$ extracted with APE smearing vs. the rescaled smearing step $\alpha_{\rm APE} n_{\rm APE}/6$. We did so for six different values of $\alpha_{\rm APE}$, namely $\alpha_{\rm APE} = 0.6, \ 0.5, \ 0.4, \ 0.3, \ 0.2$ and $0.1$. Similarly to the stout smearing presented in the previous paragraph, the topological susceptibility for all six values of $\alpha_{\rm APE}$ appears to match exactly the result of the Wilson flow. Again, this is expected for APE by considering the high similarity of the topological charge revealed in \fig{fig:comparison_topological_charge} as well as the correlation coefficient of $1.00$ noted in Tab.~\ref{fig:b40-corr2}. Like for stout smearing, one can use APE smearing instead of the Wilson flow with a well defined physical reference scale given by \eq{eq:scale_ape}.

\section{Conclusions}
\label{sec:conclusions}

In this paper, we have investigated several definitions of the topological charge. Our main conclusion is that all valid\footnote{We remind that by valid definitions we mean ones which are finite after renormalization, avoiding short-distance singularities. From a more practical point of view, also definitions should be dismissed that show large contamination by UV noise, as in the non-smeared gluonic definition, or too small value of the spectral threshold in the spectral projector method, preventing one from counting all zero modes or would-be zero modes.} definitions lead to a consistent behaviour and are highly correlated, meaning to give --to a large extent--
the same topological charge for a given configuration. The progress in recent years, in particular the introduction of the gradient flow smoothing scheme, enabled good control over the topological charge extraction and made it possible to have a well-defined field theoretic definition of the renormalized topological susceptibility and hence a comparatively cheap way to compute the topological susceptibility, including its extrapolation to the continuum limit.
This is possible, because the gradient flow provides a valid definition of a smoothing scale, which needs to be kept constant when approaching the continuum limit.
The gradient flow also makes it possible to define such a smoothing scale for other kinds of smearing methods, via a well-defined matching procedure.
Moreover, one can show that there are no short-distance singularities when the topological charge is defined at a finite flow time.
This property, in conjunction with its computational efficiency,
makes the gradient flow an excellent choice for computing the topological charge and the corresponding topological susceptibility. We see, however, no obstacle to also use other methods, such as cooling and smearing, for which the number of cooling or
smearing steps can be related analytically
to the gradient flow time.
In this way, one can define a smoothing scale also for the other smoothing schemes and perform the continuum limit by keeping it constant in physical units, which is a prerequisite for a correct continuum limit.

A warning about the usage of field theoretic definitions is provided by our follow-up analysis \cite{Alexandrou:2017bzk}, where we compared the GF definition with the spectral projector one on a wide range of ETMC's $N_f=2+1+1$ large-volume ensembles. We found that cut-off effects in the topological susceptibility from the GF are much larger than in the susceptibility from spectral projectors and can be up to 500\% at the coarsest lattice spacing of around 0.09 fm. {Nevertheless, the continuum limit is always compatible between GF and spectral projectors, and also independent of the flow time (GF) or the renormalized spectral threshold (spectral projectors), as long as these scales are fixed in physical units \cite{Giusti:2015kwf}.} We refer the reader to this paper for more details.

In fact, the numerical equivalence introduced by the matching procedure between the gradient flow and other smoothing schemes suggests that in cases where high statistics
are needed such as, for example, for the evaluation of higher moments of the topological charge in pure gauge theory, instead of
using the gradient flow, one can opt for employing either cooling or APE or stout smearing.
From our experience, we find that cooling, APE smearing or stout smearing are, respectively, 120, 20 and 30 times faster\footnote{Of course these numbers are not exact since they depend on the code details such as the level of optimization etc.} than the gradient flow for typical parameter values.

Defining the topological charge and susceptibility using the spectral projectors is another relatively new method. They constitute another theoretically clean way of defining these quantities.
However, the cost of this method is significantly larger than the one of the gradient flow.
Nevertheless, it might be the method of choice for some applications, since it yields much smaller cut-off effects than the GF, at least in the setup of our follow-up work \cite{Alexandrou:2017bzk}.
Concerning other fermionic definitions, such as the index of the overlap Dirac operator or the spectral flow of the Wilson-Dirac operator, they are theoretically very clean and provide integer values of the topological charge, but their cost is prohibitive for large-scale analyses.

In summary, we have shown in this paper that all valid definitions of the topological charge are highly correlated and, in principle, all of them can be used to analyze topological issues.
Thus, the choice for using a certain definition of the topological charge will depend on the particular problem under consideration.

\appendix

\section{Comparison of gluonic definitions}
\begin{figure}[h!]
\begin{center}
\rotatebox{0}{\hspace{-0.725cm} \includegraphics[width=10cm]{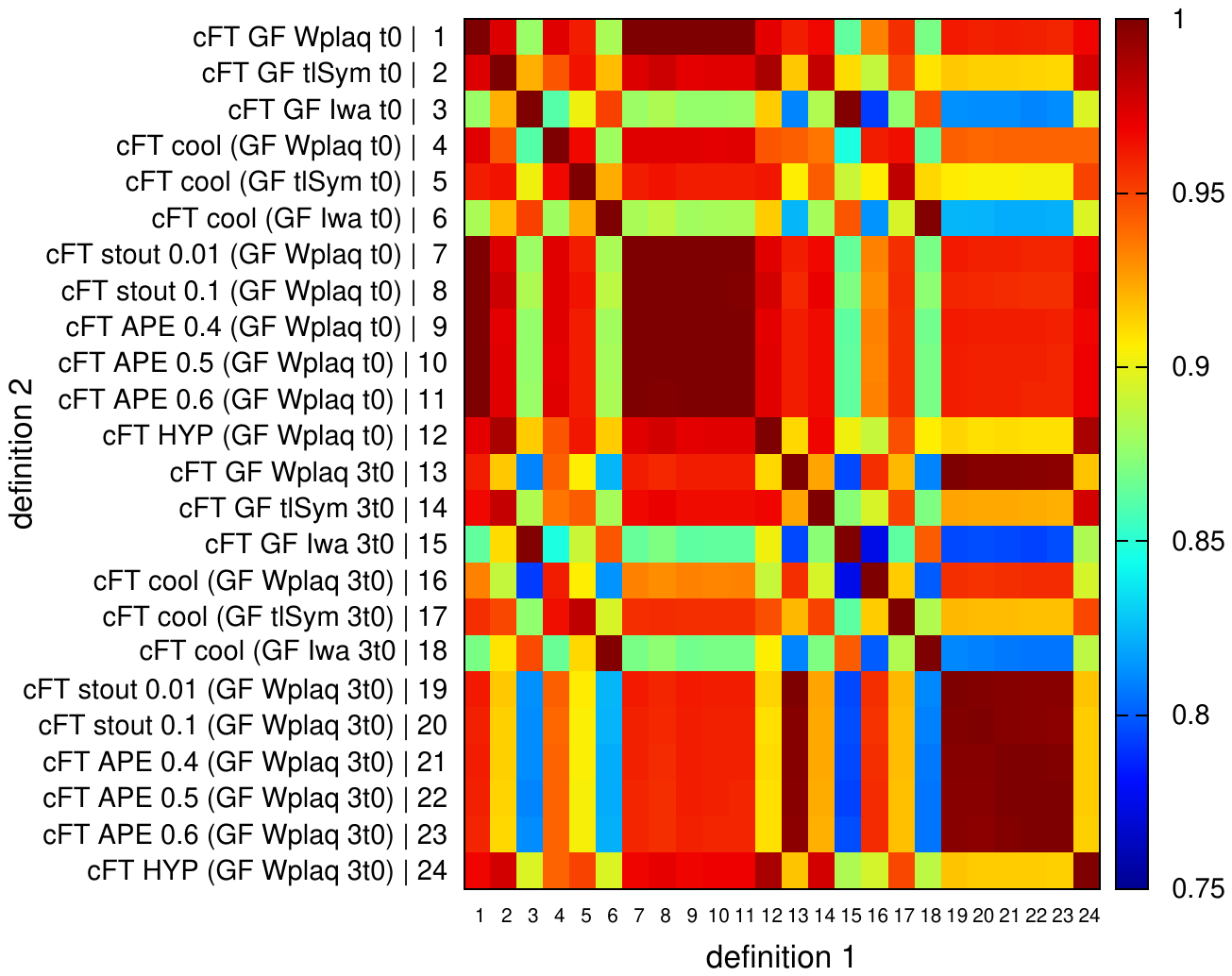}}
\end{center}
\caption{\label{fig:b40-corr4} 
Comparison of field theoretic definitions with different kinds of smoothing of UV fluctuations. The correlation between different definitions is colour-coded (note the scale is different than in Figs.~\ref{fig:b40-corr0},~\ref{fig:b40-corr1},~\ref{fig:b40-corr2},~\ref{fig:b40-corr3}).} 
\end{figure}

In this section of the Appendix, we present a comprehensive comparison of the correlation of the topological charge between different field theoretic definitions. We always use the clover discretization, but we vary the type of smoothing procedure, smoothing action (where applicable) and/or other parameters entering the definition of smoothing and flow times:
\begin{itemize}
 \item GF at flow times $t_0$, $3t_0$, three types of smoothing action (16, 18, 22, 24, 25, 27),
 \item cooling matched to GF at flow times $t_0$, $3t_0$, three types of smoothing action (28-34),
 \item stout smearing matched to GF at flow times $t_0$, $3t_0$, two values of the stout parameter $\rho_{\rm st}$ (34-37),
 \item APE smearing matched to GF at flow times $t_0$, $3t_0$, three values of the $\alpha_{\rm APE}$ parameter (38-43),
 \item HYP smearing matched to GF at flow times $t_0$, $3t_0$ (44-45).
\end{itemize}
A summary of our results in shown in Fig.~\ref{fig:b40-corr4} and Tab.~\ref{tab:corr4}.

\renewcommand{\tabcolsep}{0.1cm}
\setlength\extrarowheight{2pt}
\begin{sidewaystable*}[p!]
\vspace{16cm}  
  \centering
\begin{tiny}
  \begin{tabular}[]{c|cccccccccccccccccccccccc}
& 1 & 2 & 3 & 4 & 5 & 6 & 7 & 8 & 9 & 10 & 11 & 12 & 13 & 14 & 15 & 16 & 17 & 18 & 19 & 20 & 21 & 22 & 23 & 24 \\
\hline
1 & 1 & 0.97(0) & 0.88(1) & 0.97(0) & 0.96(0) & 0.88(1) & 1.00(0) & 1.00(0) & 1.00(0) & 1.00(0) & 1.00(0) & 0.97(0) & 0.96(0) & 0.97(0) & 0.86(1) & 0.93(0) & 0.96(0) & 0.87(1) & 0.96(0) & 0.96(0) & 0.96(0) & 0.96(0) & 0.96(0) & 0.97(0) \\
2 & 0.97(0) & 1 & 0.92(0) & 0.94(0) & 0.96(0) & 0.92(0) & 0.97(0) & 0.98(0) & 0.97(0) & 0.97(0) & 0.97(0) & 0.99(0) & 0.92(1) & 0.98(0) & 0.91(0) & 0.89(1) & 0.95(0) & 0.91(0) & 0.92(1) & 0.91(1) & 0.91(1) & 0.91(1) & 0.91(1) & 0.98(0) \\
3 & 0.88(1) & 0.92(0) & 1 & 0.86(1) & 0.90(1) & 0.95(0) & 0.88(1) & 0.88(1) & 0.88(1) & 0.88(1) & 0.88(1) & 0.91(0) & 0.81(2) & 0.88(1) & 1.00(0) & 0.79(2) & 0.88(1) & 0.95(0) & 0.81(2) & 0.81(2) & 0.81(1) & 0.81(2) & 0.81(2) & 0.90(1) \\
4 & 0.97(0) & 0.94(0) & 0.86(1) & 1 & 0.97(0) & 0.88(1) & 0.97(0) & 0.97(0) & 0.97(0) & 0.97(0) & 0.97(0) & 0.94(0) & 0.94(0) & 0.94(0) & 0.85(1) & 0.96(0) & 0.96(0) & 0.87(1) & 0.94(0) & 0.94(0) & 0.94(0) & 0.94(0) & 0.94(0) & 0.94(0) \\
5 & 0.96(0) & 0.96(0) & 0.90(1) & 0.97(0) & 1 & 0.92(0) & 0.96(0) & 0.96(0) & 0.96(0) & 0.96(0) & 0.96(0) & 0.96(0) & 0.91(1) & 0.94(0) & 0.89(1) & 0.91(1) & 0.98(0) & 0.91(0) & 0.91(1) & 0.91(1) & 0.90(1) & 0.90(1) & 0.90(1) & 0.95(0) \\
6 & 0.88(1) & 0.92(0) & 0.95(0) & 0.88(1) & 0.92(0) & 1 & 0.88(1) & 0.89(1) & 0.88(1) & 0.88(1) & 0.88(1) & 0.91(0) & 0.82(1) & 0.88(1) & 0.95(0) & 0.81(1) & 0.90(1) & 1.00(0) & 0.82(1) & 0.82(1) & 0.82(1) & 0.82(1) & 0.82(1) & 0.90(0) \\
7 & 1.00(0) & 0.97(0) & 0.88(1) & 0.97(0) & 0.96(0) & 0.88(1) & 1 & 1.00(0) & 1.00(0) & 1.00(0) & 1.00(0) & 0.97(0) & 0.96(0) & 0.97(0) & 0.87(1) & 0.93(0) & 0.96(0) & 0.87(1) & 0.96(0) & 0.96(0) & 0.96(0) & 0.96(0) & 0.96(0) & 0.97(0) \\
8 & 1.00(0) & 0.98(0) & 0.88(1) & 0.97(0) & 0.96(0) & 0.89(1) & 1.00(0) & 1 & 1.00(0) & 1.00(0) & 1.00(0) & 0.98(0) & 0.96(0) & 0.97(0) & 0.87(1) & 0.93(0) & 0.96(0) & 0.87(1) & 0.96(0) & 0.96(0) & 0.96(0) & 0.96(0) & 0.96(0) & 0.97(0) \\
9 & 1.00(0) & 0.97(0) & 0.88(1) & 0.97(0) & 0.96(0) & 0.88(1) & 1.00(0) & 1.00(0) & 1 & 1.00(0) & 1.00(0) & 0.97(0) & 0.96(0) & 0.97(0) & 0.86(1) & 0.93(0) & 0.96(0) & 0.87(1) & 0.96(0) & 0.96(0) & 0.96(0) & 0.96(0) & 0.96(0) & 0.97(0) \\
10 & 1.00(0) & 0.97(0) & 0.88(1) & 0.97(0) & 0.96(0) & 0.88(1) & 1.00(0) & 1.00(0) & 1.00(0) & 1 & 1.00(0) & 0.97(0) & 0.96(0) & 0.97(0) & 0.86(1) & 0.93(0) & 0.96(0) & 0.87(1) & 0.96(0) & 0.96(0) & 0.96(0) & 0.96(0) & 0.96(0) & 0.97(0) \\
11 & 1.00(0) & 0.97(0) & 0.88(1) & 0.97(0) & 0.96(0) & 0.88(1) & 1.00(0) & 1.00(0) & 1.00(0) & 1.00(0) & 1 & 0.97(0) & 0.96(0) & 0.97(0) & 0.86(1) & 0.93(0) & 0.96(0) & 0.87(1) & 0.96(0) & 0.96(0) & 0.96(0) & 0.96(0) & 0.96(0) & 0.97(0) \\
12 & 0.97(0) & 0.99(0) & 0.91(0) & 0.94(0) & 0.96(0) & 0.91(0) & 0.97(0) & 0.98(0) & 0.97(0) & 0.97(0) & 0.97(0) & 1 & 0.91(1) & 0.97(0) & 0.90(1) & 0.89(1) & 0.95(0) & 0.91(0) & 0.91(1) & 0.91(1) & 0.91(1) & 0.91(1) & 0.91(1) & 0.99(0) \\
13 & 0.96(0) & 0.92(1) & 0.81(2) & 0.94(0) & 0.91(1) & 0.82(1) & 0.96(0) & 0.96(0) & 0.96(0) & 0.96(0) & 0.96(0) & 0.91(1) & 1 & 0.92(0) & 0.80(2) & 0.96(0) & 0.92(0) & 0.81(1) & 1.00(0) & 1.00(0) & 1.00(0) & 1.00(0) & 1.00(0) & 0.92(0) \\
14 & 0.97(0) & 0.98(0) & 0.88(1) & 0.94(0) & 0.94(0) & 0.88(1) & 0.97(0) & 0.97(0) & 0.97(0) & 0.97(0) & 0.97(0) & 0.97(0) & 0.92(0) & 1 & 0.87(1) & 0.89(1) & 0.95(0) & 0.87(1) & 0.92(1) & 0.92(1) & 0.92(1) & 0.92(1) & 0.92(1) & 0.98(0) \\
15 & 0.86(1) & 0.91(0) & 1.00(0) & 0.85(1) & 0.89(1) & 0.95(0) & 0.87(1) & 0.87(1) & 0.86(1) & 0.86(1) & 0.86(1) & 0.90(1) & 0.80(2) & 0.87(1) & 1 & 0.77(2) & 0.86(1) & 0.94(0) & 0.80(2) & 0.80(2) & 0.80(2) & 0.79(2) & 0.80(2) & 0.88(1) \\
16 & 0.93(0) & 0.89(1) & 0.79(2) & 0.96(0) & 0.91(1) & 0.81(1) & 0.93(0) & 0.93(0) & 0.93(0) & 0.93(0) & 0.93(0) & 0.89(1) & 0.96(0) & 0.89(1) & 0.77(2) & 1 & 0.91(1) & 0.80(1) & 0.96(0) & 0.96(0) & 0.96(0) & 0.96(0) & 0.96(0) & 0.89(1) \\
17 & 0.96(0) & 0.95(0) & 0.88(1) & 0.96(0) & 0.98(0) & 0.90(1) & 0.96(0) & 0.96(0) & 0.96(0) & 0.96(0) & 0.96(0) & 0.95(0) & 0.92(0) & 0.95(0) & 0.86(1) & 0.91(1) & 1 & 0.88(1) & 0.92(0) & 0.92(1) & 0.92(1) & 0.92(1) & 0.92(1) & 0.95(0) \\
18 & 0.87(1) & 0.91(0) & 0.95(0) & 0.87(1) & 0.91(0) & 1.00(0) & 0.87(1) & 0.87(1) & 0.87(1) & 0.87(1) & 0.87(1) & 0.91(0) & 0.81(1) & 0.87(1) & 0.94(0) & 0.80(1) & 0.88(1) & 1 & 0.81(1) & 0.81(1) & 0.81(1) & 0.81(1) & 0.81(1) & 0.89(1) \\
19 & 0.96(0) & 0.92(1) & 0.81(2) & 0.94(0) & 0.91(1) & 0.82(1) & 0.96(0) & 0.96(0) & 0.96(0) & 0.96(0) & 0.96(0) & 0.91(1) & 1.00(0) & 0.92(1) & 0.80(2) & 0.96(0) & 0.92(0) & 0.81(1) & 1 & 1.00(0) & 1.00(0) & 1.00(0) & 1.00(0) & 0.92(1) \\
20 & 0.96(0) & 0.91(1) & 0.81(2) & 0.94(0) & 0.91(1) & 0.82(1) & 0.96(0) & 0.96(0) & 0.96(0) & 0.96(0) & 0.96(0) & 0.91(1) & 1.00(0) & 0.92(1) & 0.80(2) & 0.96(0) & 0.92(1) & 0.81(1) & 1.00(0) & 1 & 1.00(0) & 1.00(0) & 1.00(0) & 0.91(1) \\
21 & 0.96(0) & 0.91(1) & 0.81(1) & 0.94(0) & 0.90(1) & 0.82(1) & 0.96(0) & 0.96(0) & 0.96(0) & 0.96(0) & 0.96(0) & 0.91(1) & 1.00(0) & 0.92(1) & 0.80(2) & 0.96(0) & 0.92(1) & 0.81(1) & 1.00(0) & 1.00(0) & 1 & 1.00(0) & 1.00(0) & 0.91(0) \\
22 & 0.96(0) & 0.91(1) & 0.81(2) & 0.94(0) & 0.90(1) & 0.82(1) & 0.96(0) & 0.96(0) & 0.96(0) & 0.96(0) & 0.96(0) & 0.91(1) & 1.00(0) & 0.92(1) & 0.79(2) & 0.96(0) & 0.92(1) & 0.81(1) & 1.00(0) & 1.00(0) & 1.00(0) & 1 & 1.00(0) & 0.91(1) \\
23 & 0.96(0) & 0.91(1) & 0.81(2) & 0.94(0) & 0.90(1) & 0.82(1) & 0.96(0) & 0.96(0) & 0.96(0) & 0.96(0) & 0.96(0) & 0.91(1) & 1.00(0) & 0.92(1) & 0.80(2) & 0.96(0) & 0.92(1) & 0.81(1) & 1.00(0) & 1.00(0) & 1.00(0) & 1.00(0) & 1 & 0.91(1) \\
24 & 0.97(0) & 0.98(0) & 0.90(1) & 0.94(0) & 0.95(0) & 0.90(0) & 0.97(0) & 0.97(0) & 0.97(0) & 0.97(0) & 0.97(0) & 0.99(0) & 0.92(0) & 0.98(0) & 0.88(1) & 0.89(1) & 0.95(0) & 0.89(1) & 0.92(1) & 0.91(1) & 0.91(0) & 0.91(1) & 0.91(1) & 1 \\
\end{tabular}
\end{tiny}
\caption{Comparison of field theoretic definitions with different kinds of smoothing of UV fluctuations. The numbers correspond to the numbering given in Fig.~\ref{fig:b40-corr4}. We give the correlation coefficient between different definitions and its error (0 means that the error is smaller than 0.005).}
  \label{tab:corr4}
\end{sidewaystable*}

We have already discussed the correlations within the class of gradient flow definitions.
We now concentrate on comparisons for different groups of definitions and also between the groups.

If one uses cooling as the smoothing procedure, the relation between different cooling times (matched to different flow times) and different smoothing actions for the cooling procedure is very similar to the one for corresponding cases for GF (i.e.\ with the same smoothing actions).
Using stout smearing, one observes that the smoothing parameter $\rho_{\rm st}\in\{0.01, 0.1\}$ has no effect on the correlations.
This is natural, since the stout smearing procedure is basically equivalent to the gradient flow with an appropriate step.
If this step is small enough, one expects that the results are exact, i.e.\ the gauge fields evolve according to the continuous gradient flow equations, without any flow time discretization effects.
For APE smearing, we also note that the APE parameter $\alpha_{\rm APE}$ has practically no effect on the resulting correlations (we also checked other values than the ones listed in Tab.~\ref{tab:defs}).
However, if the number of APE steps is varied, keeping the $\alpha$ parameter fixed, the correlation decreases from 1 to around 0.95.
This is qualitatively and quantitatively similar to the GF case and further demonstrates that the matching between GF and APE (or other kinds of smoothing procedure) is very robust.
The slight decrease of correlation when going from flow time $t_0$ to $3t_0$ demonstrates that at flow time $t_0$, one still has not reached the plateau of $Q$, i.e. the values of $Q$, at least for some gauge field configurations, still change when increasing the flow time.
However, this effect is much smaller for HYP smearing, where the correlation between the values of $Q$ corresponding to numbers of HYP smearing steps matched to flow times $t_0$ and $3t_0$ is 99\%, as compared to typically 95\% when using other kinds of smoothing.

Finally, we discuss correlations between different kinds of smoothers.
We already argued that GF and stout smearing are equivalent, hence the correlation is perfect if the number of stout smearing steps is matched to the flow time.
The correlation between GF and APE smearing is also close to 100\% (for matched smoothing scales), while the one between GF/cooling (with the same smoothing action) and GF/HYP smearing is 97\%.
If one compares definitions at unmatched smoothing scales (e.g. GF at flow time $t_0$ with APE at a number of steps corresponding to $3t_0$), one obviously observes a significant decrease of correlation.
It is also worth to mention that while for stout, APE or HYP smearing, the notion of a smoothing action does not make sense, still taking into account the way they are constructed, they correspond more to GF with the Wilson plaquette smoother, rather than to GF with more complicated smoothing actions.
This implies that the correlation between the values of $Q$ for these smearing types is high with respect to GF with the Wilson plaquette smoothing action (97\%-100\%), but decreases to a large extent when comparing to GF (or cooling) with the tree-level Symanzik or Iwasaki smoother, to around 90\% and 80\%, respectively.
The very lowest correlation (77\%) is observed when comparing GF with the Iwasaki smoother to cooling with the tree-level Symanzik improved action, both at flow time $3t_0$.
This correlation is actually even significantly smaller than the correlation of both these cases alone to index-type (fermionic) definitions.

\section{Comparison of different smoothing actions and flow times for gradient flow}
\label{sec:gf1}
\begin{figure}[h!]
\begin{center}
\rotatebox{0}{\hspace{-0.725cm} \includegraphics[width=10cm]{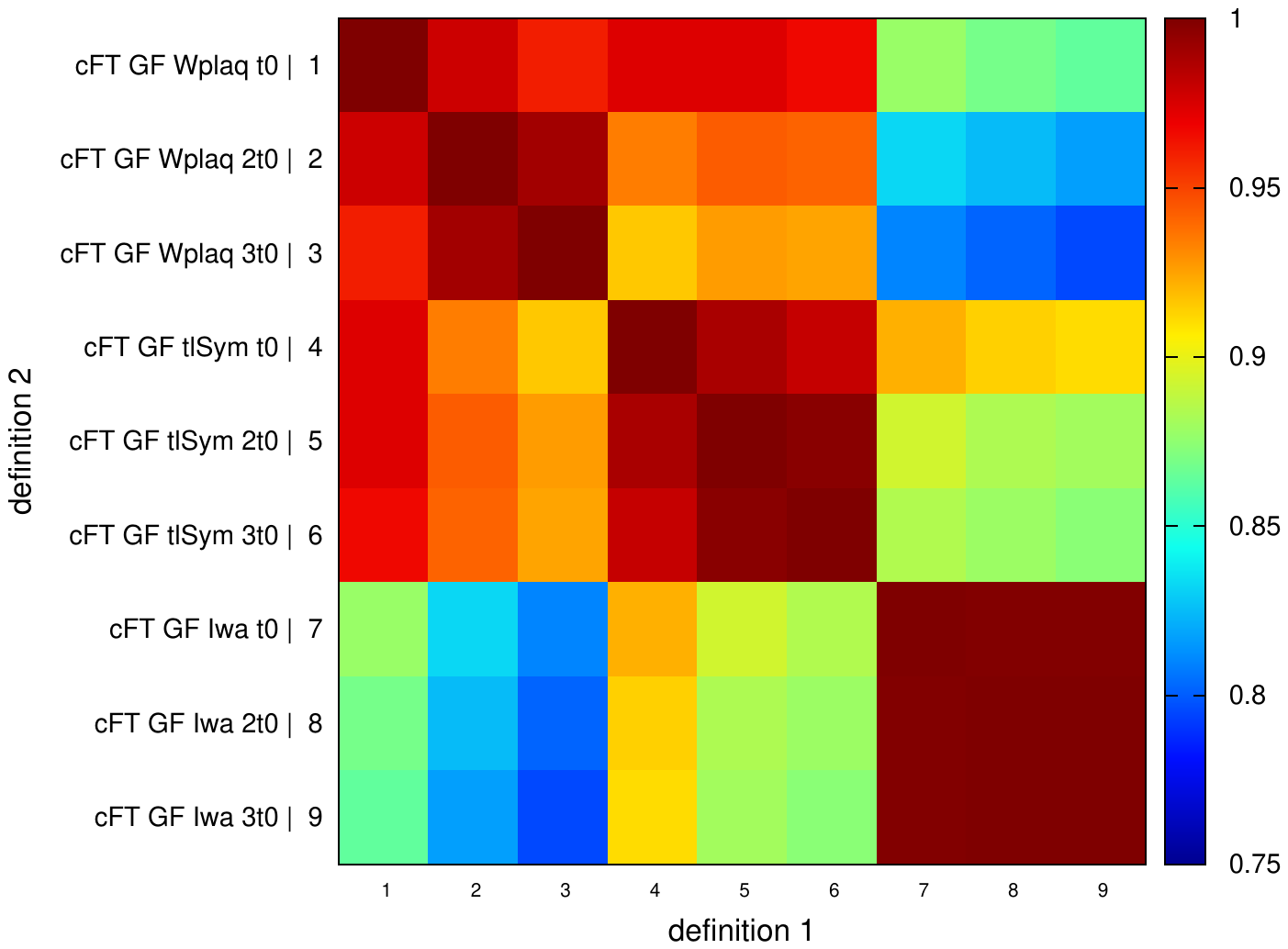}}
\end{center}
\caption{\label{fig:b40-corr2} 
Comparison of field theoretic definitions with GF smoothing and different smoothing actions and flow times. The correlation between different definitions is colour-coded {(note the scale is different than in Figs.~\ref{fig:b40-corr0},~\ref{fig:b40-corr1},~\ref{fig:b40-corr3},~\ref{fig:b40-corr4})}.} 
\end{figure}

The aim of this section is to provide a comparison of the correlation of the topological charge extracted using, different smoothing actions for the gradient flow, as well as different flow times.
The included cases are:
\begin{itemize}
 \item Wilson plaquette smoothing action, flow times $t_0$, $2t_0$ and $3t_0$ (16, 17, 18),
 \item tree-level Symanzik improved smoothing action, flow times $t_0$, $2t_0$ and $3t_0$ (22, 23, 24),
 \item Iwasaki smoothing action, flow times $t_0$, $2t_0$ and $3t_0$ (25, 26, 27).
\end{itemize}
A summary of our findings is given in Fig.~\ref{fig:b40-corr2} and Tab.~\ref{tab:corr2}.

\begin{table*}[h!]
  \centering
\begin{scriptsize}
  \begin{tabular}[]{c|ccccccccc}
& 1 & 2 & 3 & 4 & 5 & 6 & 7 & 8 & 9 \\
\hline
1 & 1 & 0.98(0) & 0.96(0) & 0.97(0) & 0.97(0) & 0.97(0) & 0.88(1) & 0.87(1) & 0.86(1) \\
2 & 0.98(0) & 1 & 0.99(0) & 0.93(0) & 0.94(0) & 0.94(0) & 0.83(1) & 0.82(1) & 0.82(2) \\
3 & 0.96(0) & 0.99(0) & 1 & 0.92(1) & 0.93(0) & 0.92(0) & 0.81(2) & 0.80(2) & 0.80(2) \\
4 & 0.97(0) & 0.93(0) & 0.92(1) & 1 & 0.99(0) & 0.98(0) & 0.92(0) & 0.91(0) & 0.91(0) \\
5 & 0.97(0) & 0.94(0) & 0.93(0) & 0.99(0) & 1 & 1.00(0) & 0.89(1) & 0.88(1) & 0.88(1) \\
6 & 0.97(0) & 0.94(0) & 0.92(0) & 0.98(0) & 1.00(0) & 1 & 0.88(1) & 0.88(1) & 0.87(1) \\
7 & 0.88(1) & 0.83(1) & 0.81(2) & 0.92(0) & 0.89(1) & 0.88(1) & 1 & 1.00(0) & 1.00(0) \\
8 & 0.87(1) & 0.82(1) & 0.80(2) & 0.91(0) & 0.88(1) & 0.88(1) & 1.00(0) & 1 & 1.00(0) \\
9 & 0.86(1) & 0.82(2) & 0.80(2) & 0.91(0) & 0.88(1) & 0.87(1) & 1.00(0) & 1.00(0) & 1 \\
\end{tabular}
\end{scriptsize}
\caption{Comparison of field theoretic definitions with GF smoothing and different smoothing actions and flow times. The numbers correspond to the numbering given in Fig.~\ref{fig:b40-corr2}. We give the correlation coefficient between different definitions and its error (0 means that the error is smaller than 0.005).}
  \label{tab:corr2}
\end{table*}

As expected, when the smoothing action is fixed, correlations decrease with increasing difference between corresponding flow times.
However, the decrease is very slight and practically invisible in the case of the Iwasaki smoother.
This suggests that increasing the flow time has very small effect on the values of $Q$ and the effect is almost absent for the Iwasaki case.
When comparing different smoothing actions, one notices that while Wilson plaquette is still very much correlated with tree-level Symanzik improved (92\%-97\%), the correlation with respect to the Iwasaki smoothing action drops down significantly (to 80\%-88\%).
The correlation of tree-level Symanzik improved to Iwasaki is larger than Wilson plaquette vs. Iwasaki, but still smaller than the one with respect to Wilson plaquette.

\begin{figure}[h!]
\begin{center}
\rotatebox{0}{\hspace{-0.725cm} \includegraphics[width=10cm]{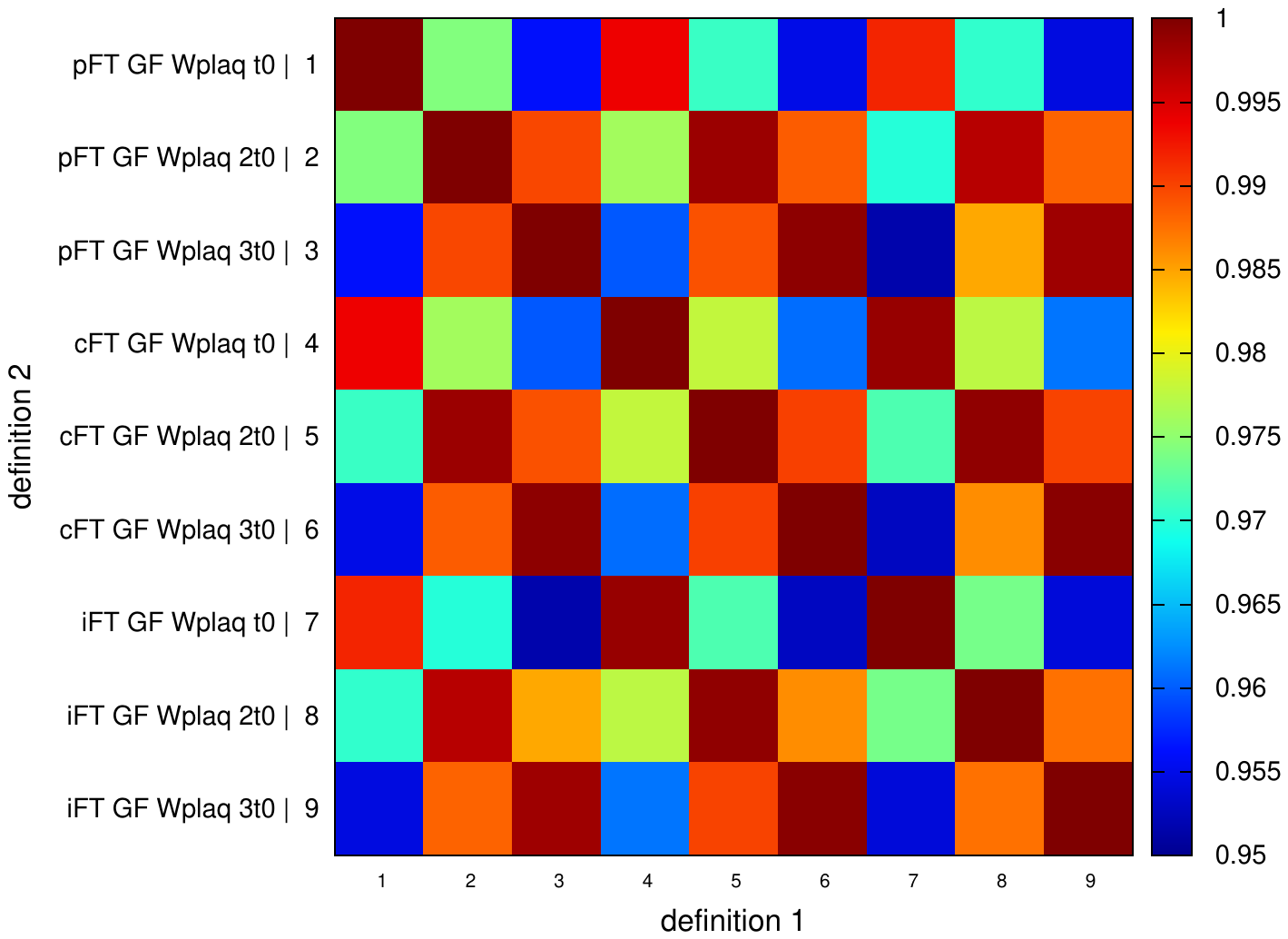}}
\end{center}
\caption{\label{fig:b40-corr3} 
Comparison of field theoretic definitions with GF smoothing and different discretizations of the topological charge operator. The correlation between different definitions is colour-coded {(note the scale is different than in Figs.~\ref{fig:b40-corr0},~\ref{fig:b40-corr1},~\ref{fig:b40-corr2},~\ref{fig:b40-corr4})}.} 
\end{figure}

\appendix{Comparison of different discretizations of the topological charge operator}
\label{sec:gf2}
In this subsection, we make another comparison of field theoretic definitions, using in all cases the gradient flow with the Wilson plaquette smoothing action, but different discretizations of the topological charge operator and different flow times:
\begin{itemize}
 \item plaquette discretization, flow times $t_0$, $2t_0$ and $3t_0$ (13, 14, 15),
 \item clover discretization, flow times $t_0$, $2t_0$ and $3t_0$ (16, 17, 18),
 \item improved discretization (clover + rectangles), flow times $t_0$, $2t_0$ and $3t_0$ (19, 20, 21).
\end{itemize}
A summary of our findings is given in Fig.~\ref{fig:b40-corr3} and Tab.~\ref{tab:corr3}.

\begin{table*}[h!]
  \centering
\begin{scriptsize}
  \begin{tabular}[]{c|ccccccccc}
& 1 & 2 & 3 & 4 & 5 & 6 & 7 & 8 & 9 \\
\hline
1 & 1 & 0.97(0) & 0.96(0) & 0.99(0) & 0.97(0) & 0.95(0) & 0.99(0) & 0.97(0) & 0.95(0) \\
2 & 0.97(0) & 1 & 0.99(0) & 0.98(0) & 1.00(0) & 0.99(0) & 0.97(0) & 1.00(0) & 0.99(0) \\
3 & 0.96(0) & 0.99(0) & 1 & 0.96(0) & 0.99(0) & 1.00(0) & 0.95(0) & 0.98(0) & 1.00(0) \\
4 & 0.99(0) & 0.98(0) & 0.96(0) & 1 & 0.98(0) & 0.96(0) & 1.00(0) & 0.98(0) & 0.96(0) \\
5 & 0.97(0) & 1.00(0) & 0.99(0) & 0.98(0) & 1 & 0.99(0) & 0.97(0) & 1.00(0) & 0.99(0) \\
6 & 0.95(0) & 0.99(0) & 1.00(0) & 0.96(0) & 0.99(0) & 1 & 0.95(0) & 0.99(0) & 1.00(0) \\
7 & 0.99(0) & 0.97(0) & 0.95(0) & 1.00(0) & 0.97(0) & 0.95(0) & 1 & 0.97(0) & 0.95(0) \\
8 & 0.97(0) & 1.00(0) & 0.98(0) & 0.98(0) & 1.00(0) & 0.99(0) & 0.97(0) & 1 & 0.99(0) \\
9 & 0.95(0) & 0.99(0) & 1.00(0) & 0.96(0) & 0.99(0) & 1.00(0) & 0.95(0) & 0.99(0) & 1 \\
\end{tabular}
\end{scriptsize}
\caption{Comparison of field theoretic definitions with GF smoothing and different discretizations of the topological charge operator. The numbers correspond to the numbering given in Fig.~\ref{fig:b40-corr3}. We give the correlation coefficient between different definitions and its error (0 means that the error is smaller than 0.005).}
  \label{tab:corr3}
\end{table*}

The correlations between different discretizations at a fixed flow time are almost perfect (99\%-100\%) and decrease with an increase of the flow time difference.
However, even the largest difference, the one between the simple plaquette discretization at flow time $t_0$ and the improved one (clover and rectangle terms) at flow time $3t_0$ yields a very high correlation (95\%).
This behaviour is totally consistent with expectations.

\begin{acknowledgements}
We thank the ETM Collaboration for generating ensembles of gauge field configurations that we have used in this work and for a very enjoyable collaboration.
We are grateful to Giancarlo Rossi for careful reading of the manuscript and useful suggestions.
AA has been supported by an internal program of the University of Cyprus under the name of BARYONS. 
KC was supported in part by the National Science Centre (Poland) grant SONATA BIS
2016/22/E/ST2/00013. 
AD acknowledges support by the Emmy Noether Programme of the DFG
(German Research Foundation), grant WA 3000/1-1.
This work was supported in part by the Helmholtz International Center for
FAIR within the framework of the LOEWE program launched by the State of
Hesse.
Numerical simulations were carried out at different machines of the Leibniz Rechenzentrum (Garching), at the Pozna\'n Supercomputing and Networking Center, at the LOEWE-CSC and the FUCHS-CSC high performance computers of the Goethe University Frankfurt (HPC-Hessen, funded by the State Ministry of Higher Education, Research and the Arts), at the PAX cluster at DESY Zeuthen and at HPC clusters at the ITP Bern partly financed by the SNSF.
We thank members of these supercomputing institutions for support.
\end{acknowledgements}

\end{document}